\documentclass[prd,amsmath,amssymb,nofootinbib,superscriptaddress,showkeys,reprint]{revtex4-2}
\usepackage{graphicx}
\usepackage{booktabs}
\usepackage{array}
\usepackage{multirow}
\usepackage{dcolumn}
\usepackage{bm}
\usepackage{cancel}

\usepackage[caption=false]{subfig}
\usepackage[table]{xcolor}
\definecolor{MAGENTA}{named}{magenta}
\usepackage[colorlinks=true, linkcolor=blue, citecolor=blue, urlcolor=blue]{hyperref}
\usepackage{orcidlink}
\usepackage{siunitx}
\newcolumntype{B}{>{\bfseries}c}

\begin{document}

\title{ Black Hole Solution in Anti-de Sitter Space with a New Cloud of Strings Surrounded by Dark Matter Halo }

\author{Faizuddin Ahmed\orcidlink{0000-0003-2196-9622}} 
\email{faizuddinahmed15@gmail.com; fahmed@rgu.ac} 
\affiliation{Department of Physics, The Assam Royal Global University, Guwahati, 781035, Assam, India}

\author{Edilberto O. Silva\orcidlink{0000-0002-0297-5747}}
\email{edilberto.silva@ufma.br (Corresp. author)}
\affiliation{Departamento de F\'{\i}sica, Universidade Federal do Maranh\~{a}o, 65085-580 S\~{a}o Lu\'{\i}s, Maranh\~{a}o, Brazil}

\date{\today}

\begin{abstract}
This paper presents a comprehensive theoretical study of a Schwarzschild-like Anti-de Sitter (AdS) black hole (BH) influenced by a new cloud of strings (NCS) and a dark matter halo (DMH) characterized by a Dehnen-type density profile. We analyze the geodesic motion of both massless and massive test particles, highlighting how the NCS and DMH parameters affect the effective potentials, photon spheres, circular orbits, BH shadow, and the innermost stable circular orbit (ISCO) of test particles. Additionally, we investigate the thermodynamic behavior of the BH  in an extended phase space by deriving key quantities, such as the Hawking temperature, the equation of state (EoS), Gibbs free energy, internal energy, and the specific heat capacity. Our results show that the presence of NCS and DMH induces significant modifications in both the dynamical and thermodynamical behavior of the BH, including shifts in the Hawking-Page transition and divergences in heat capacity, thereby reshaping the phase structure of the BH.
\end{abstract}
\maketitle

\section{Introduction}\label{Sec:I}

In general relativity (GR), black holes (BHs) emerge naturally as exact solutions to Einstein’s field equations, with the earliest being the Schwarzschild solution~\cite{Schwarzschild1916}. Over time, this family has expanded to include rotating (Kerr)~\cite{Kerr1963}, charged (Reissner-Nordström)~\cite{Reissner1916, Nordstrom1918}, and cosmological (de Sitter and anti-de Sitter)~\cite{Carter1973,HawkingPage1983} generalizations. These solutions have significantly advanced our understanding of spacetime singularities, event horizons, and causal structure~\cite{Penrose1965,Hawking1970}. Although proposed more than a century ago, BHs have only recently been confirmed observationally through two breakthroughs: the direct imaging of a supermassive BH by the Event Horizon Telescope (EHT)~\cite{EHT2019a,EHT2019b}, and the detection of gravitational waves from binary BH mergers by the LIGO-VIRGO collaboration~\cite{Abbott2016a,Abbott2016b}. These observations have not only validated key predictions of GR in the strong-field regime but also marked the dawn of gravitational-wave astronomy and BH imaging as precision tools for testing fundamental physics~\cite{Barack2019,Berti2015}.

Despite this progress, several questions remain open. These include the true nature of spacetime singularities, the microscopic origin of BH entropy~\cite{Bekenstein1973, Hawking1975}, the information paradox~\cite{Hawking1976,Mathur2009}, and the role of BHs in quantum gravity frameworks such as string theory and loop quantum gravity~\cite{StromingerVafa1996,Ashtekar2005}. Additionally, understanding how BHs interact with surrounding fields or matter, such as accretion disks, magnetic fields, dark matter halos, and cosmic defects, remains a central challenge~\cite{ZeldovichNovikov1971,Letelier1979}. In this context, modifications to the classical BH spacetimes, such as those involving surrounding perfect fluid dark matter (PFDM)~\cite{Xu2018,Konoplya2019} or topological structures like clouds of strings~\cite{Letelier1979,BHCloud2021}, offer promising avenues to explore deviations from idealized models. These extended configurations provide more realistic descriptions of astrophysical BHs and allow for direct comparison with current and future observational data from instruments like LISA, SKA, and the next-generation EHT~\cite{LISA2023,EHT2022}. Therefore, BH physics continues to serve as a unique and rich testing ground for theories of gravity, offering critical insights into both classical and quantum aspects of spacetime.

Research using the Dehnen-type dark matter (DM) halo model has provided more profound insights into the interactions between black holes (BHs) and DM. For instance, investigations into how varying DM density slopes affect the survival of star clusters following gas expulsion have been conducted \cite{AA31}. Additionally, stellar distributions have been studied using both Plummer and Dehnen profiles, revealing distinct behaviors in their central cusps. Dehnen-type DM halo solutions have also been employed to analyze ultra-faint dwarf galaxies \cite{AA33}, while new black hole solutions embedded within such halos have recently been proposed \cite{AA32,AA34}. These works explore various phenomena, including thermodynamics, null geodesics \cite{AA32}, and constraints on halo parameters \cite{AA42}. More recently, the impact of DM halos on observable features such as quasinormal modes, photon sphere radius, black hole shadows \cite{AA43,AA44}, and gravitational waveforms from periodic orbits \cite{AA45} has been investigated within the BH-Dehnen halo framework \cite{AA34}. Collectively, these studies are essential for advancing our understanding of how DM environments influence black hole spacetimes and their observable signatures.

The study of dark matter (DM) remains a pivotal challenge in contemporary physics, as it constitutes approximately 27\% of the universe’s total energy density, yet its fundamental nature remains elusive. Unraveling the properties of DM is critical not only for explaining galaxy formation and the large-scale structure of the cosmos but also for understanding its interplay with dark energy, which is responsible for the universe’s accelerated expansion. Within this framework, the perfect fluid dark matter (PFDM) model has emerged as a compelling approach to investigating DM, especially in the vicinity of black holes (BHs). Unlike conventional particle-based DM models, the PFDM paradigm treats dark matter as a continuous, non-viscous fluid characterized by specific equations of state that govern its dynamical behavior. This fluid description enables a novel perspective on how DM influences spacetime geometry around BHs and modifies their physical properties. For example, Qiao et al.~\cite{CKQ2023} applied the PFDM model to analyze DM clustering around BHs, revealing significant deviations in BH metrics compared to classical vacuum solutions. Notably, the PFDM framework predicts modifications to various black hole solutions, including Kerr \cite{ZX2018,ZH2018,MR2019}, Schwarzschild \cite{JR2021,BH2025-1}, Reissner-Nordström \cite{ZH2019}, Bardeen \cite{HXZ2021}, and Euler-Heisenberg BHs \cite{SJM202024,BH2025-2}. These altered metrics lead to measurable effects across several astrophysical and gravitational phenomena. For instance, gravitational waveforms emitted by BHs embedded in PFDM halos exhibit distinct signatures compared to standard predictions \cite{SS2021,LL2022}, potentially providing new observational windows into the nature of DM. Further investigations into the thermodynamics and stability of PFDM-modified BHs have revealed novel phase transition behaviors \cite{SH2020, AK2023, AS2024}, which are absent in vacuum solutions. These studies deepen our understanding of BH microphysics in DM environments and offer insights into the fundamental interactions governing these systems. The PFDM model also impacts the propagation of light and matter near BHs. Gravitational lensing analyses \cite{FA2021} indicate altered deflection angles and lensing patterns due to the presence of the PFDM halo. At the same time, accretion disc dynamics around BHs are likewise affected, modifying observable electromagnetic spectra \cite{MH2023}. Shadow images of BHs, a key observational target, have been shown to vary significantly under PFDM influence \cite{SH2019,TC2021}, providing testable predictions for current and future high-resolution observations. Moreover, the PFDM environment modifies key parameters such as photon sphere radii and deflection angles \cite{FA2022}, and influences the quasinormal mode spectra that characterize BH ringdown signals \cite{HC2024}. These alterations extend to greybody factors, which affect BH radiation and particle emission rates \cite{MM2022}, as well as to the thermodynamic properties of BH horizons \cite{HA2023,RN2023,GR2023}. Recent work also explores the changes to event horizon structures caused by PFDM \cite{YF2024}, highlighting the broad-reaching consequences of treating DM as a perfect fluid. Collectively, the PFDM model presents a promising and versatile alternative to traditional dark matter theories, particularly in elucidating how DM influences black hole spacetimes and related observables across galactic and cosmological scales. By linking cosmological DM properties with strong-field gravity phenomena, PFDM enriches our toolkit for probing the dark sector and advancing our comprehension of the universe’s most enigmatic components.

In this work, we explore the Schwarzschild-AdS BH spacetime modified by two physically motivated external components: a generalized cloud of strings and a surrounding perfect fluid dark matter (PFDM) distribution. These modifications are not merely theoretical constructs; both the cloud of strings and PFDM have well-established foundations in cosmology and astrophysics. String clouds can be viewed as topological defects that may have formed during early universe phase transitions, while PFDM plays a vital role in explaining the flat rotation curves of galaxies and the large-scale structure of the universe. Incorporating these structures into the BH geometry enables a more realistic modeling of BH environments, particularly those influenced by PFDM and cosmic defects. We begin by formulating a modified spacetime metric that captures the combined effects of the cloud of strings and the PFDM. We then analyze how these modifications impact the underlying geometry. Subsequently, we study the geodesic structure of this spacetime by examining both null (photon) and timelike (massive particle) geodesics. This allows us to probe the influence of the surrounding matter on key observational phenomena such as gravitational lensing, photon spheres, BH shadows, orbital stability, and test particle motion with innermost stable circular orbits (ISCO). Special attention is also given to the analysis of photon rings, which play a central role in determining the observable shadow of the BH. These features have become increasingly relevant due to direct imaging of supermassive BHs by the Event Horizon Telescope (EHT). Furthermore, we investigate the thermodynamic properties of the modified BH solution. We derive key thermodynamic quantities, including the Hawking temperature, entropy, Gibbs free energy, and specific heat. Within the framework of extended phase space thermodynamics, where the cosmological constant is treated as a thermodynamic pressure, we formulate the first law and derive the corresponding equation of state. Finally, we analyze the thermodynamic stability of the system by studying the behavior of the heat capacity and identifying possible phase transitions, such as the Hawking-Page transition. Our results show how the presence of the string cloud and PFDM modifies the thermodynamic phase structure, potentially giving rise to critical phenomena analogous to those observed in van der Waals fluids.

This paper is organized as follows: In Section \ref{Sec:II}, we present the background geometry of an AdS BH incorporating a new cloud of strings and surrounded by a dark matter halo. In Section \ref{Sec:III}, we investigate the geodesic motion of both massless photons and massive test particles in the vicinity of the BH. We also analyze the topological features of photon rings and discuss the implications that follow. In Section \ref{Sec:IV}, we examine the thermodynamic properties of the BH system by treating the BH mass as enthalpy and deriving the associated thermodynamic variables. Furthermore, we explore a generalized form of the first law of thermodynamics and introduce additional thermodynamic parameters. Finally, in Section \ref{Sec:V}, we provide our concluding remarks and discuss the main findings of the study.

\section{Background Geometry: Schwarzschild-AdS BH spacetime with a NCS and a Dark-Matter Halo}\label{Sec:II}

Throughout this section, we set $8\pi G=1=c$, so Einstein’s equations read $G_{\mu\nu}+\Lambda g_{\mu\nu}=T_{\mu\nu}$. Here, we consider a Schwarzschild-like BH spacetime surrounded by a DM halo characterized by a Dehnen-type density profile. Moreover, the BH solution is coupled with a cloud of strings. In Ref.~\cite{UU}, the authors presented the BH spacetime involving the DM distribution. The resulting spacetime metric describing the BH-DM solution 
\begin{align}
ds^2&=-\left(1-\frac{2M}{r}-8\pi\,\rho_s\,r_s^2\,\ln\!\left(1+\frac{r_s}{r}\right)\right)\,dt^2\notag \\&+\left(1-\frac{2M}{r}-8\pi\,\rho_s\,r_s^2\,\ln\!\left(1+\frac{r_s}{r}\right)\right)^{-1}\,dr^2\notag\\&+r^{2}\,\left(d\theta ^{2}+\sin ^{2}\theta d\phi ^{2}\right).\label{aa1}
\end{align}
Here, $\rho_s$ and $r_s$ are the characteristic density and characteristic scale of the DM halo, respectively.

Assuming there is no direct coupling between the new cloud of strings and the DM halo, we consider a static and spherically symmetric geometry described by the following line element:
\begin{eqnarray}
ds^2=-f(r)\,dt^2+\frac{dr^2}{f(r)}+r^{2}\left(d\theta ^{2}+\sin ^{2}\theta d\phi ^{2}\right),\label{metric}
\end{eqnarray}
with metric function
\begin{align}
f(r)&=1-\frac{2M}{r}-8\pi\,\rho_s\,r_s^2\,\ln\!\left(1+\frac{r_s}{r}\right)\notag\\&+\frac{\lvert \alpha\rvert\, b^2}{r^2}\,{}_2F_1\left(-\frac{1}{2},-\frac{1}{4},\frac{3}{4},-\frac{r^4}{b^4}\right)+\frac{r^2}{\ell^2_p}.\label{function-1}
\end{align}
Here $(|\alpha|,b)$ represents the NCS parameters \cite{Letelier1979,GA,AK,arxiv1,arxiv2}. It is noted that the procedure for incorporating NCS into BH solutions can be followed as outlined in \cite{AK, arxiv1}.

\begin{figure}[ht!]
\includegraphics[width=0.87\linewidth]{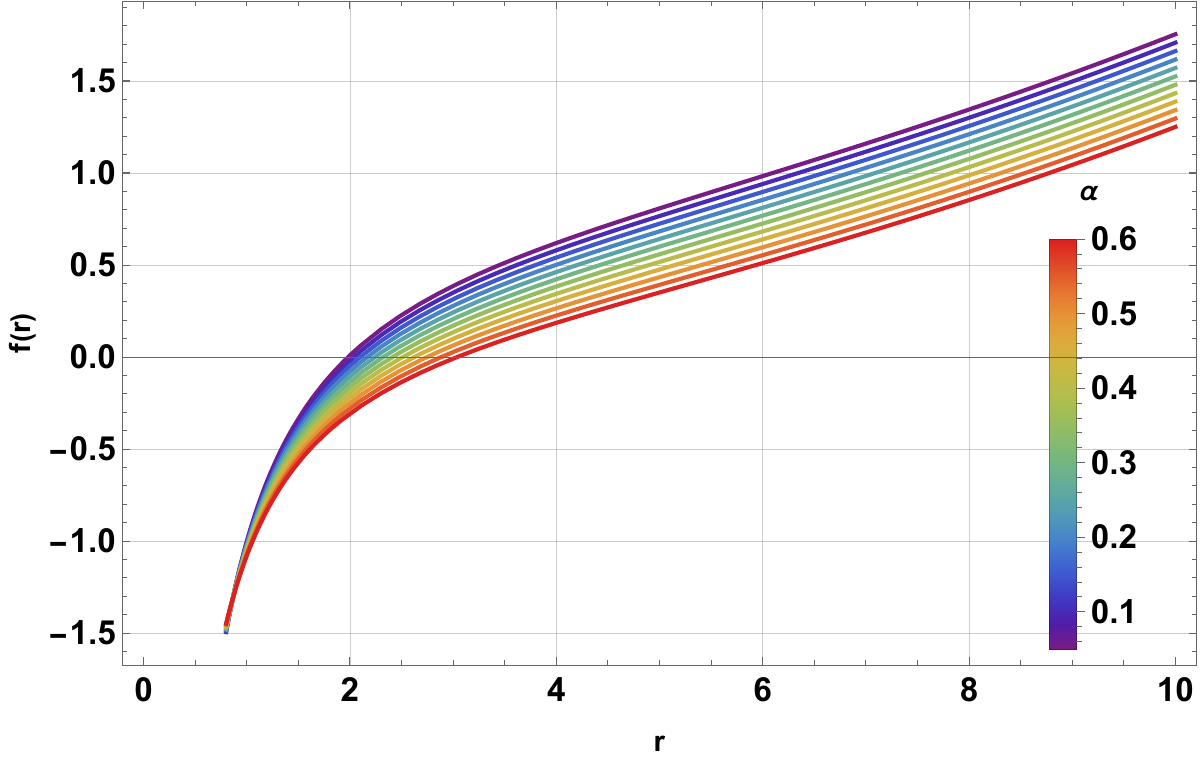}\\
(i) $r_s=0.2,\rho_s=0.02,b=0.5$\\
\includegraphics[width=0.86\linewidth]{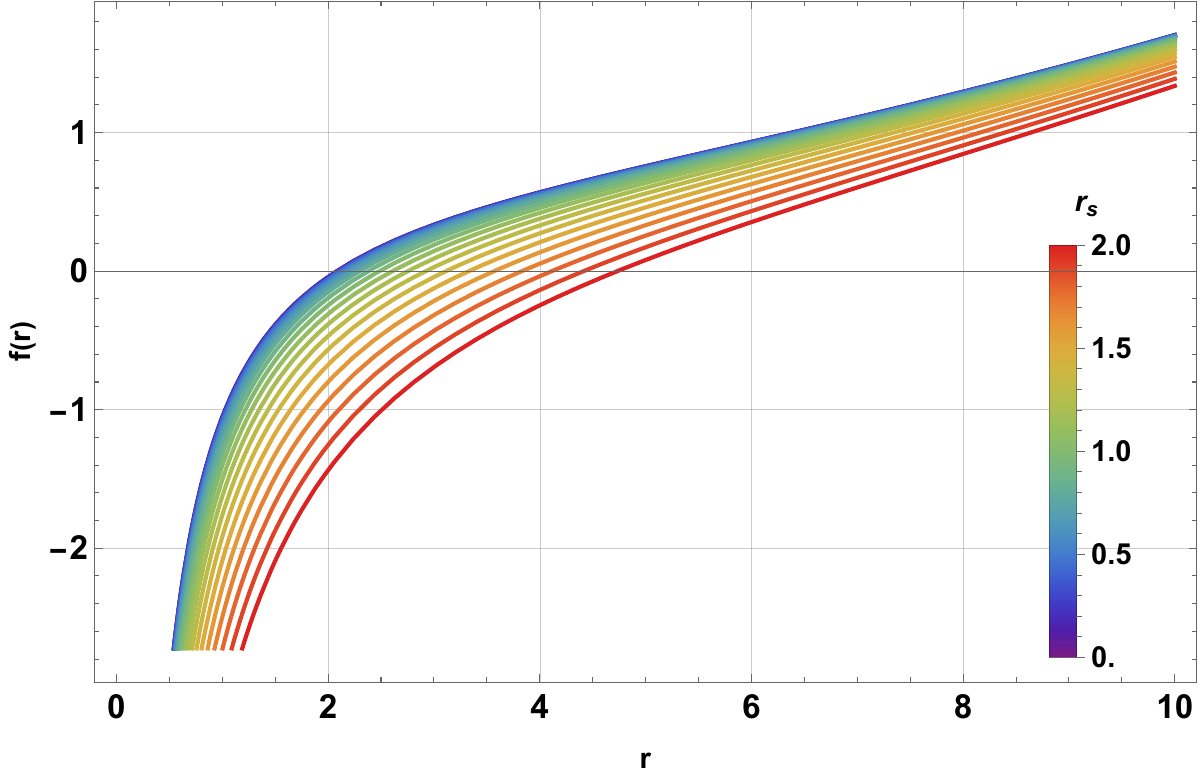}\\
(ii) $\alpha=0.1,\rho_s=0.02,b=0.5$\\
\includegraphics[width=0.86\linewidth]{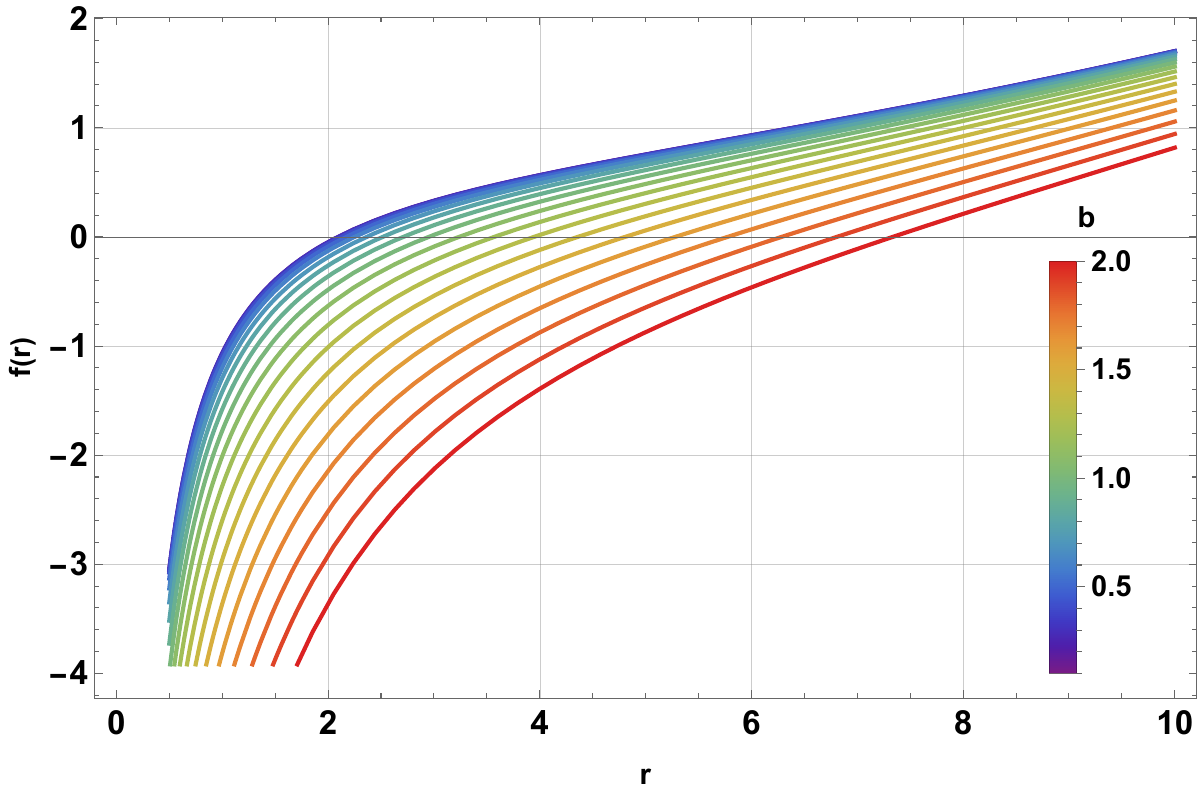}\\
(iii) $\alpha=0.1,\rho_s=0.05,r_s=0.2$\\
\caption{\footnotesize The metric function $f(r)$ as a radial coordinate $r$ for various values of NCS parameter $\alpha$ and the halo radius $r_s$. Here $M=1,\,\rho_s=0.02,\,\ell_p=10,\,b=0.5$.}
\label{fig:metric}
\end{figure}

In the absence of a DM halo profile, {\it i.e.}, when \( r_s \to 0 \) and \( \rho_s \to 0 \), the metric function simplifies to the form
\begin{align}
f(r) = 1 - \frac{2\,M}{r} + \frac{|\alpha|\, b^2}{r^2} \, {}_2F_1\left(-\frac{1}{2}, -\frac{1}{4}, \frac{3}{4}, -\frac{r^4}{b^4}\right) + \frac{r^2}{\ell^2_p}, \label{function-2}
\end{align}
where \({}_2F_1\) denotes the Gauss hypergeometric function. The solution with this function (\ref{function-2}) represents an AdS BH configuration with new string clouds and has been discussed in detail in \cite{GA}.

Furthermore, in the limit \( b \to 0 \) and \( |\alpha| = \alpha \), the metric function \( f(r) \) simplifies to the form:
\begin{equation}
f(r) = 1 - \alpha - \frac{2M}{r} - 8\pi\,\rho_s\,r_s^2\,\ln\!\left(1+\frac{r_s}{r}\right) + \frac{r^2}{\ell^2_p}, \label{function-3}
\end{equation}
which corresponds to the Letelier BH solution in AdS space-time with a DM halo \cite{arxiv}. Notably, in the limit \( \alpha \to 0 \), this function further reduces to the BH solution surrounded by a DM halo, as previously reported in \cite{UU}.

\section{Geodesic Analysis of BH}\label{Sec:III}

Geodesic motion offers the most direct description of how photons and test particles move in the strong gravitational fields of black holes, revealing the underlying spacetime geometry through observable phenomena. In static, spherically symmetric spacetimes, the inherent symmetries result in conserved quantities, such as energy and angular momentum. These conserved quantities classify particle dynamics into bound, plunging, and scattering trajectories for massive particles, and into capture or escape trajectories for photons \cite{wald}. For timelike geodesics, stable circular orbits can exist down to the innermost stable circular orbit (ISCO); below this point, even small perturbations lead to an inevitable plunge into the black hole. Phenomena such as periapsis precession and strong gravitational light deflection emerge naturally from this framework, and form the theoretical basis for relativistic precession of stellar orbits, motion of hotspots in accretion flows, and gravitational lensing. In the case of null geodesics, there exists an unstable photon region, commonly referred to as the photon sphere in spherically symmetric spacetimes, which serves as a separatrix between photons that escape to infinity and those that fall into the black hole \cite{chandra}. The critical impact parameters associated with this photon region define the boundary of the black hole shadow and shape the characteristics of strong gravitational lensing. Overall, geodesic analysis provides a powerful bridge between spacetime geometry and observational data, offering a unified language for interpreting lensing features, black hole shadows, relativistically broadened spectral lines, and variability in emission near compact objects.

\subsection{Effective Potential}

Considering a test particle moving in the vicinity of the gravitational field generated by the BH space-time given in Eq.~\eqref{metric}. Expressing the line-element (\ref{metric}) as  $ds^2=g_{\mu\nu}\,dx^{\mu}\,dx^{\nu}$, where $g_{\mu\nu}$ is the metric tensor, the Lagrangian density function can be written as,
\begin{align}
    \mathcal{L}&=\frac{1}{2}\,g_{\mu\nu}\,\frac{dx^{\mu}}{d\lambda}\,\frac{dx^{\nu}}{d\lambda}\notag\\&=\frac{1}{2}\,\left[-f(r)\,\dot{t}^2+\frac{\dot{r}^2}{f(r)}+r^{2}\left(\dot{\theta}^{2}+\sin ^{2}\theta\, \dot{\phi} ^{2}\right)\right]=\frac{\epsilon}{2}.\label{bb1}
\end{align}
Here, $\dot{x}^{\mu}=\frac{dx^{\mu}}{d\lambda}$ with $\lambda$ is an affine parameter. Here, $\epsilon=0$ for a null geodesic and $-1$ for a time-like.

The generalized momenta can be defined  as
\begin{equation}
    p_{\mu}=\frac{\partial \mathcal{L}}{\partial \dot{x}^{\mu}}.\label{bb2}
\end{equation}

Using Eq.~(\ref{bb1}), we can obtain the following components as,
\begin{align}
p_t&=\frac{\partial \mathcal{L}}{\partial \dot{t}}=-f(r)\,\dot{t}=-\mathrm{E}, \label{bb3}\\
p_r&=\frac{\partial \mathcal{L}}{\partial \dot{r}}=\frac{\dot{r}}{f(r)},\label{bb4}\\
p_{\theta}&=\frac{\partial \mathcal{L}}{\partial \dot{\theta}}=r^2\,\dot{\theta},\label{bb5}\\
p_{\phi}&=\frac{\partial \mathcal{L}}{\partial \dot{\phi}}=r^2\,\sin^2 \theta\,\dot{\phi}=\mathrm{L},\label{bb6}
\end{align}
Physically, $\mathrm{E}$ is the total conserved energy as measured at infinity, while $\mathrm{L}$ is the conserved angular momentum about the symmetry axis.

Without loss of generality, we restrict the motion of test particles in the equatorial plane, defined by $\theta=\pi/2$ and $\dot{\theta}=0$. Using the normalization condition $g_{\mu\nu}\,\dot{x}^{\mu}\,\dot{x}^{\nu}=\epsilon$, the particle motion can be obtained as 
\begin{align}
\dot{r}^2+V_\text{eff}(r)=\mathrm{E}^2,\label{bb7}
\end{align}
where $V_\text{eff}(r)$ is the effective potential of the system and is given by
\begin{align}
V_\text{eff}(r)&=\left(-\epsilon+\frac{\mathrm{L}^2}{r^2}\right)\,f(r)\notag\\&
=\left(-\epsilon+\frac{\mathrm{L}^2}{r^2}\right)\,\Bigg[1-\frac{2M}{r}-8\pi\,\rho_s\,r_s^2\,\ln\!\left(1+\frac{r_s}{r}\right)\notag\\&+\frac{\lvert \alpha\rvert\, b^2}{r^2}\,{}_2F_1\left(-\frac{1}{2},-\frac{1}{4},\frac{3}{4},-\frac{r^4}{b^4}\right)+\frac{r^2}{\ell^2_p}\Bigg].\label{bb8}
\end{align}
Obviously, when $\epsilon=0$, it corresponds to the effective potential governing photon dynamics, and when $\epsilon=-1$, it corresponds to the effective potential governing timelike geodesics.

One can see that BH mass $M$, the curvature radius $\ell_p$, the new cloud of strings parameters $(\alpha, b)$, and the DMH parameters $(r_s,\rho_s)$ significantly influence the curvature of space-time. Consequently, these parameters reshape the effective potential, which in turn governs the dynamics of photons and massive test particles compared to the standard Schwarzschild BH cases.

\begin{figure}[ht!]
    \centering
    \includegraphics[width=0.88\linewidth]{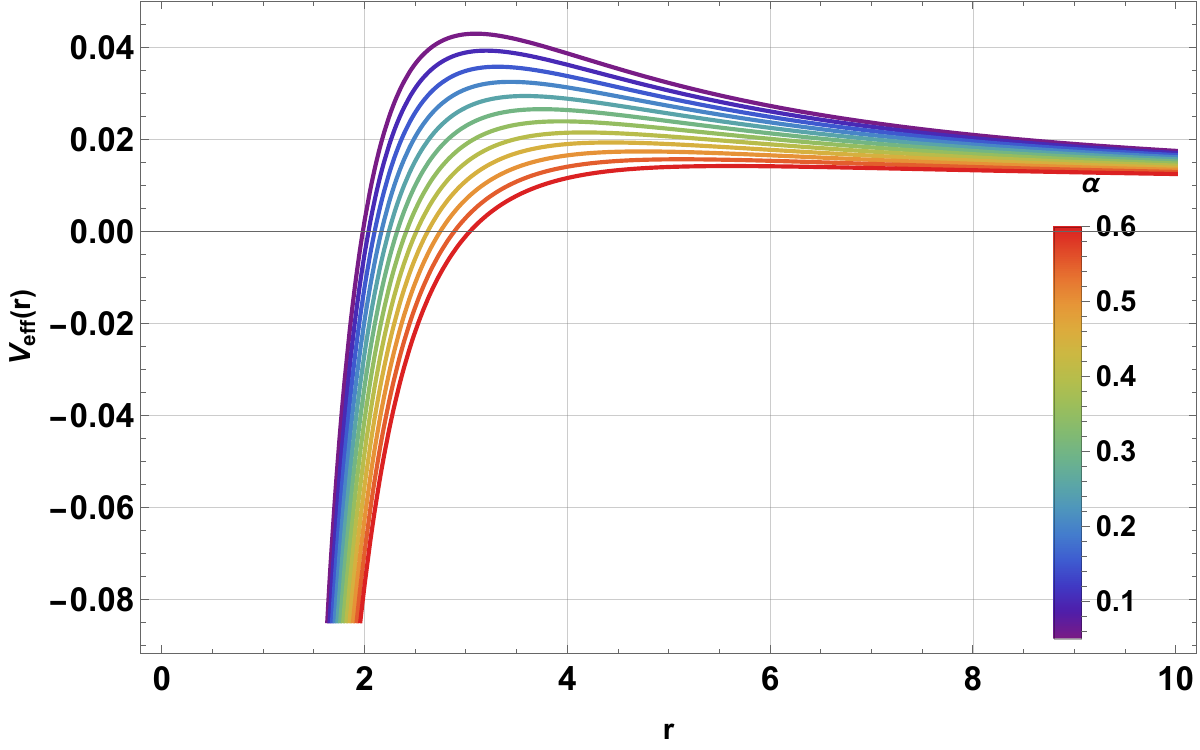}\\
    (i) $r_s=0.2\,b=0.5,\rho_s=0.02$ \\
    \includegraphics[width=0.88\linewidth]{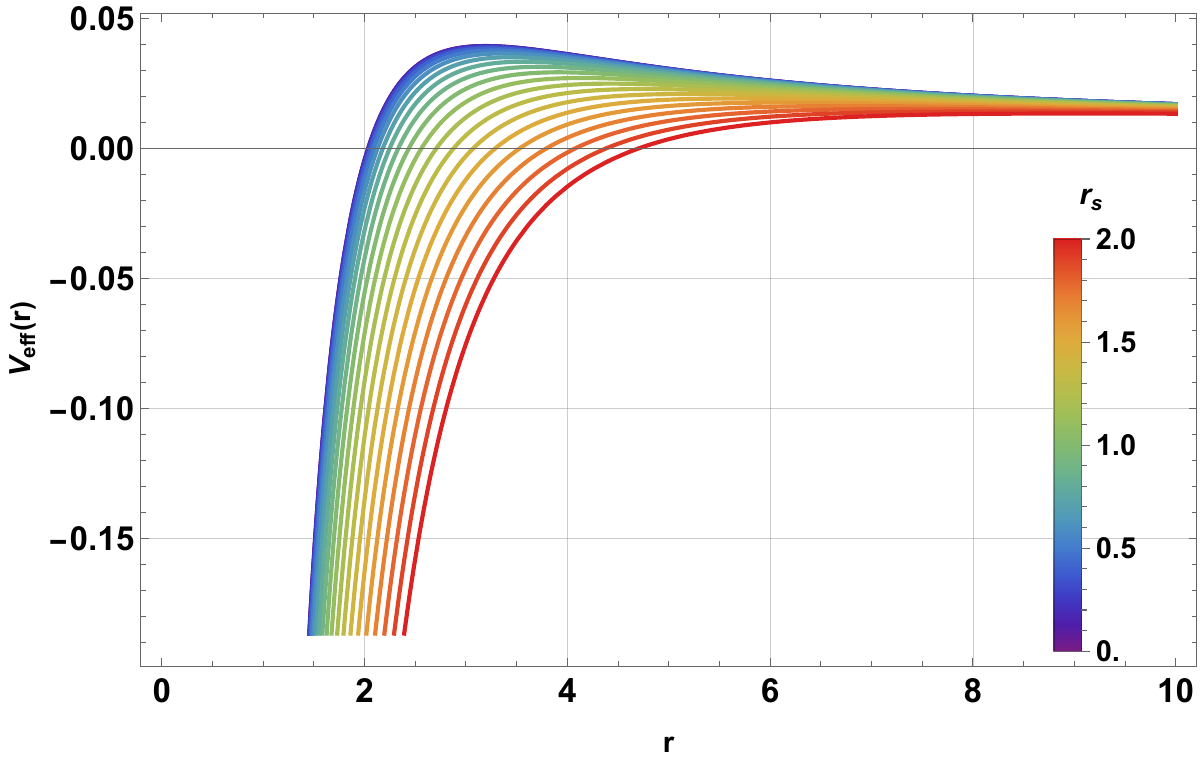}\\
    (ii) $\alpha=0.1,\rho_s=0.02,b=0.5$\\
    \includegraphics[width=0.88\linewidth]{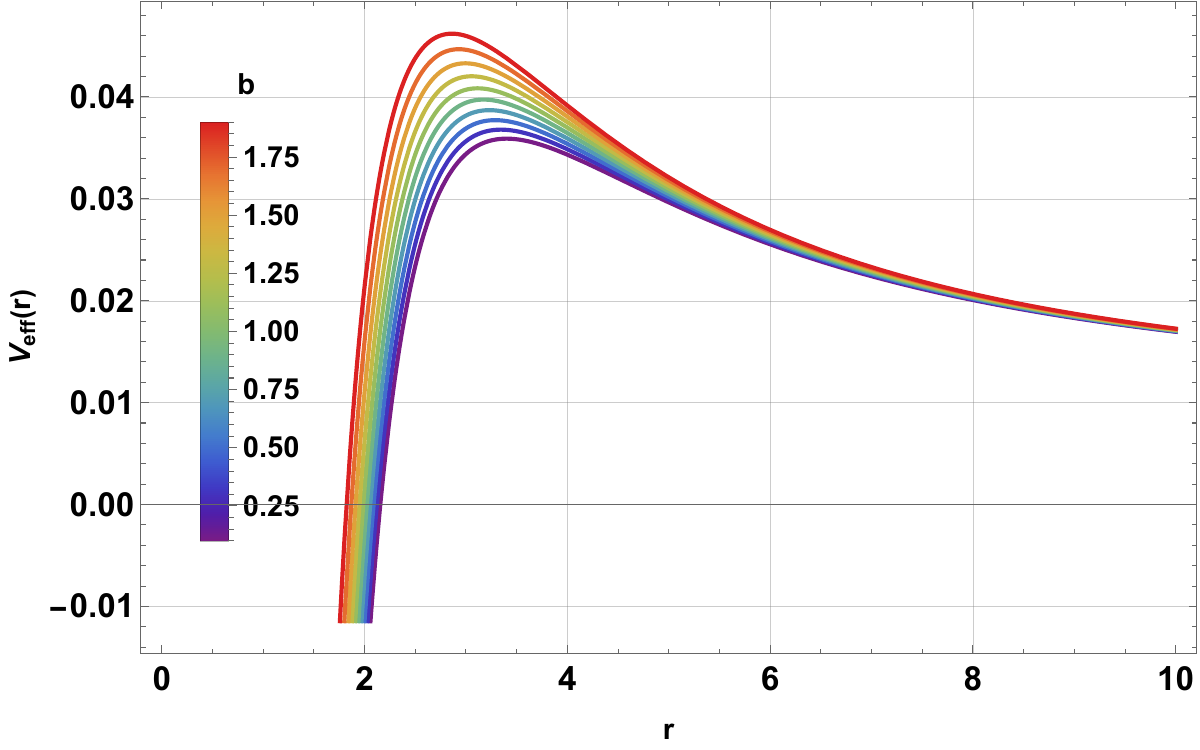}\\
    (iii) $\alpha=0.1,r_s=0.5,\,\rho_s=0.02$\\
    \caption{\footnotesize Behavior of the effective potential governs the photon dynamics as a function of $r$ for various values of NCS parameter $\alpha$ and the halo radius $r_s$. Here $M=1,\,\ell_p=10,\,\mathrm{L}=1$.}
    \label{fig:potential-null}
\end{figure}

\subsection{\large {\bf Constraint on parameters from EHT observations}}

The Event Horizon Telescope (EHT) collaboration has successfully published observational data for the supermassive BHs M87* \cite{EHT2019a,EHT2019b,EHT2,EHT3,EHT4,EHT5} and Sgr A* \cite{EHT2022,EHT8}. These groundbreaking observations show the characteristic BH "shadow" a central dark region formed due to the intense gravitational field, which prevents photons from escaping. These results provide critical visual evidence for exploring BH physics in the strong-field regime of gravity. Moreover, the features of BH shadows offer powerful experimental tools for testing general relativity and various alternative theories of gravity. For instance, several studies have used shadow properties to constrain modified gravity models \cite{AA,OY} and to place bounds on BH model parameters \cite{YU,QT,NUM}.

In this section, we use the EHT observational data to perform a parameter space constraint analysis of the Schwarzschild-like BH with a DMr halo and a cloud of strings (SH-DMH-CS). By treating the SH-DMH-CS as a theoretical candidate for M87* and Sgr A*, we systematically compare model predictions with EHT observations to place stringent constraints on its parameter space. This analysis allows us to determine the viable ranges for the dark matter halo and cloud of strings parameters, thereby evaluating the astrophysical plausibility of such black holes in light of current observational data.

In spherically symmetric spacetimes, photon motion is governed by an effective potential that determines whether photons fall into the black hole, escape to infinity, or orbit temporarily within the unstable photon sphere \cite{chandra,wald}. This instability leads to significant observational effects, especially in strong gravitational lensing, where light can loop around the black hole multiple times \cite{perlick}. The photon sphere’s critical impact parameter sets the angular size of the black hole shadow. 

For the motion of photons, they follow null geodesics, $\varepsilon=0$, the effective potential from Eq. (\ref{bb8}) reduces as
\begin{align}
V_\text{eff}(r)&=\frac{\mathrm{L}^2}{r^2}\,\Bigg(1-\frac{2M}{r}-8\pi\,\rho_s\,r_s^2\,\ln\!\left(1+\frac{r_s}{r}\right)\notag\\&+\frac{\lvert \alpha\rvert\, b^2}{r^2}\,{}_2F_1\left(-\frac{1}{2},-\frac{1}{4},\frac{3}{4},-\frac{r^4}{b^4}\right)+\frac{r^2}{\ell^2_p}\Bigg).\label{bb10}
\end{align}
The potential given in Eq.~(\ref{bb10}) governs the dynamics of photons. With the help of this potential, we will discuss photon trajectories, the effective radial force experienced by photons, the photon sphere, and the shadow cast by the BH, and analyze the outcomes.

\begin{center}
\large{\bf I.\, Photon Trajectories}
\end{center}

Photon trajectories describe the paths followed by massless particles, such as photons, through spacetime, particularly under the influence of gravity. Photons move along null geodesics, which are curves for which the spacetime interval is zero. These paths are profoundly influenced by the curvature of spacetime caused by massive objects or compact objects.

The equation of the orbit using Eqs. (\ref{bb5}) and (\ref{bb6}) and finally employing (\ref{bb10}) is given by
\begin{align}
\left(\frac{1}{r^2}\,\frac{dr}{d\phi}\right)^2+\frac{1}{r^2}&=\frac{1}{\beta^2}-\frac{1}{\ell^2_p}+\frac{2M}{r^3}+\frac{8\pi\,\rho_s\,r^2_s}{r^2}\,\mbox{ln}\left(1+\frac{r_s}{r}\right)\notag\\&-\frac{1}{r^2}\,\frac{\lvert \alpha\rvert\, b^2}{r^2}\,{}_2F_1\left(-\frac{1}{2},-\frac{1}{4},\frac{3}{4},-\frac{r^4}{b^4}\right).\label{bb11}
\end{align}
Performing a transformation to a new variable via $r=1/u$ into the above equation results
\begin{align}
\left(\frac{du}{d\phi}\right)^2+u^2&=\frac{1}{\beta^2}-\frac{1}{\ell^2_p}+2\,M\, u^3+8\pi\,\rho_s\,r^2_s\,u^2\,\mbox{ln}(1+r_s\,u)\notag\\&-u^4\,\lvert \alpha\rvert\, b^2\,{}_2F_1\left(-\frac{1}{2},-\frac{1}{4},\frac{3}{4},-\frac{1}{b^4\,u^4}\right).\label{bb12}
\end{align}
Differentiating both sides w. r. to $\phi$ and after simplification results
\begin{align}
\frac{d^2 u}{d\phi^2} + u &= 3 M u^2 + 8\pi \rho_s r_s^2 \left[ u \ln(1 + r_s u) + \frac{u^2 r_s}{2(1 + r_s u)} \right] \notag\\&
- 2\, |\alpha|\, b^2\, u^3 \, {}_2F_1\left(-\tfrac{1}{2}, -\tfrac{1}{4}, \tfrac{3}{4}, -\frac{1}{b^4 u^4} \right)\nonumber\\ 
&- \frac{|\alpha|}{3\, b^2\, u} \, {}_2F_1\left(\tfrac{1}{2}, \tfrac{3}{4}, \tfrac{7}{4}, -\frac{1}{b^4 u^4} \right).\label{bb13}
\end{align}
Equation (\ref{bb13}) represents a nonlinear second-order differential equation governing photon trajectories in the given gravitational field. It is evident that the DMH profile, characterized by $(r_s,\rho_s)$, together with the NCS parameters $(|\alpha|,b)$, significantly influences the photon trajectories and consequently modifies the geodesic paths of light propagating around the BH.

\begin{figure*}[tbhp]
  \centering
  \includegraphics[width=0.85\linewidth]{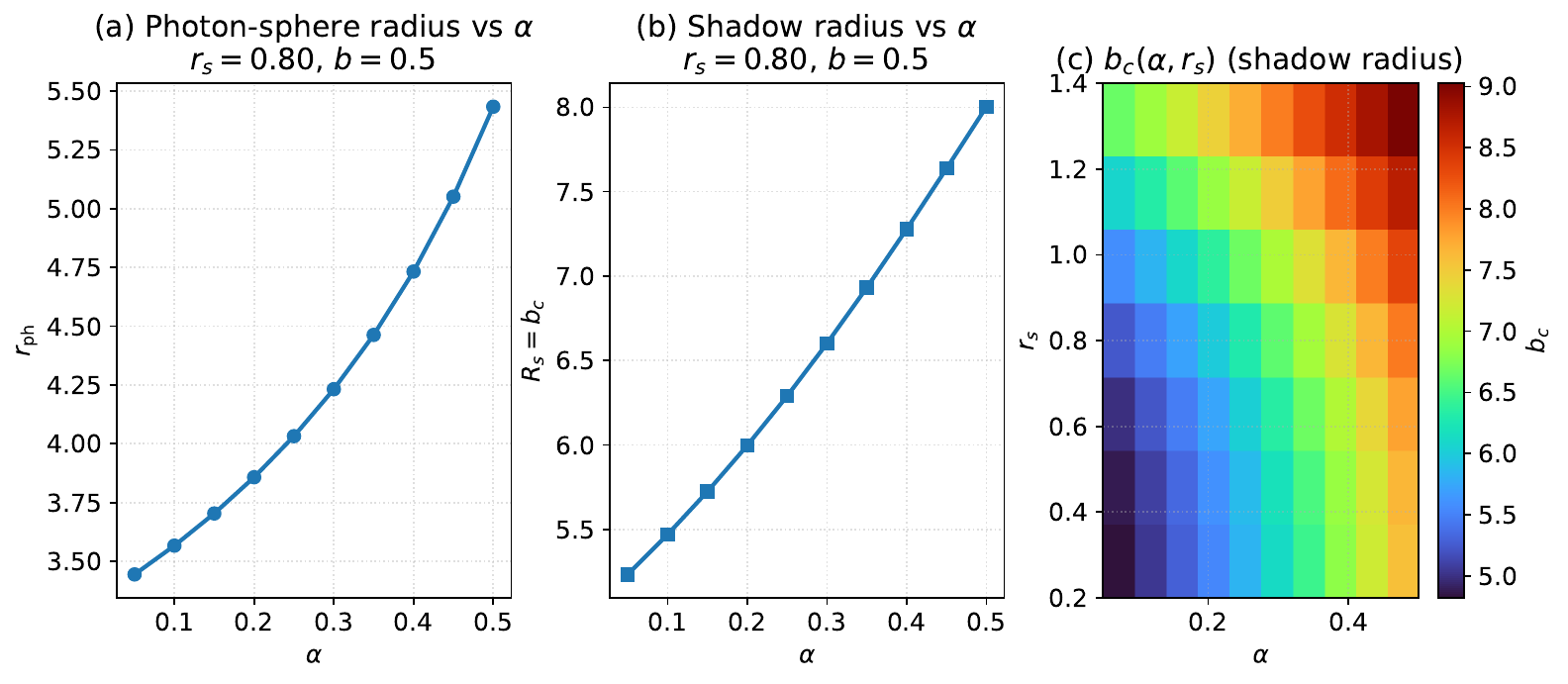}
  \caption{Critical scales from the full model. 
  (a) Photon–sphere radius $r_{\rm ph}$ as a function of the nonlocal parameter $\alpha$ at fixed halo scale $r_s=0.80$ and coupling $b=0.50$. 
  (b) Shadow radius $R_s=b_c$ (the critical impact parameter) for the same setup. 
  (c) Heatmap of $b_c(\alpha,r_s)$ showing how the shadow size varies jointly with $\alpha$ and $r_s$. 
  Unless stated otherwise, the background density remains constant.}
  \label{fig:criticals}
\end{figure*}
Figure~\ref{fig:criticals} summarizes how the critical lensing scales respond to the nonlocal parameter $\alpha$ and to the halo length scale $r_s$ in the general model. 
In Figs. \ref{fig:criticals}(a) and \ref{fig:criticals}(b) consider a representative halo ($r_s=0.80$) and coupling $b=0.50$, and show that both the photon–sphere radius $r_{\rm ph}$ and the shadow radius $R_s=b_c$ increase monotonically with $\alpha$. 
This reflects the $\alpha$–dependent terms in the effective potential: as $\alpha$ grows, the unstable circular null orbit moves outward, which in turn pushes the critical impact parameter to larger values. Figure \ref{fig:criticals}(c) maps the critical impact parameter $b_c$ across the $(\alpha,r_s)$ plane. 
At fixed $r_s$, $b_c$ increases with $\alpha$ (consistent with \ref{fig: criticals}(b)); at fixed $\alpha$, larger $r_s$ also leads to a larger $b_c$, indicating that more extended halos enlarge the shadow. 
Together, the three panels capture the coherent trend that both the photon sphere and the shadow expand when either the nonlocal coupling is strengthened (larger $\alpha$) or the halo scale increases (larger $r_s$), while all other physical inputs are held constant.

\begin{figure}[tbhp]
  \centering
  \includegraphics[width=0.85\linewidth]{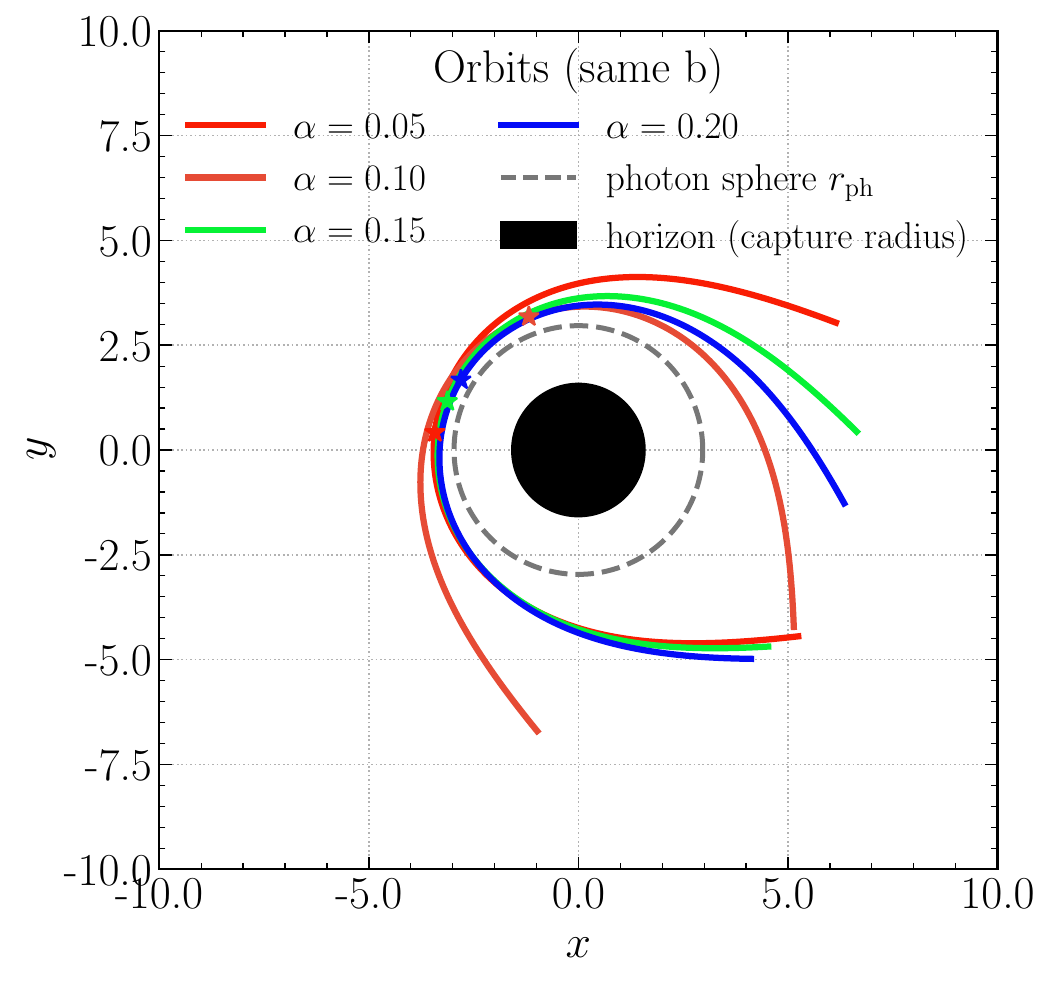}
  \caption{Light--ray trajectories from the full Eq.~(\ref{bb13}) for four values of the nonlocal parameter
  $\alpha\in\{0.05,\,0.10,\,0.15,\,0.20\}$, computed at fixed mass/halo/background and common $b$.
  The dashed circle marks the photon sphere $r_{\rm ph}$ (representative when the four cases are overplotted), and the black disk indicates the effective horizon/capture radius $r_{\rm cap}$.
  Colored star markers (one per curve) indicate the pericentre $r_{\min}$ and are plotted in the same color as their corresponding orbit.
  Physical parameters held fixed across all curves: $M=1.0$, $\ell_p=10.0$, $r_s=0.50$, $\rho_s=2.0\times10^{-2}$, $b=1.50$.
  For each $\alpha$ the pericentre is chosen close to the photon sphere, $r_{\min}\simeq1.003\,r_{\rm ph}$.}
  \label{fig:multi_alpha_orbits}
\end{figure}

Figure~\ref{fig:multi_alpha_orbits} shows single--arc light trajectories obtained from the full Eq.~(\ref{bb13}) for four values of the nonlocal parameter $\alpha$, with all other physical inputs fixed.
For each $\alpha$ we determine the photon-sphere radius $r_{\rm ph}$ and the corresponding critical slope that separates capture from scattering, and we launch rays just above this threshold so that the pericentre lies near $r_{\rm ph}$, $r_{\min}\simeq1.003\,r_{\rm ph}$.
The dashed circle indicates $r_{\rm ph}$ (drawn in a representative way when curves are overplotted) and the black disk denotes the effective capture radius $r_{\rm cap}$.
The colored star symbols mark $r_{\min}$ for each case; each star uses the same color as its orbit to facilitate cross-identification.

A clear monotonic trend emerges: as $\alpha$ increases (at fixed $b$ and background), the pericentre shifts outward and the total deflection decreases, yielding wider exit angles and shorter whirls around $r_{\rm ph}$.
This behavior reflects the $\alpha$-dependent contributions in Eq.~(\ref{bb13}), which effectively soften the net attraction in the strong-field regime while preserving the qualitative structure of null geodesics near the photon sphere.

\begin{figure}[tbhp]
\centering
\includegraphics[width=0.85\linewidth]{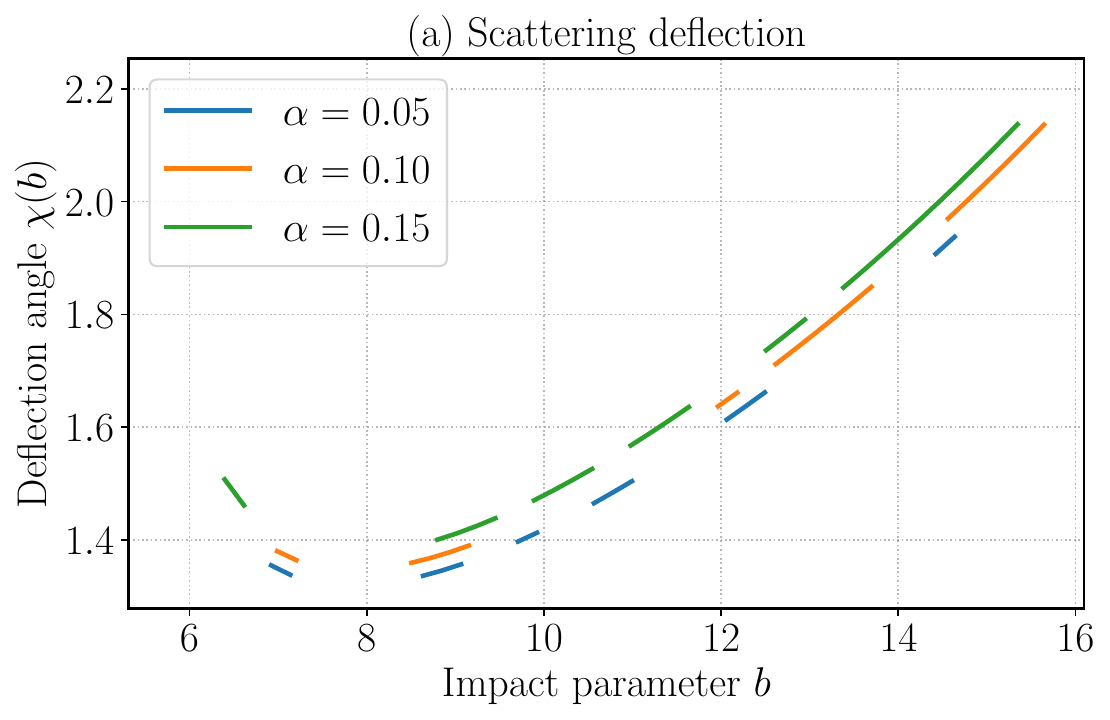}
\includegraphics[width=0.85\linewidth]{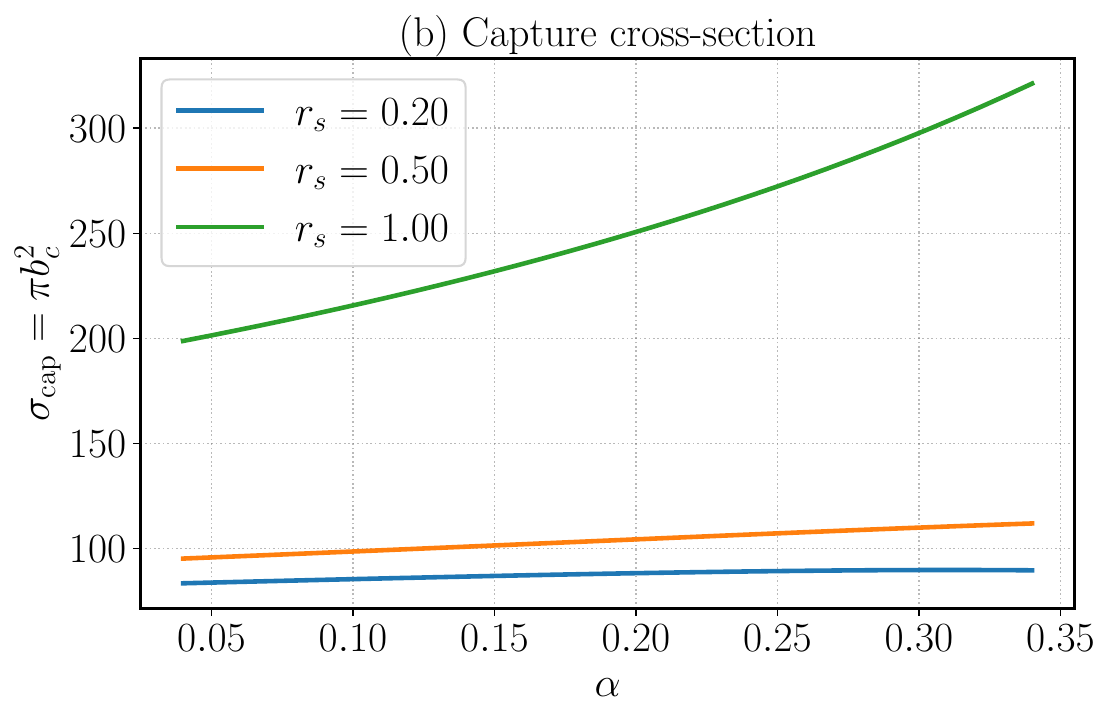}
\caption{\footnotesize
Scattering deflection and capture cross-section in the full model (hypergeometric NCS $+$ Dehnen halo $+$ AdS).
(a): deflection angle $\chi(b)$ vs.\ impact parameter $b$ for $\alpha\in\{0.05,0.10,0.15\}$ at fixed halo scale (e.g.\ $r_s=0.20$) and the remaining parameters as in this section. Curves are shown only for $b>b_c(\alpha)$, where $b_c$ is the critical impact parameter for photon capture; as $b\!\downarrow\!b_c$, the deflection grows rapidly, whereas for $b\!\gg\!b_c$ it becomes slowly varying. Increasing $\alpha$ (stronger NCS sector) raises $b_c$ and yields larger $\chi(b)$ at fixed $b$. (b): capture cross–section $\sigma_{\rm cap}(\alpha)=\pi b_c(\alpha)^2$ scanned over $\alpha$ for three halo scales $r_s\in\{0.20,\,0.50,\,1.00\}$. Both a stronger string cloud (larger $\alpha$) and a more extended halo (larger $r_s$) increase $b_c$ and hence enlarge $\sigma_{\rm cap}$, quantifying how the NCS/halo sectors deepen the effective potential felt by null geodesics and make capture more likely at fixed asymptotic conditions.}
  \label{fig:deflection-capture}
\end{figure}

In this geometry, null geodesics experience three competing ingredients
(Fig.~\ref{fig:deflection-capture}(a)-(b)): the AdS curvature (which
governs the large-$r$ tail), the Dehnen halo (attractive, controlled by
$\rho_s$ and $r_s$), and the hypergeometric NCS sector (an effective
angular-deficit-like contribution governed by $\alpha$ and $b$). 
Figure \ref{fig:deflection-capture}(a) organizes the scattering outcome in terms of the impact parameter.
The threshold $b_c(\alpha)$, set by the unstable photon orbit, marks the onset
of capture; as $b\!\downarrow\!b_c$ the deflection angle $\chi(b)$ grows
sharply, while for $b\!\gg\!b_c$ the curve flattens as the trajectory probes
mostly the asymptotic region. At fixed halo parameters, increasing $\alpha$
strengthens the NCS contribution, shifts the photon sphere outward, and
enhances bending, hence the systematic ordering
$\chi_{\alpha=0.15}(b)\!>\!\chi_{\alpha=0.10}(b)\!>\!\chi_{\alpha=0.05}(b)$
across most of the plotted range. Panel (b) recasts the same physics in terms
of the capture cross-section $\sigma_{\rm cap}=\pi b_c^2$. For all three halos
scales $r_s$ displayed, $\sigma_{\rm cap}$ grows with $\alpha$, and at fixed
$\alpha$ it increases with $r_s$ (Fig. \ref{fig:deflection-capture}(b)); both trends follow from a deeper effective
potential and a larger critical cylinder in impact-parameter space. Together,
The two panels quantify how the NCS and halo sectors control light
propagation: they raise the deflection for near-critical rays and expand the
capture basin, effects that dovetail with the thermodynamic shifts (e.g.,
larger spinodal windows and modified Gibbs structures) discussed elsewhere in
this section.

\begin{figure}[ht!]
    \centering
    \includegraphics[width=0.86\linewidth]{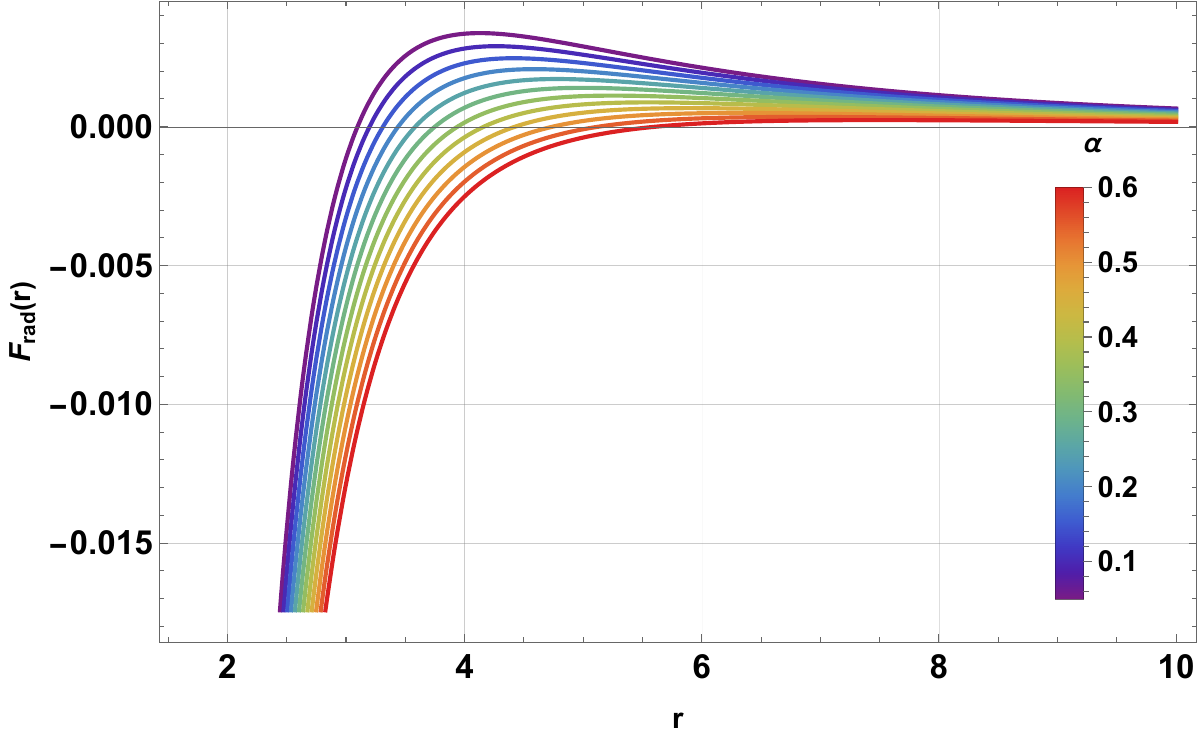}\\
    (i) $r_s=0.2\,b=0.5,\rho_s=0.02$\\
    \includegraphics[width=0.86\linewidth]{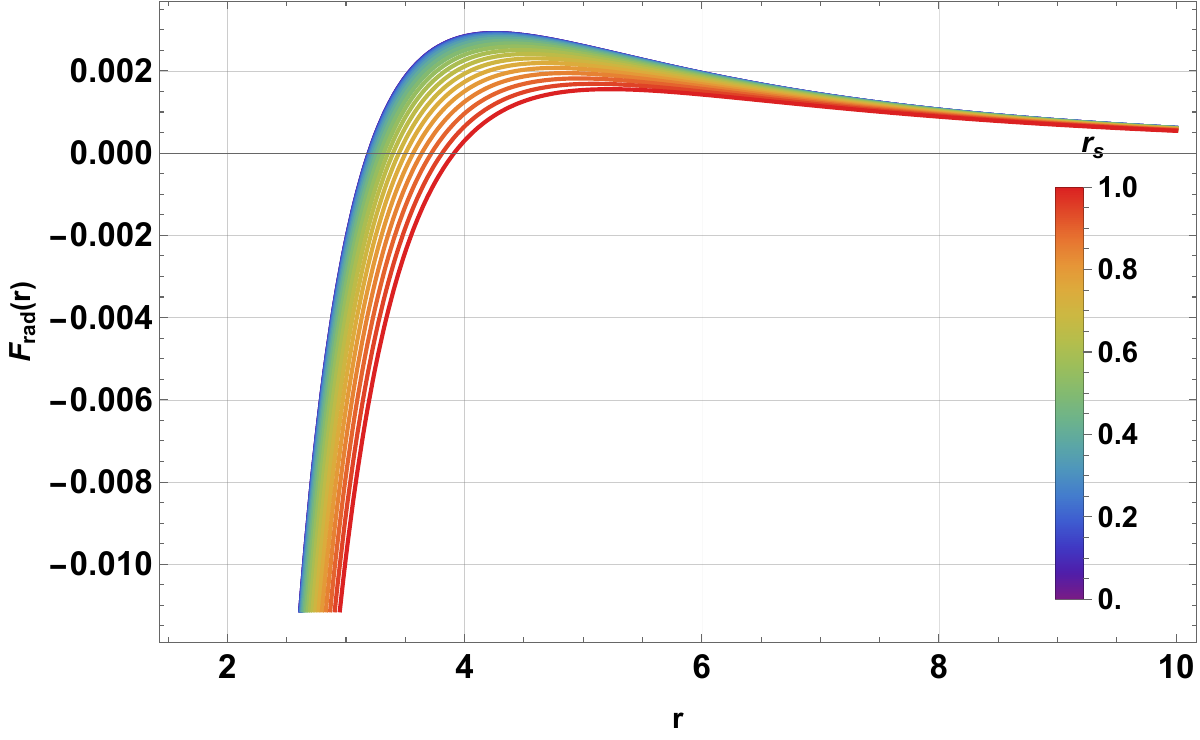}\\
    (ii) $\alpha=0.1,\rho_s=0.02,b=0.5$\\
    \includegraphics[width=0.86\linewidth]{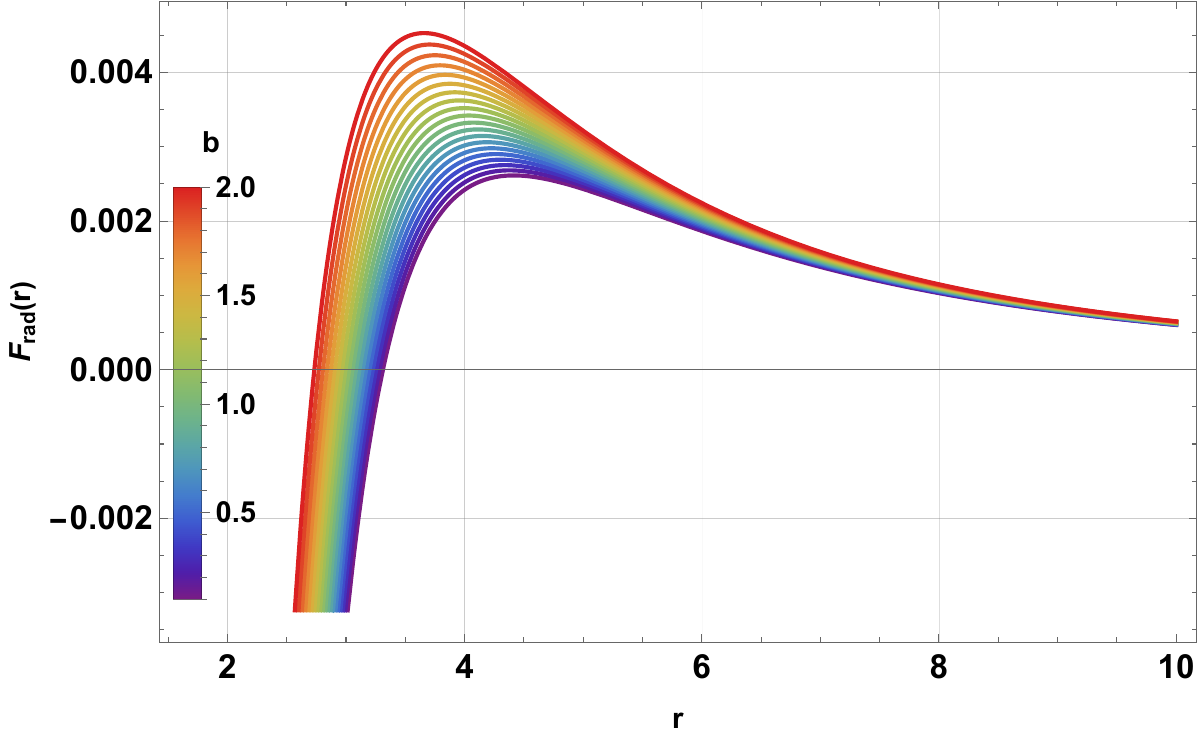}\\
    (iii) $\alpha=0.1,r_s=0.2,\,\rho_s=0.02$\\
    \caption{\footnotesize Behavior of the effective radial force as a function of $r$ for various values of NCS parameter $\alpha$ and the halo radius $r_s$. Here $M=1,\,\ell_p=10,\,\mathrm{L}=1$.}
    \label{fig:force}
\end{figure}

\begin{center}
\large{\bf II.\, Effective Radial Force Experiences by Photons }
\end{center}

Photons follow null geodesics, and their motion in a static, spherically symmetric spacetime can be described using an effective radial potential derived from conserved energy and angular momentum. While not a Newtonian force, the gradient of this potential reflects spacetime curvature and determines whether photons escape, fall into the black hole, or form circular orbits around it. Circular photon orbits occur where this effective force vanishes, defining the photon sphere \cite{chandra,wald}. 

One may define an effective radial force associated with the effective potential as
\begin{equation}
    F_\text{rad}=-\frac{1}{2}\,\frac{dV_{\rm eff}}{dr}.\label{bb14}
\end{equation}
Substituting the effective potential given in Eq. (\ref{bb10}), we find the following expression:
\begin{align}
F_\text{rad}&=\frac{\mathrm{L}^2}{r^3}\Bigg[
1 - \frac{3M}{r}
-8\pi \rho_s r_s^2 \ln\left(1 + \frac{r_s}{r} \right)
- \frac{4\pi \rho_s r_s^3}{r + r_s}\notag\\&
+ \frac{2|\alpha| b^2}{r^2} \, {}_2F_1\left( -\tfrac{1}{2}, -\tfrac{1}{4}, \tfrac{3}{4}, -\frac{r^4}{b^4} \right)\notag\\&
+ \frac{|\alpha| r^2}{3b^2} \, {}_2F_1\left( \tfrac{1}{2}, \tfrac{3}{4}, \tfrac{7}{4}, -\frac{r^4}{b^4} \right)
\Bigg].\label{bb15}
\end{align}
One can see that BH mass $M$, the curvature radius $\ell_p$, the NCS parameters $(|\alpha|,b)$, and the DMH profile characterized by parameters $(r_s,\rho_s)$ significantly influence the effective radial force experienced by the photons in the gravitational field.

The effective radial force framework offers an intuitive link between the abstract geodesic equations and tangible astrophysical phenomena such as gravitational lensing, BH shadows, and the behavior of light near compact objects. This approach underscores how the geometry of spacetime governs photon dynamics and provides a direct connection to observational signatures that probe general relativity in the strong-field regime \cite{perlick,frolov}.

In Figure~\ref{fig:force}, we present plots showing the behavior of the effective radial force as a function of $r$, by varying NCS parameters \(\alpha\), \(b\), and the halo radius \(r_s\), while keeping all other parameters fixed. Panels (i) and (ii) illustrate that the radial force decreases with increasing \(\alpha\) and \(r_s\), respectively. In contrast, panel (iii) shows that this radial force increases as the parameter \(b\) increases.

\begin{table}[ht!]
\centering
\scriptsize
\setlength{\tabcolsep}{6pt}
\renewcommand{\arraystretch}{1.4}
\begin{tabular}{|c|c|c|c|c|c|c|c|}
\hline
$\alpha (\downarrow)  \backslash r_s (\rightarrow) $ & 0.2 & 0.4 & 0.6 & 0.8 & 1.0 & 1.2 & 1.4 \\
\hline
\multicolumn{8}{c}{\textbf{Photon sphere radius $r_{\text{ph}}$}} \\
\hline
0.05 & 3.097 & 3.138 & 3.244 & 3.444 & 3.769 & 4.253 & 4.935 \\
0.10 & 3.198 & 3.242 & 3.354 & 3.567 & 3.912 & 4.426 & 5.150 \\
0.15 & 3.312 & 3.358 & 3.477 & 3.704 & 4.071 & 4.619 & 5.392 \\
0.20 & 3.439 & 3.488 & 3.616 & 3.858 & 4.251 & 4.837 & 5.664 \\
0.25 & 3.584 & 3.636 & 3.773 & 4.032 & 4.455 & 5.085 & 5.974 \\
0.30 & 3.749 & 3.805 & 3.953 & 4.232 & 4.688 & 5.368 & 6.328 \\
0.35 & 3.940 & 4.001 & 4.160 & 4.463 & 4.957 & 5.695 & 6.737 \\
0.40 & 4.162 & 4.228 & 4.402 & 4.732 & 5.272 & 6.078 & 7.216 \\
0.45 & 4.425 & 4.497 & 4.688 & 5.051 & 5.644 & 6.532 & 7.784 \\
0.50 & 4.741 & 4.820 & 5.031 & 5.433 & 6.092 & 7.076 & 8.466 \\
\hline
\multicolumn{8}{c}{\textbf{Shadow radius $R_s$}} \\
\hline
0.05 & 4.822 & 4.871 & 4.999 & 5.234 & 5.591 & 6.071 & 6.649 \\
0.10 & 5.043 & 5.094 & 5.227 & 5.470 & 5.838 & 6.326 & 6.904 \\
0.15 & 5.283 & 5.336 & 5.474 & 5.725 & 6.101 & 6.594 & 7.167 \\
0.20 & 5.544 & 5.599 & 5.741 & 5.998 & 6.380 & 6.874 & 7.437 \\
0.25 & 5.826 & 5.883 & 6.028 & 6.290 & 6.676 & 7.166 & 7.712 \\
0.30 & 6.132 & 6.189 & 6.338 & 6.602 & 6.987 & 7.466 & 7.988 \\
0.35 & 6.461 & 6.519 & 6.668 & 6.932 & 7.311 & 7.773 & 8.261 \\
0.40 & 6.813 & 6.871 & 7.019 & 7.279 & 7.645 & 8.081 & 8.529 \\
0.45 & 7.187 & 7.244 & 7.388 & 7.638 & 7.984 & 8.386 & 8.786 \\
0.50 & 7.578 & 7.632 & 7.769 & 8.003 & 8.321 & 8.680 & 9.026 \\
\hline
\end{tabular}
\caption{\footnotesize Photon sphere radius $r_{\text{ph}}$ and shadow radius $R_s$ for varying NCS parameter $\alpha$ and the halo radius $r_s$.  Here $M=1,\,\rho_s=0.02,\,b=0.5,\,\ell_p=10$.}
\label{tab:1}
\end{table}

\begin{table}[ht!]
\centering
\scriptsize
\setlength{\tabcolsep}{6pt}
\renewcommand{\arraystretch}{1.4}
\begin{tabular}{|c|c|c|c|c|c|c|c|}
\hline
$b (\downarrow) \backslash r_s (\rightarrow)$ & 0.2 & 0.4 & 0.6 & 0.8 & 1.0 & 1.2 & 1.4 \\
\hline
\multicolumn{8}{c}{\textbf{Photon sphere radius $r_{\text{ph}}$}} \\
\hline
0.20 & 3.283 & 3.327 & 3.439 & 3.653 & 3.999 & 4.514 & 5.240 \\
0.40 & 3.227 & 3.270 & 3.383 & 3.595 & 3.941 & 4.455 & 5.180 \\
0.60 & 3.170 & 3.213 & 3.326 & 3.538 & 3.883 & 4.396 & 5.120 \\
0.80 & 3.113 & 3.156 & 3.268 & 3.481 & 3.825 & 4.337 & 5.060 \\
1.00 & 3.055 & 3.099 & 3.211 & 3.422 & 3.766 & 4.278 & 5.000 \\
1.20 & 2.997 & 3.040 & 3.152 & 3.363 & 3.707 & 4.218 & 4.940 \\
1.40 & 2.936 & 2.979 & 3.092 & 3.303 & 3.646 & 4.157 & 4.879 \\
1.60 & 2.873 & 2.916 & 3.029 & 3.241 & 3.585 & 4.096 & 4.817 \\
1.80 & 2.806 & 2.850 & 2.963 & 3.176 & 3.521 & 4.033 & 4.755 \\
2.00 & 2.733 & 2.778 & 2.893 & 3.108 & 3.454 & 3.968 & 4.691 \\
\hline
\multicolumn{8}{c}{\textbf{Shadow radius $R_s$}} \\
\hline
0.20 & 5.142 & 5.192 & 5.323 & 5.561 & 5.921 & 6.399 & 6.965 \\
0.40 & 5.076 & 5.127 & 5.259 & 5.501 & 5.866 & 6.351 & 6.924 \\
0.60 & 5.009 & 5.061 & 5.195 & 5.440 & 5.810 & 6.301 & 6.884 \\
0.80 & 4.942 & 4.994 & 5.130 & 5.377 & 5.752 & 6.251 & 6.842 \\
1.00 & 4.873 & 4.926 & 5.063 & 5.314 & 5.694 & 6.199 & 6.800 \\
1.20 & 4.802 & 4.856 & 4.995 & 5.249 & 5.634 & 6.147 & 6.757 \\
1.40 & 4.730 & 4.784 & 4.925 & 5.183 & 5.573 & 6.094 & 6.712 \\
1.60 & 4.654 & 4.710 & 4.853 & 5.114 & 5.510 & 6.039 & 6.667 \\
1.80 & 4.576 & 4.632 & 4.778 & 5.043 & 5.446 & 5.982 & 6.621 \\
2.00 & 4.492 & 4.550 & 4.698 & 4.969 & 5.378 & 5.924 & 6.574 \\
\hline
\end{tabular}
\caption{\footnotesize Photon sphere radius $r_{\text{ph}}$ and shadow radius $R_s$ for NCS parameter $b$ and the halo radius $r_s$.  Here $M=1,\,\rho_s=0.02,\,\alpha=0.1,\,\ell_p=10$.}
\label{tab:2}
\end{table}

\begin{center}
\large{\bf III.\, Photon Sphere and BH shadows}
\end{center}

The photon sphere marks the boundary between capture and escape: photons with smaller impact parameters fall into the black hole, while those with larger ones escape. Orbits at the photon sphere are highly unstable; small perturbations push photons inward or outward. This instability defines the apparent size of the BH shadow, as observed by the Event Horizon Telescope.
 
For circular null orbits of radius $r$=const., the conditions $\dot{r}=0$ and $\ddot{r}=0$ must be satisfied. The first condition simplifies to 
\begin{equation}
\mathrm{E}^2=V_\text{eff}(r)\label{bb15a}
\end{equation}
which gives us the critical impact parameter for photons. This parameter using Eq. (\ref{bb10}) is given by
\begin{widetext}
\begin{equation}
b_c=\frac{\mathrm{L}_\text{ph}}{\mathrm{E}_\text{ph}}=\frac{r}{\sqrt{1-\frac{2M}{r}-8\pi\,\rho_s\,r_s^2\,\mbox{ln} {\left(1+\frac{r_s}{r}\right)}+\frac{\lvert \alpha\rvert\, b^2}{r^2}\,{}_2F_1\left(-\frac{1}{2},-\frac{1}{4},\frac{3}{4},-\frac{r^4}{b^4}\right)+\frac{r^2}{\ell^2_p}}}\Bigg{|}_{r=\mbox{const.}}.\label{bb16}
\end{equation} 
\end{widetext}
Noted that if $b\,(=\mathrm{L}/\mathrm{E}) < b_{c}$, the photon is captured by the BH and inevitably crosses the event horizon. If $b > b_{c}$, the photon is scattered back to infinity, experiencing gravitational deflection. If $b = b_{c}$, the photon asymptotically approaches the photon sphere, orbiting in an unstable circular trajectory. Thus, the critical impact parameter $b_{c}$ acts as the dividing line between capture and escape, making it a fundamental quantity in defining the apparent BH shadow as seen by distant observers. In practice, $b_{c}$ corresponds to the shadow radius, while $\beta$ labels individual photon trajectories relative to this boundary \cite{perlick}.

Now, we focus on an important feature of the BH called the apparent shadow size cast by the BH. The radius of the BH shadow is equal to the critical impact parameter for a photon when it traverses in unstable circular orbits. This is defined by
\begin{widetext}
\begin{equation}
R_s=b_c=\frac{r_\text{ph}}{\sqrt{1-\frac{2M}{r_\text{ph}}-8\pi\,\rho_s\,r_s^2\,\mbox{ln} {\left(1+\frac{r_s}{r_\text{ph}}\right)}+\frac{\lvert \alpha\rvert\, b^2}{r^2_\text{ph}}\,{}_2F_1\left(-\frac{1}{2},-\frac{1}{4},\frac{3}{4},-\frac{r^4_\text{ph}}{b^4}\right)+\frac{r_\text{ph}^2}{\ell^2_p}}}.\label{bb17}
\end{equation}
\end{widetext}
Now, we determine the photon sphere radius $r=r_\text{ph}$ using the second condition for circular null orbits as stated earlier. This condition $\ddot{r}=0$ implies that 
\begin{equation}
\frac{dV_\text{eff}(r)}{dr}\Bigg{|}_{r=r_\text{ph}}=0.\label{bb18}
\end{equation}
Substituting potential (\ref{bb10}) into the above relation results 
\begin{align}
&1 - \frac{3M}{r}
-8\pi \rho_s r_s^2 \ln\left(1 + \frac{r_s}{r} \right)
- \frac{4\pi \rho_s r_s^3}{r + r_s}
+ \frac{2|\alpha| b^2}{r^2} \notag\\& \times {}_2F_1\left( -\tfrac{1}{2}, -\tfrac{1}{4}, \tfrac{3}{4}, -\frac{r^4}{b^4} \right)
+ \frac{|\alpha| r^2}{3b^2} \, {}_2F_1\left( \tfrac{1}{2}, \tfrac{3}{4}, \tfrac{7}{4}, -\frac{r^4}{b^4} \right)=0.\label{bb19}
\end{align}
Equation (\ref{bb19}) represents an infinite polynomial equation in the radial coordinate $r$, for which obtaining an exact analytical solution is highly challenging. Nevertheless, the photon sphere radius $r = r_{\text{ph}}$ can be determined numerically by assigning suitable values to the parameters appearing in the polynomial equation.

\begin{figure}[ht!]
\centering
\includegraphics[width=0.85\linewidth]{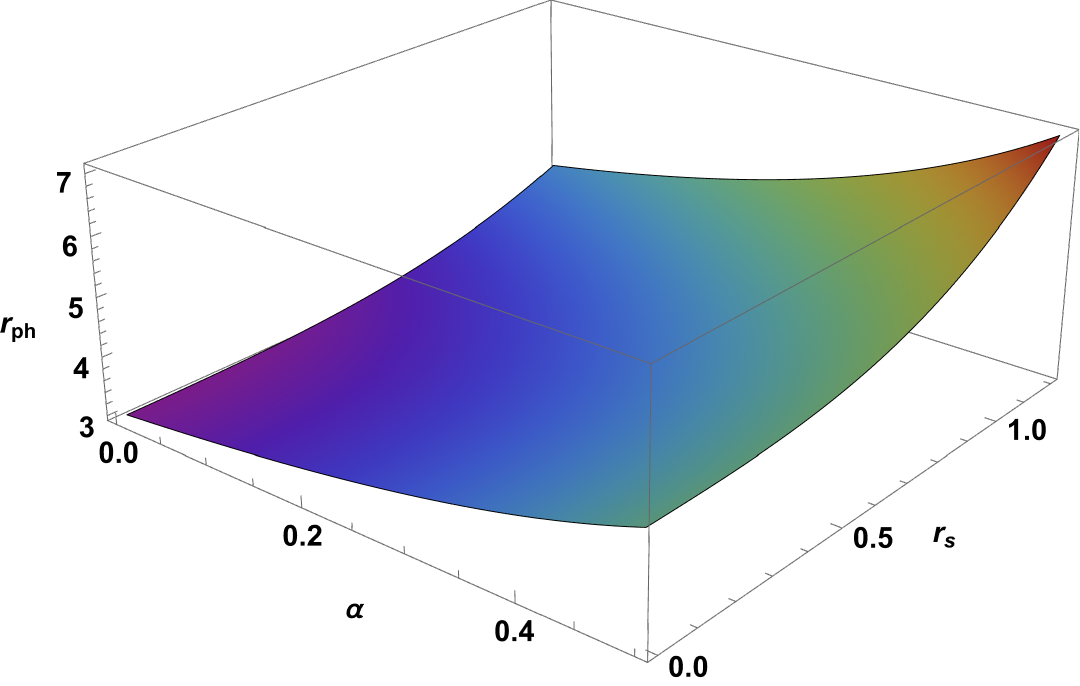}\\
(i) $b=0.5$\\
\includegraphics[width=0.85\linewidth]{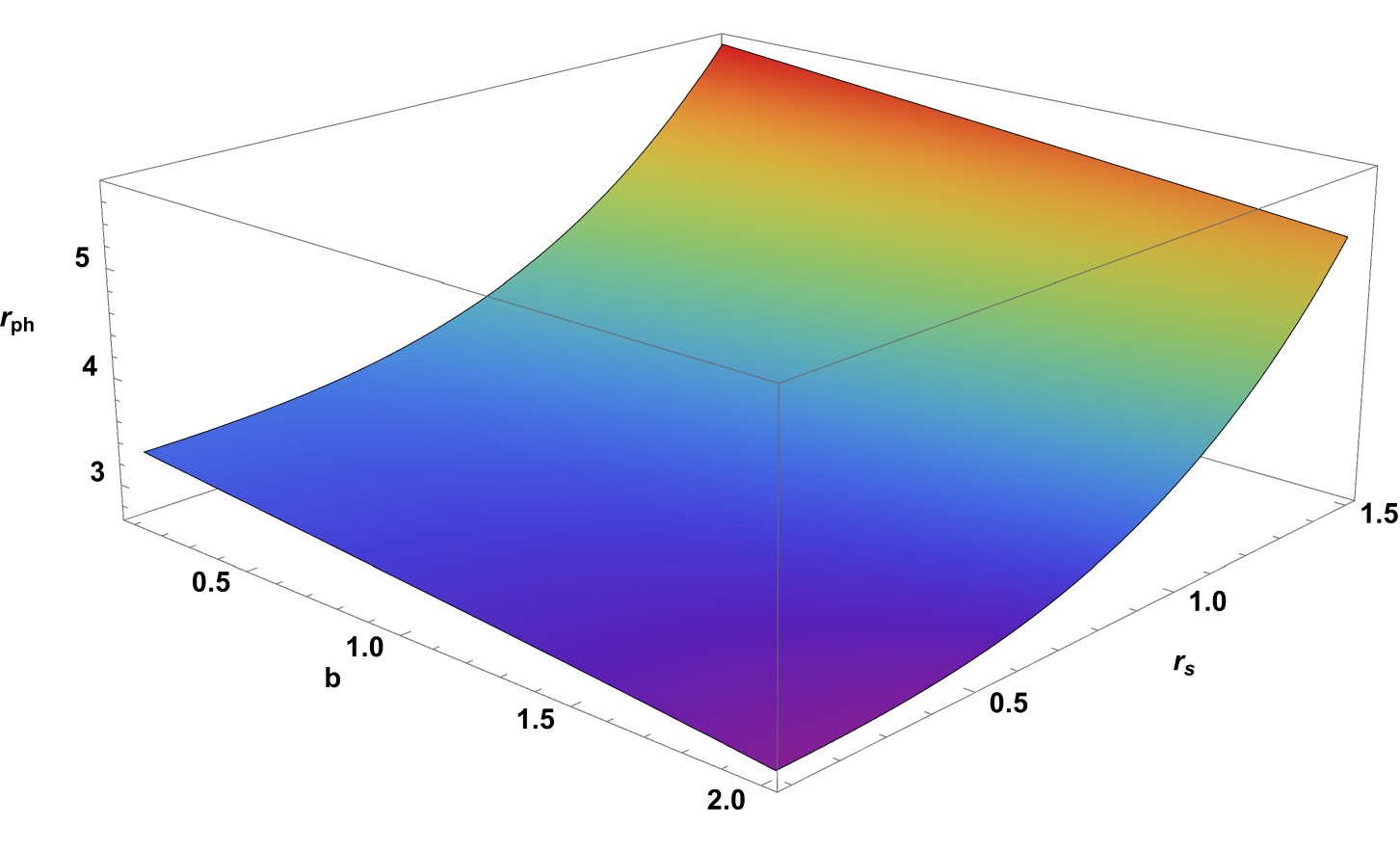}\\
(ii) $\alpha=0.1$\\
\caption{\footnotesize 3D plot of the photon sphere radius $r_\text{ph}$ as a function of $(\alpha,r_s)$ and $(b,r_s)$. Here $M=1,\,\rho_s=0.02,\ell_p=10$.}
\label{fig:3D-plot-1}
\end{figure}

\begin{figure}[ht!]
    \centering
    \includegraphics[width=0.85\linewidth]{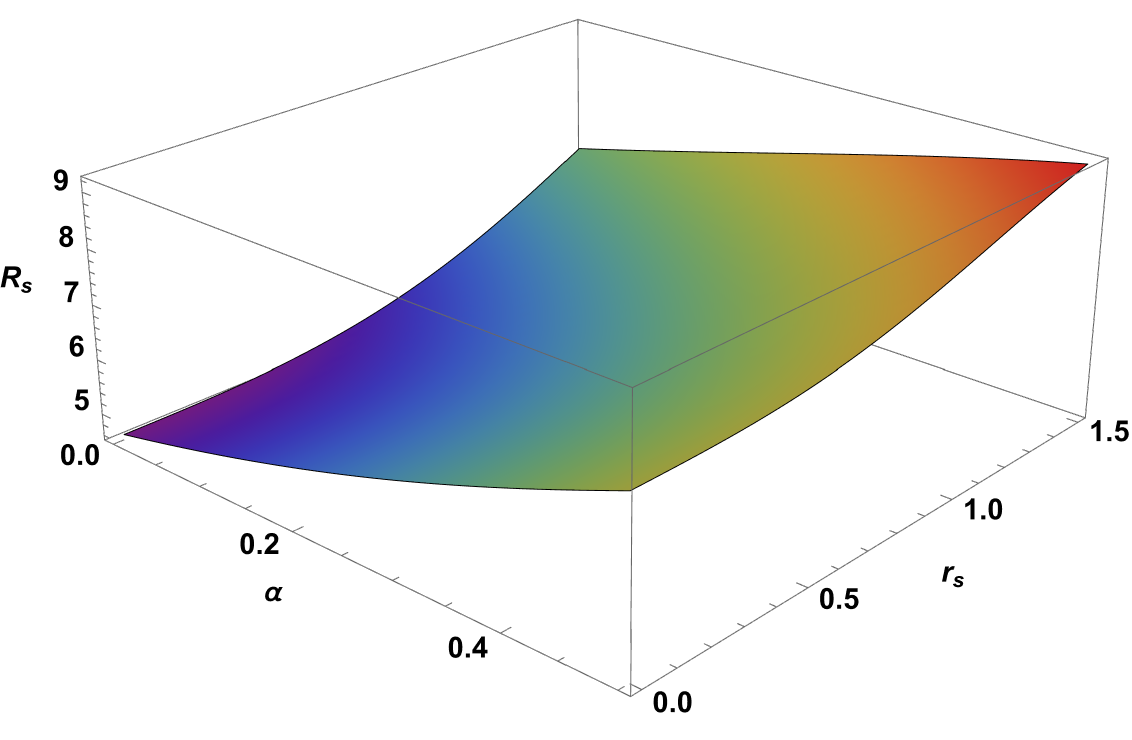}\\
    (i) $b=0.5$\\
    \includegraphics[width=0.85\linewidth]{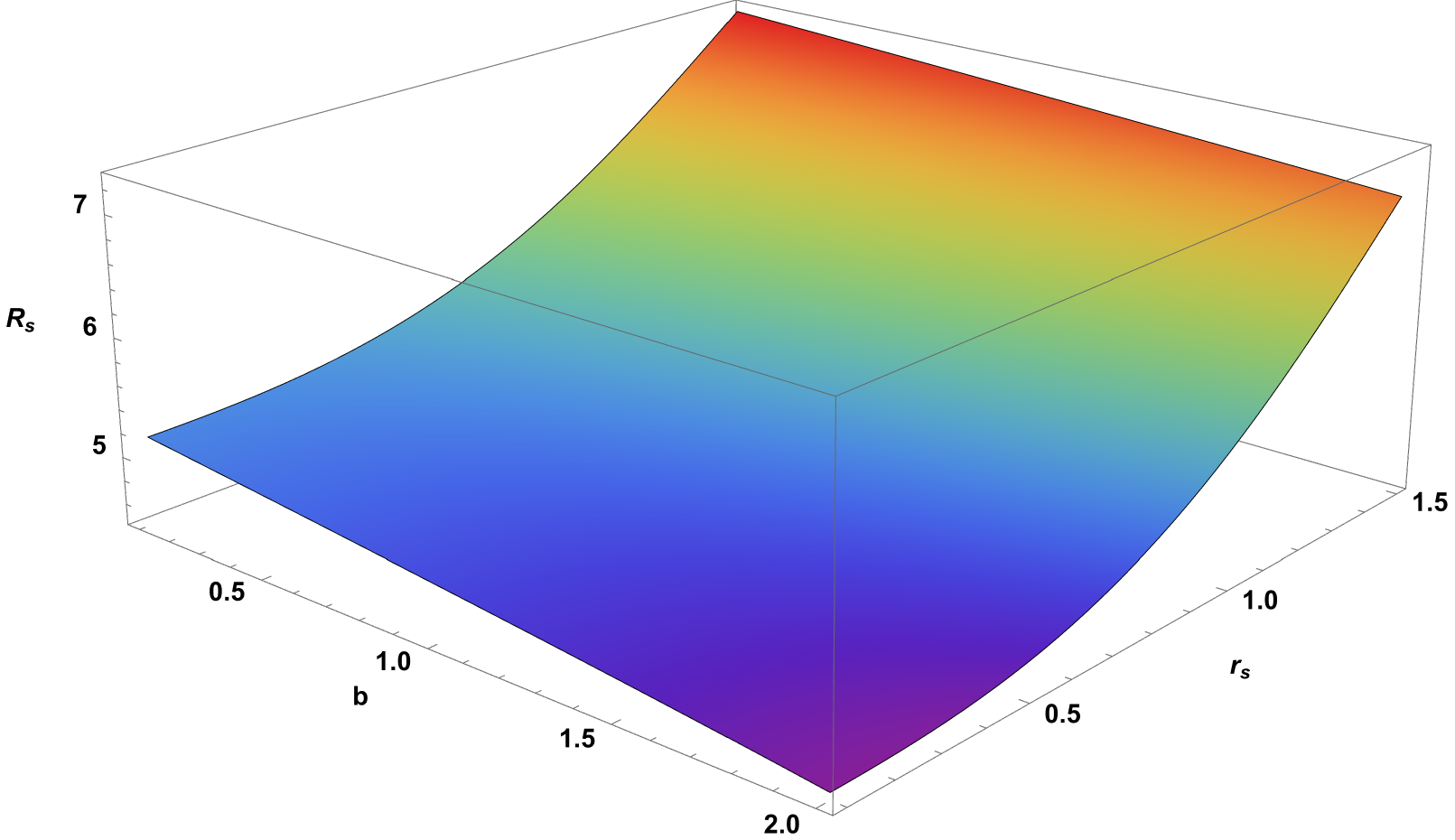}\\
    (ii) $\alpha=0.1$\\
    \caption{\footnotesize 3D plot of the shadow size $R_s$ as a function of $(\alpha,r_s)$ and $(b,r_s)$. Here $M=1,\,\rho_s=0.02,\ell_p=10$.}
    \label{fig:3D-plot-2}
\end{figure}

\begin{figure}[tbhp]
    \centering
    \includegraphics[width=0.75\linewidth]{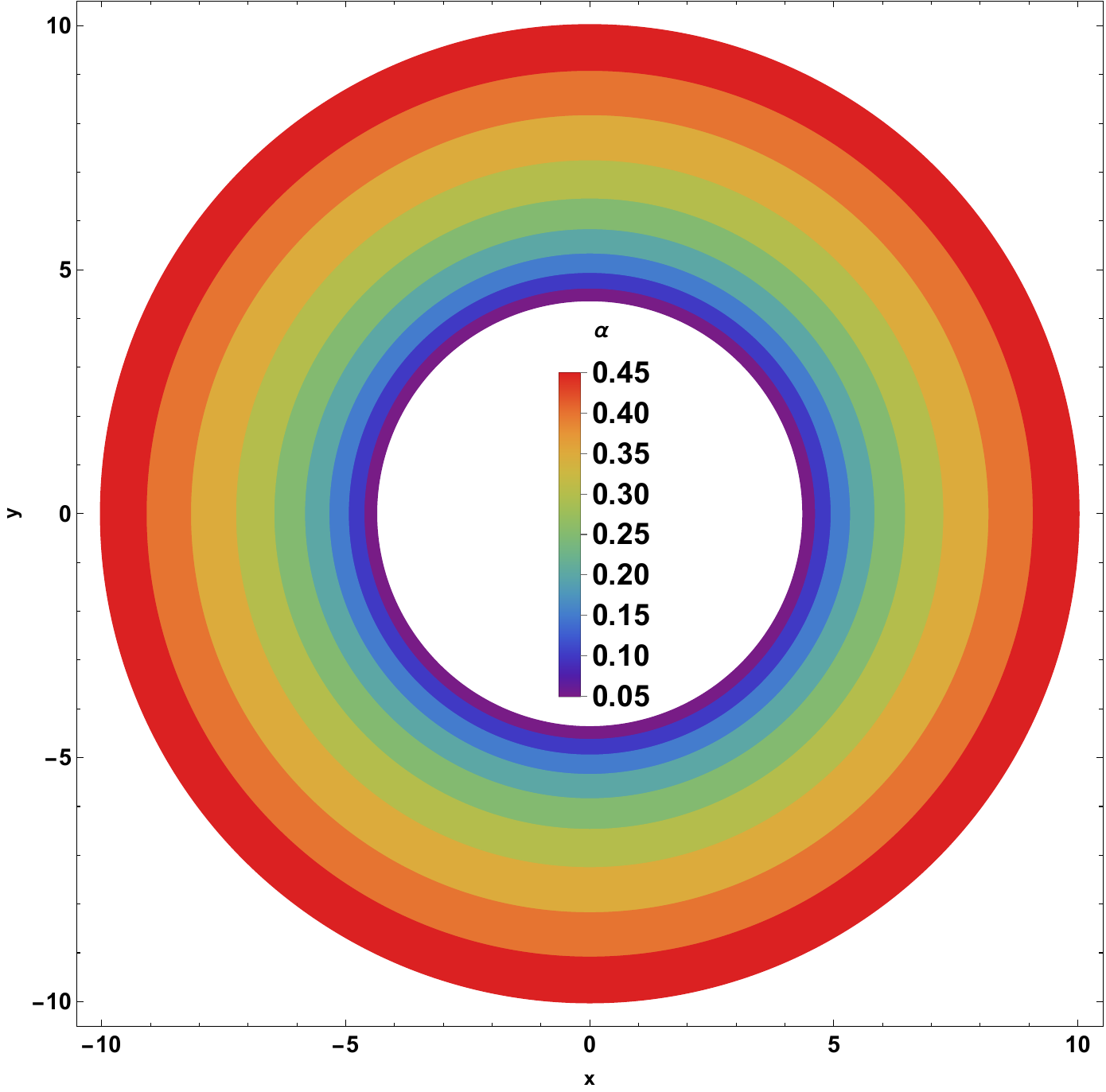}\\
     (i) $b=0.5,\,r_s=0.2$\\ 
    \includegraphics[width=0.75\linewidth]{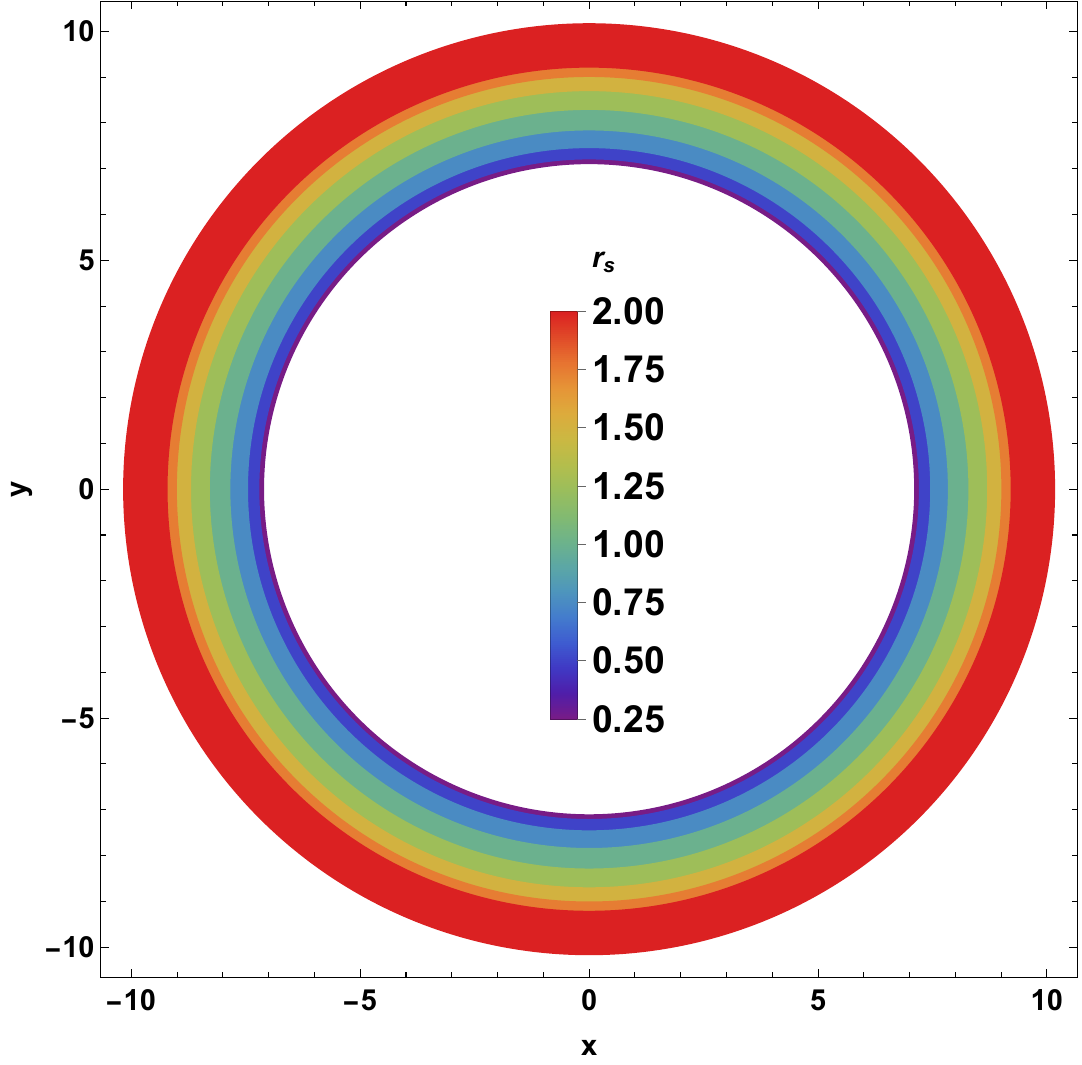}\\
    (ii) $\alpha=0.5=b$ \\
    \includegraphics[width=0.75\linewidth]{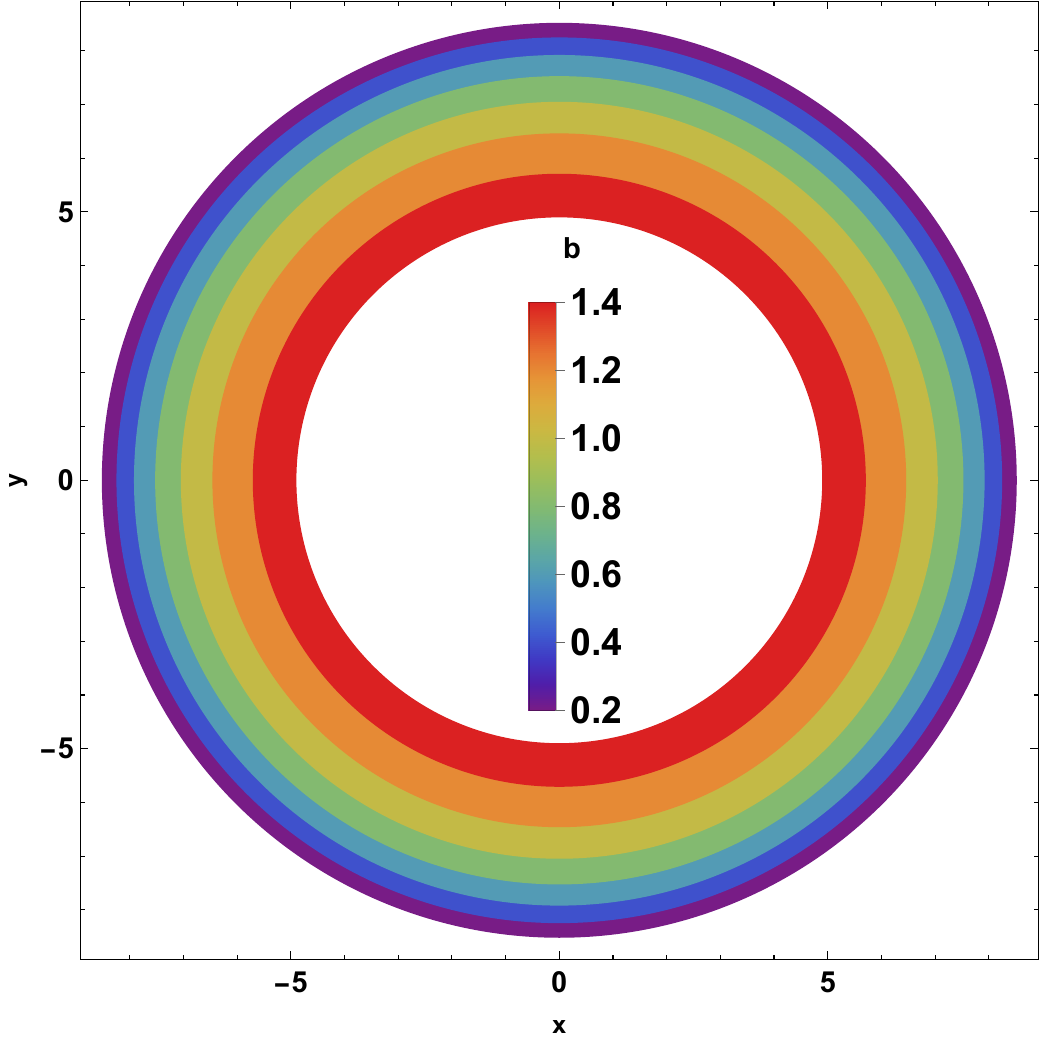}\\
     (iii) $\alpha=0.5=r_s$\\
    \caption{\footnotesize Shadow rings for different values of $\alpha$, $r_s$ and $b$. Here $M=1\,\rho_s=0.02,\,\ell_p=10$.}
    \label{fig:shadow-ring}
\end{figure}

In Tables \ref{tab:1}-\ref{tab:2}, we presented numerical values of the photon sphere radius $r_\text{ph}$ and the BH shadow size $R_s$ by varying the NCS parameters $(|\alpha|, b)$ and the DMH radius $r_s$, while keeping other parameters fixed. In both Tables, we have observed that both $r_\text{ph}$ and $R_s$ increase gradually by increasing $r_s$, while keeping NCS parameters $(|\alpha|, b)$ fixed. A similar trend can be observed by increasing $(|\alpha|, b)$ for a particular value of $r_s$.

Combined with the EHT observational results for the supermassive BHs M87* and Sgr A*, the observed shadow radius of M87* is approximately $R_{s}^{\text{M87*}} \approx (5.5 \pm 0.75)\,M$~\cite{AA2,CB}, while for Sgr A* it is $R_{s}^{\text{Sgr A*}} \approx (4.885 \pm 0.335)\,M$~\cite{SV}. Our numerical results, presented in Tables~\ref{tab:1}-\ref{tab:2}, show that the observational data from both M87* and Sgr A* can place meaningful constraints on the parameter space of the selected BH model, specifically on $|\alpha|, b$, and $r_s$.

In Figure~\ref{fig:3D-plot-1}, we present a 3D plot of the photon sphere radius \( r_\text{ph} \) as a function of the parameter pairs \( (\alpha, r_s) \) and \( (b, r_s) \).

Similarly, in Figure~\ref{fig:3D-plot-2}, we display a 3D plot of the shadow radius \( R_s \) as a function of the same parameter combinations used for the photon sphere.

In Figure~\ref{fig:shadow-ring}, we present the BH shadow rings for different values of the parameter \(\alpha, r_s \) and \( b\).

\begin{figure}[ht!]
\centering
\includegraphics[width=0.88\linewidth]{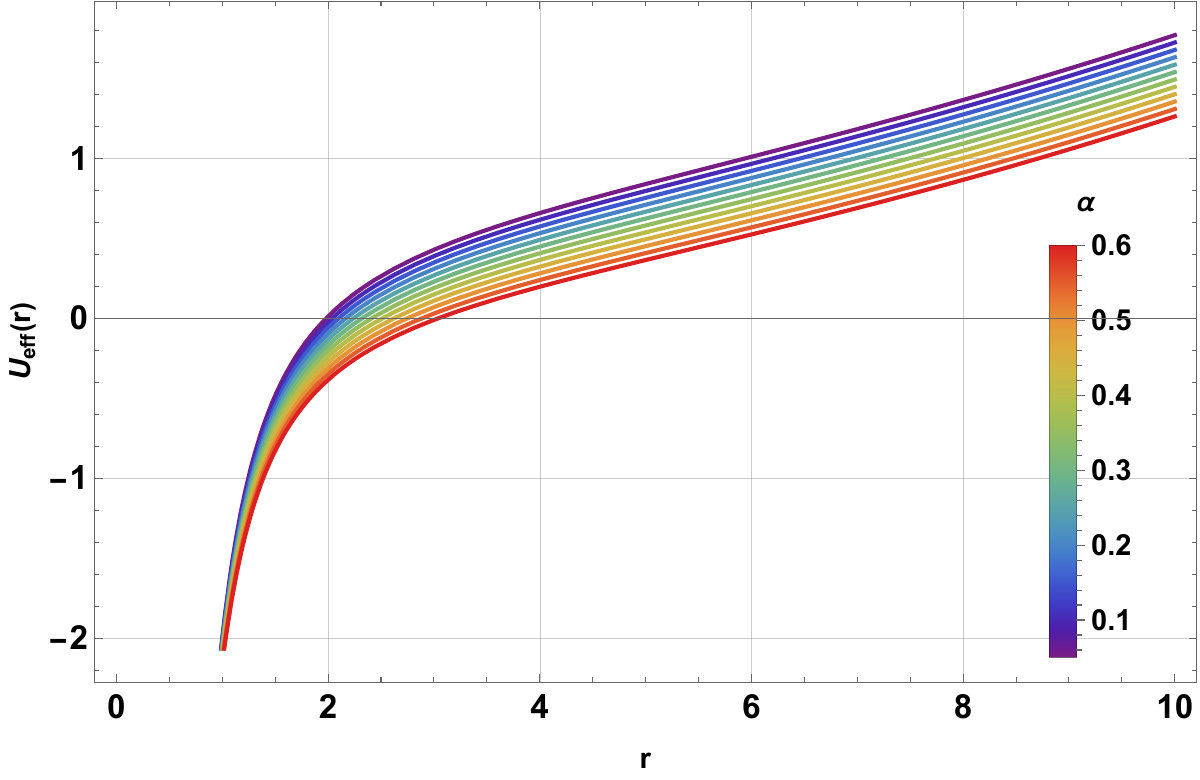}\\
(i) $r_s=0.2,\,\rho_s=0.02,\,b=0.5$ \\
\includegraphics[width=0.88\linewidth]{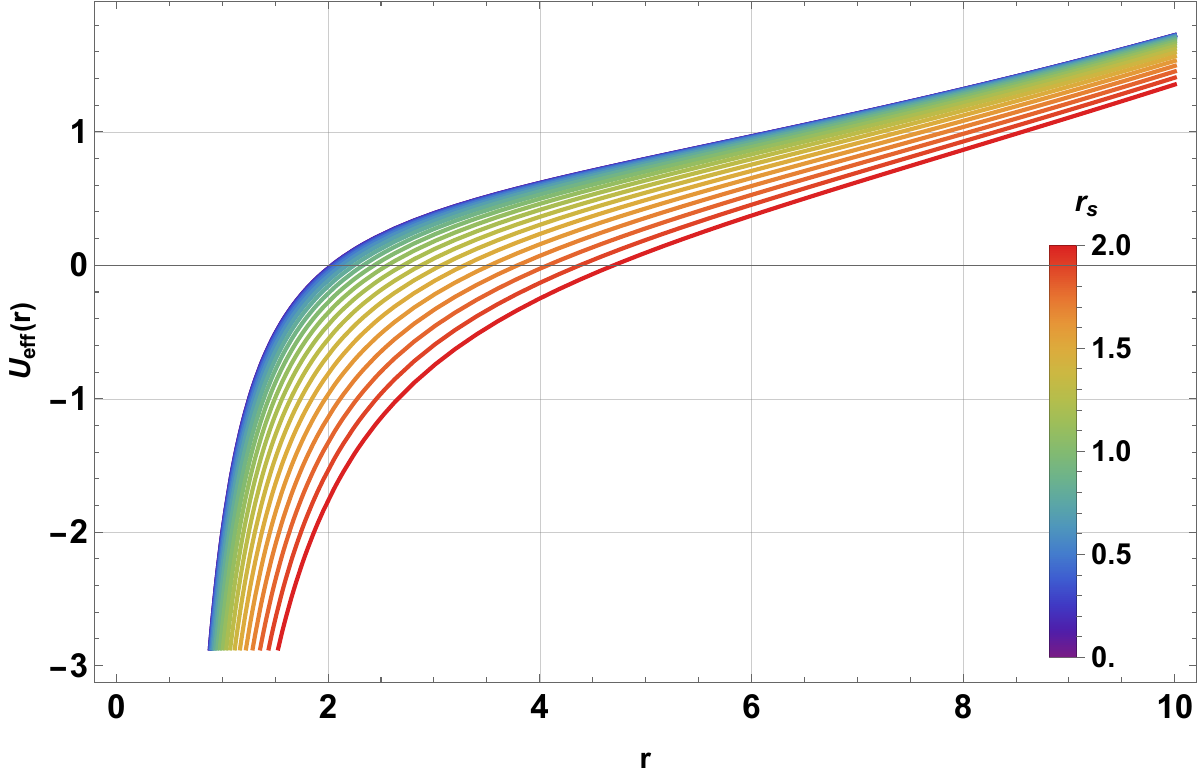}\\
(ii) $\alpha=0.1,\,\rho_s=0.02,\,b=0.5$\\
\includegraphics[width=0.88\linewidth]{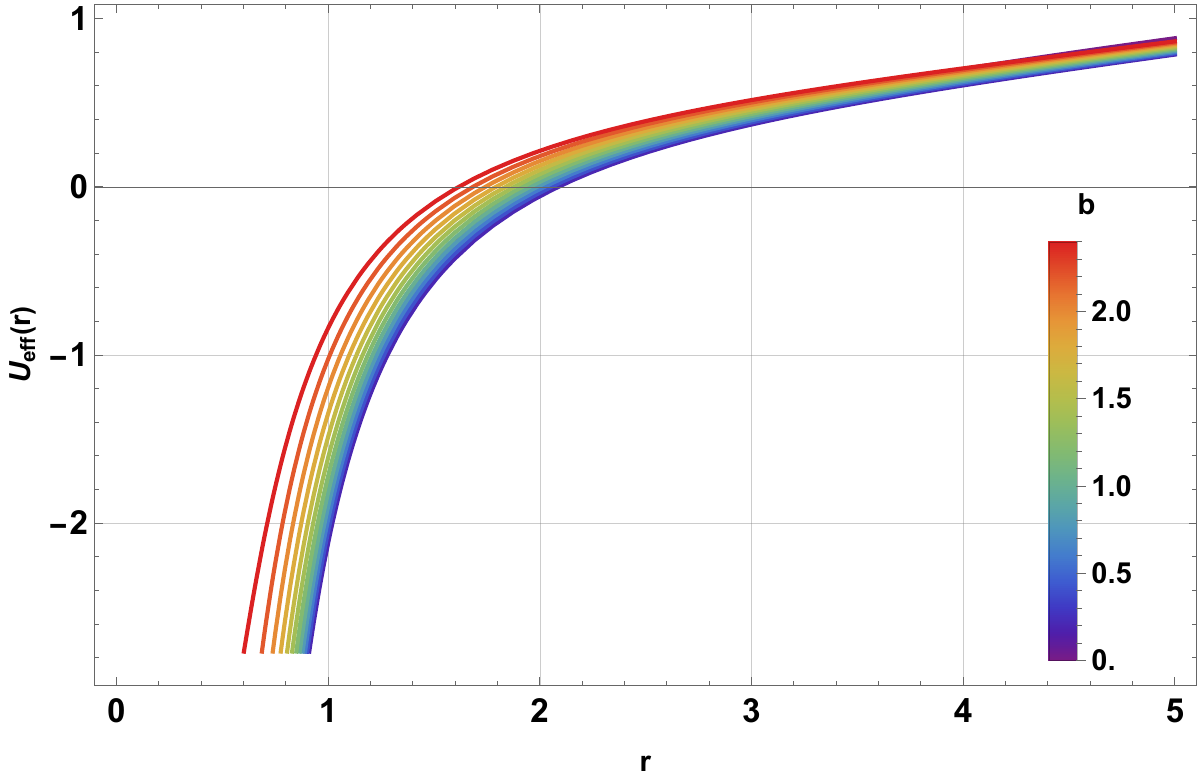}\\
(ii) $\alpha=0.1,\,\rho_s=0.02,\,r_s=0.2$\\
\caption{\footnotesize Behavior of the effective potential for time-like particles as a function of $r$ for different values of NCS parameters $(\alpha,b)$ and the halo $r_s$. Here, $M=1,\,\ell_p=10,\,\mathcal{L}=1$.}
\label{fig:potential-timelike}
\end{figure}

\subsection{\large {\bf Test Particles Dynamics}}

The dynamics of test particles around BHs in external fields offer vital insights into the spacetime structure and strong gravity effects. In particular, the innermost stable circular orbit (ISCO) is crucial for understanding accretion efficiency, electromagnetic emissions, and the motion of compact binaries. The ISCO radius, sensitive to the background geometry, serves as a probe for distinguishing BH solutions and testing deviations from general relativity. ISCO properties are linked to astrophysical observations, including X-ray spectra, quasi-periodic oscillations (QPOs), and gravitational wave signals from extreme mass-ratio inspirals (EMRIs). External fields, like dark matter halos and clouds of strings, can significantly modify ISCO characteristics, impacting observable phenomena and offering a valuable window into BH physics.

\begin{figure}[ht!]
\centering
\includegraphics[width=0.85\linewidth]{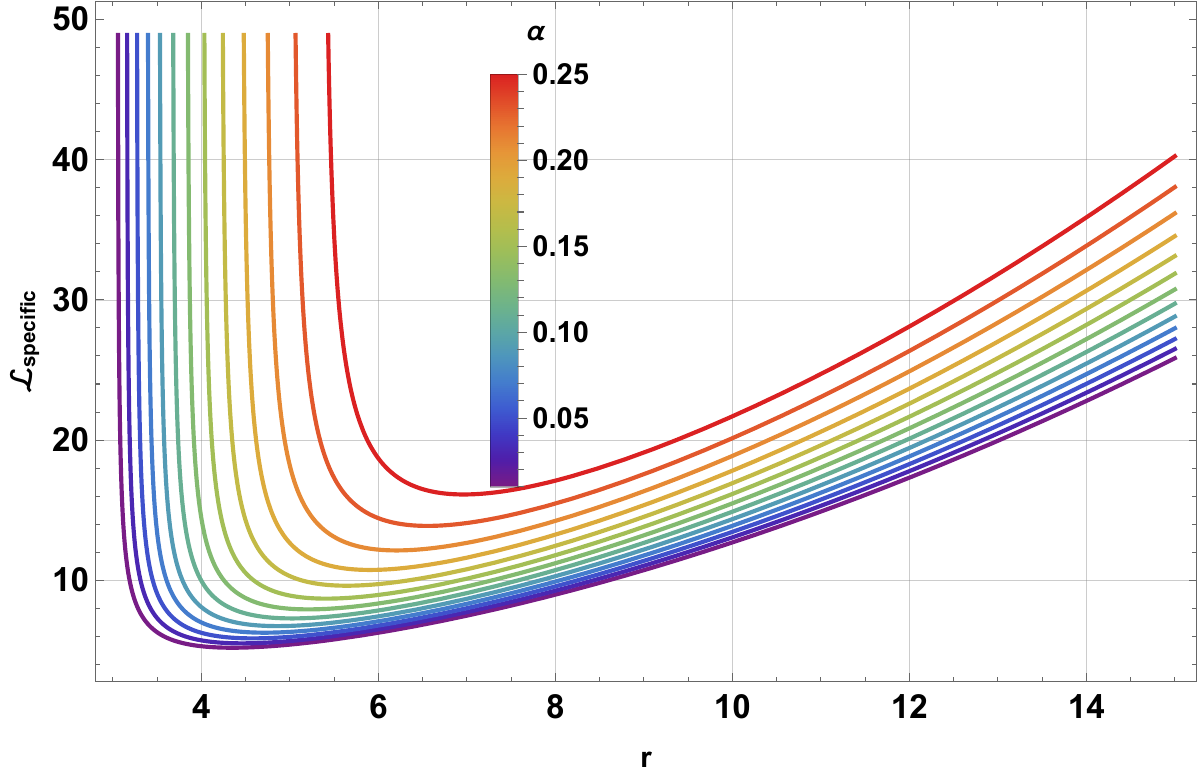}\\
(i) $r_s=0.2,\,b=0.5$ \\
\includegraphics[width=0.85\linewidth]{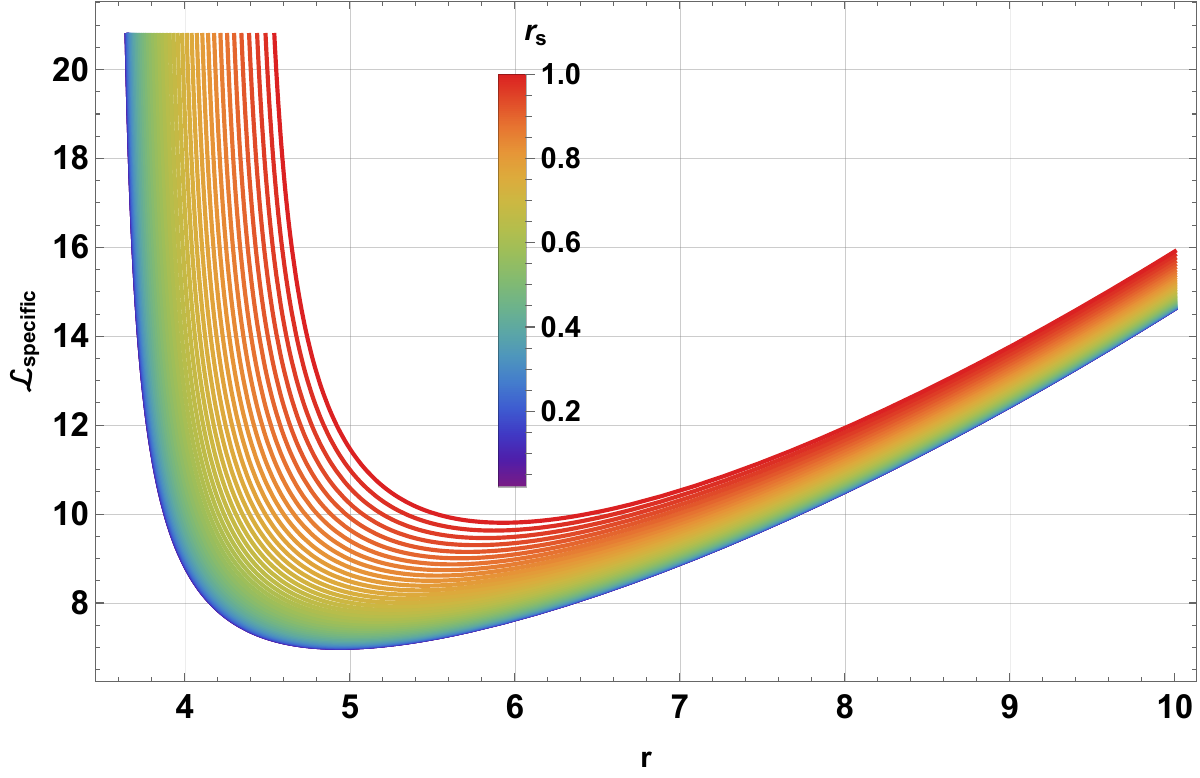}\\
(ii) $\alpha=0.1,\,b=0.5$ \\
\includegraphics[width=0.85\linewidth]{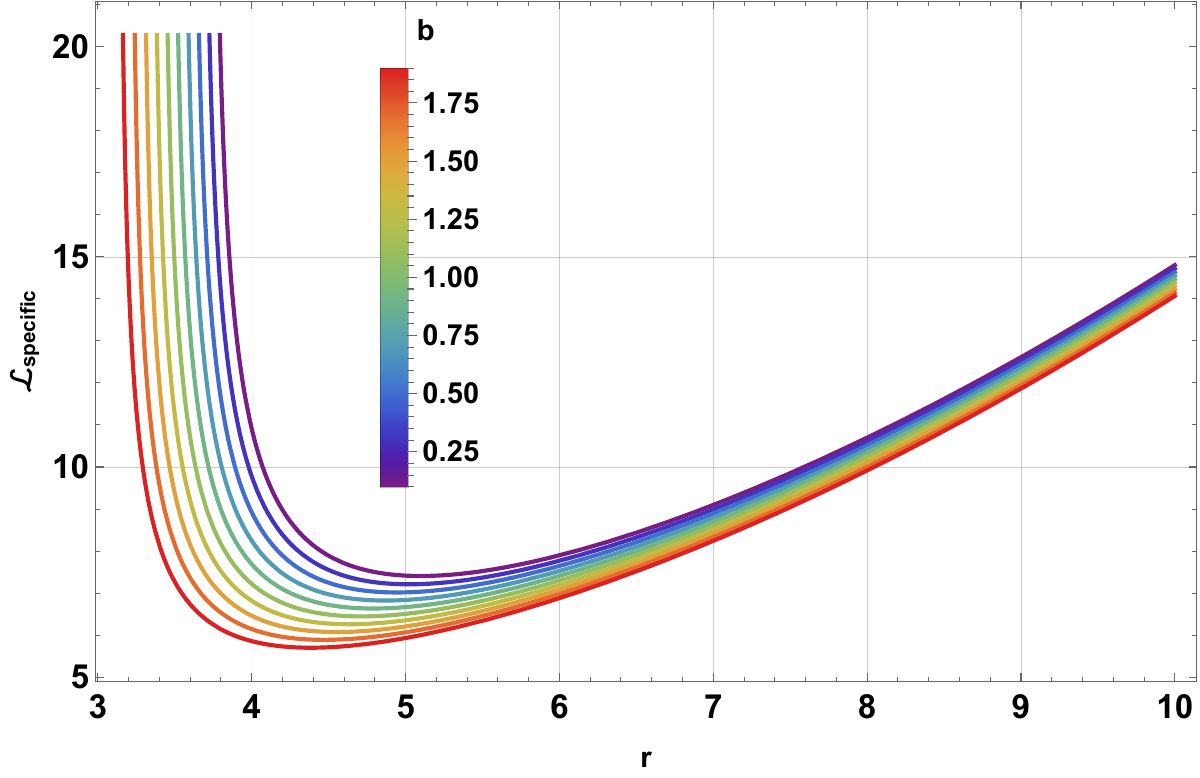}\\
(iii) $r_s=0.2,\,\rho_s=0.02$ \\
\caption{\footnotesize Behavior of the specific angular momentum. $M=1,\,\rho_s=0.01,\,\ell_p=10$}
\label{fig:momentum}
\end{figure}

\begin{figure}[ht!]
\centering
\includegraphics[width=0.85\linewidth]{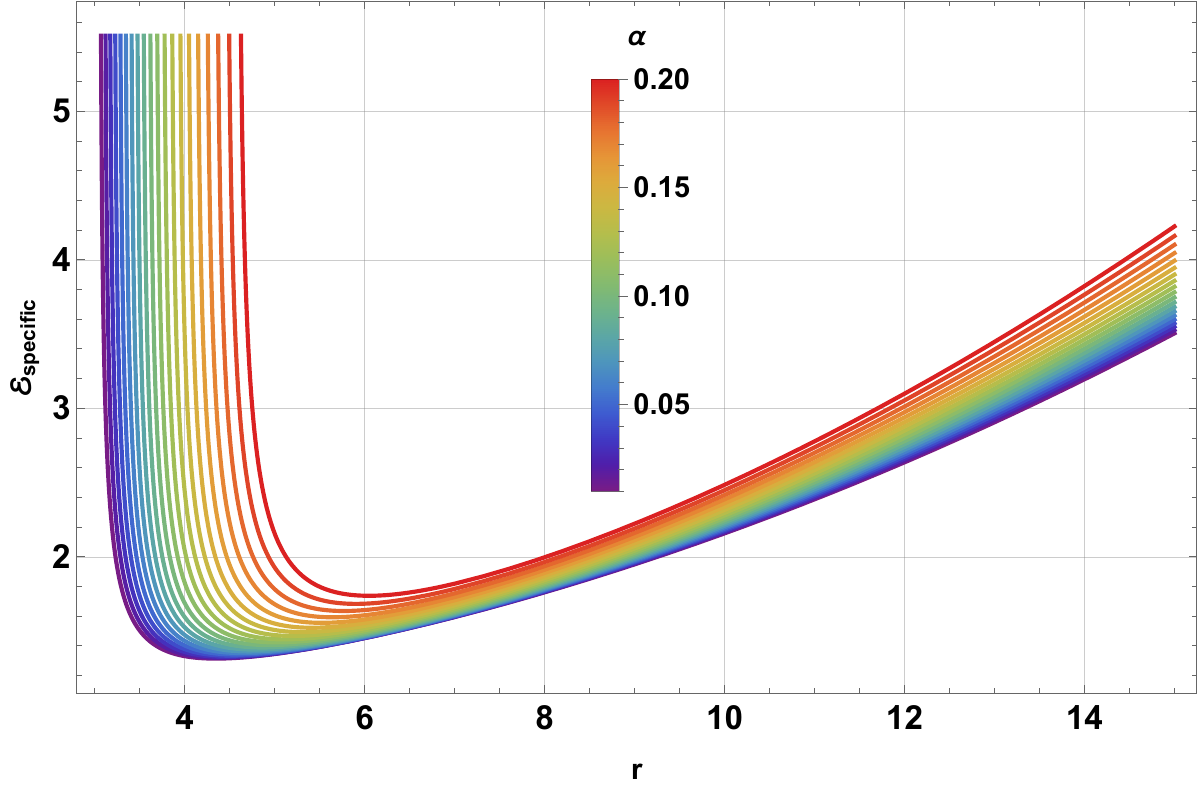}\\
(i) $r_s=0.2,\,b=0.5$\\
\includegraphics[width=0.85\linewidth]{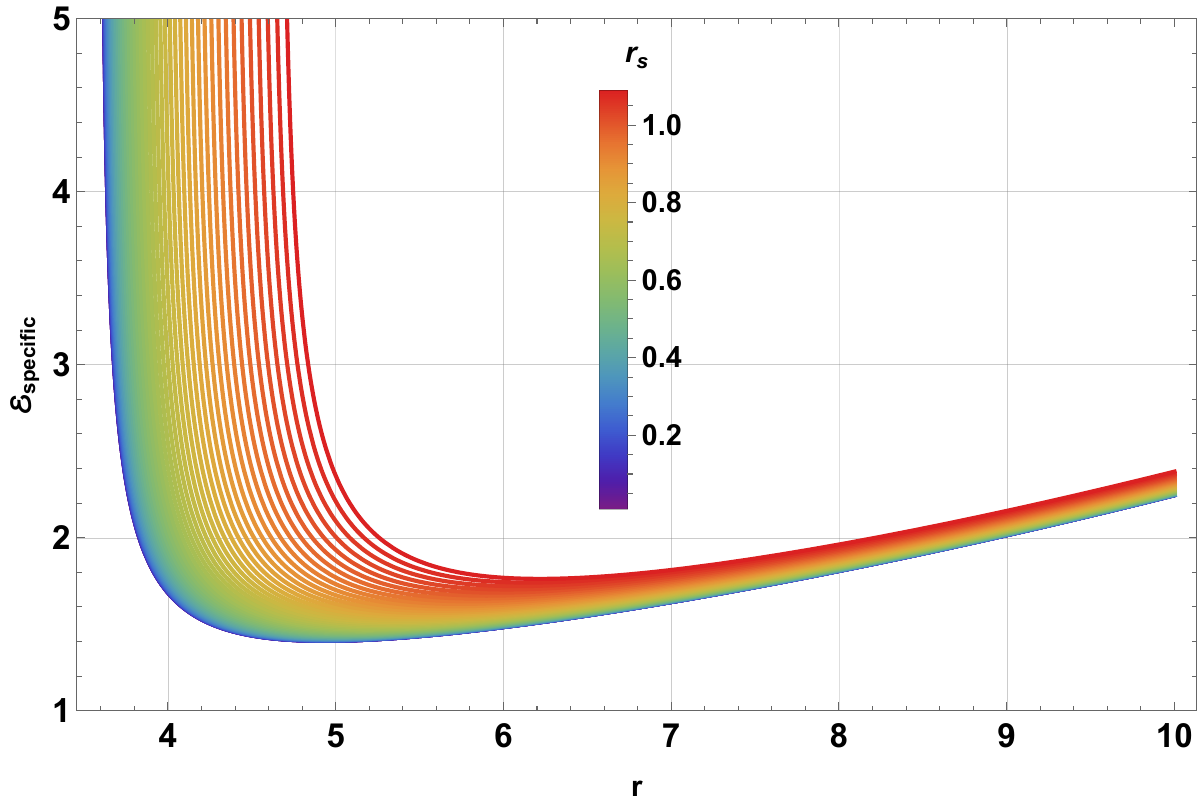}\\
(ii) $\alpha=0.1,\,b=0.5$\\
\includegraphics[width=0.85\linewidth]{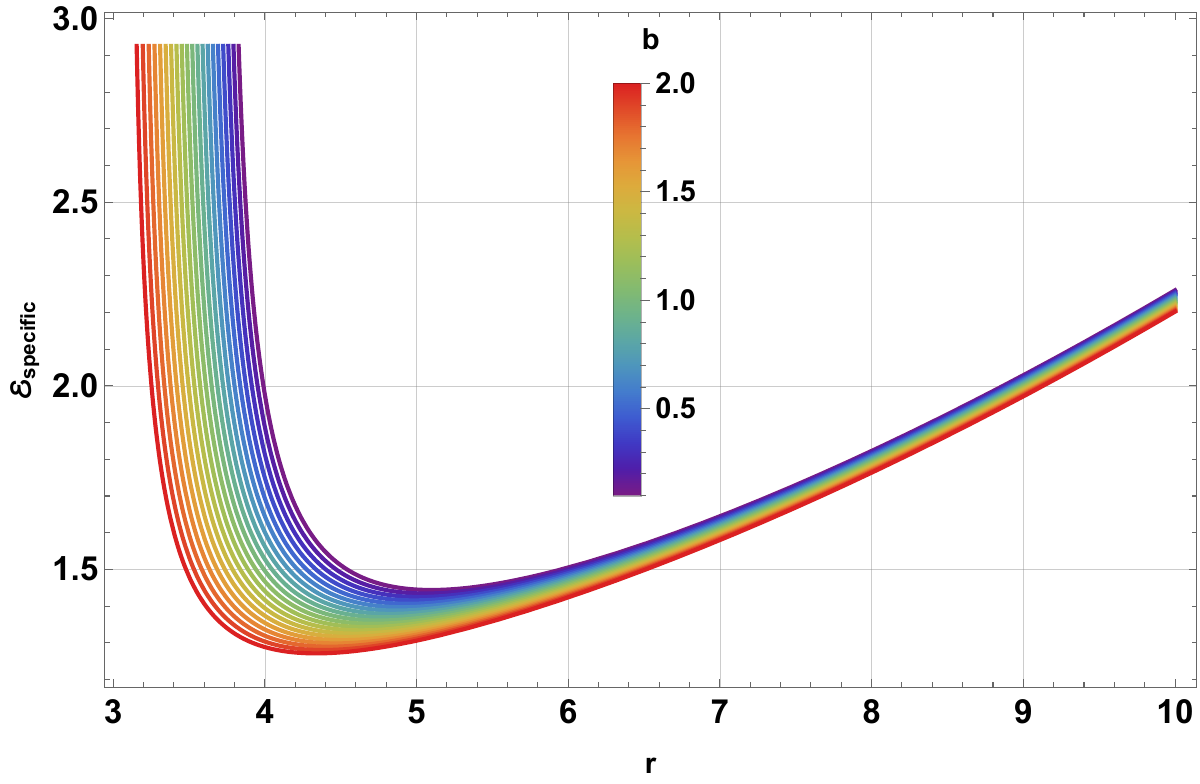}\\
(iii) $\alpha=0.1,\,r_s=0.2$\\
\caption{\footnotesize Behavior of the specific energy. Here $M=1,\,\rho_s=0.01,\,\ell_p=10$.}
\label{fig:energy}
\end{figure}

\begin{center}
\large{\bf I.\, Effective Potential }
\end{center}

The motion of time-like particles in the given gravitational field can be studied in a manner analogous to the previous section. We consider the motion of the test particles in the equatorial plane, defined by $\theta=\pi/2$. Under this assumption, the previous derivation yields the following equations: ~(\ref{bb3}), (\ref{bb4}), and (\ref{bb6}) can be rewritten as
\begin{align}
    \left(\frac{dt}{d\tau}\right)&=\frac{\mathrm{E}}{m\,f(r)}=\frac{\mathcal{E}}{f(r)},\label{dd1}\\
    \left(\frac{dr}{d\tau}\right)^2&=\frac{\mathrm{E}^2}{m^2}-\left(1+\frac{\mathrm{L}^2}{m^2\,r^2}\right)\,f(r)=\mathcal{E}^2-\left(1+\frac{\mathcal{L}^2_0}{r^2}\right)\,f(r),\label{dd2}\\
    \left(\frac{d\phi}{d\tau}\right)&=\frac{\mathrm{L}}{m\,r^2}=\frac{\mathcal{L}_0}{r^2},\label{dd3}
\end{align}
where $\mathcal{E}=\mathrm{E}/m$ and $\mathcal{L}_0=\mathrm{L}/m$ respectively, are the energy and angular momentum of time-like particles per unit mass. 

Writing the equation of motion (\ref{dd2}) as
\begin{equation}
    \left(\frac{dr}{d\tau}\right)^2+U_\text{eff}(r)=\mathcal{E}^2,\label{dd4}
\end{equation}
the effective potential of the system is given by
\begin{align}
U_\text{eff}(r)&=\left(1+\frac{\mathcal{L}^2_0}{r^2}\right)\,\Bigg[1-\frac{2M}{r}-8\pi\,\rho_s\,r_s^2\,\mbox{ln} {\left(1+\frac{r_s}{r}\right)}\notag\\&+\frac{\lvert \alpha\rvert\, b^2}{r^2}\,{}_2F_1\left(-\frac{1}{2},-\frac{1}{4},\frac{3}{4},-\frac{r^4}{b^4}\right)+\frac{r^2}{\ell^2_p}\Bigg].\label{dd5}
\end{align}

In Fig. \ref{fig:potential-timelike}, we depict the effective potential of massive test particles as a function of $r$ by varying the NCS parameters $(\alpha,b)$ and the halo radius $r_s$. In panel (i)-(ii), we observe that potential decreases with increasing $\alpha$ and $r_s$. While in pane (iii), this potential increases with increasing $b$.

\begin{center}
\large{\bf II.\, Innermost Stable Circular Orbits: ISCO}
\end{center}

For the motion of time-like particles in circular orbits of radius $r=r_0$, the conditions $\dot{r}=0$ and $\ddot{r}=0$ must be satisfied. These conditions using Eq. (\ref{dd4}) simplify as
\begin{align}
\mathcal{E}^2&=U_\text{eff}(r)\notag\\&=\left(1+\frac{\mathcal{L}^2}{r^2}\right)\,\Bigg[1-\frac{2M}{r}-8\pi\,\rho_s\,r_s^2\,\mbox{ln} {\left(1+\frac{r_s}{r}\right)}\notag\\&+\frac{\lvert \alpha\rvert\, b^2}{r^2}\,{}_2F_1\left(-\frac{1}{2},-\frac{1}{4},\frac{3}{4},-\frac{r^4}{b^4}\right)+\frac{r^2}{\ell^2_p}\Bigg],\label{dd6}
\end{align}
and
\begin{equation}
\frac{dU_\text{eff}}{dr}\Bigg{|}_{r=\mbox{const.}}=0.\label{dd7}
\end{equation}

\begin{figure}[ht!]
    \centering
    \includegraphics[width=0.85\linewidth]{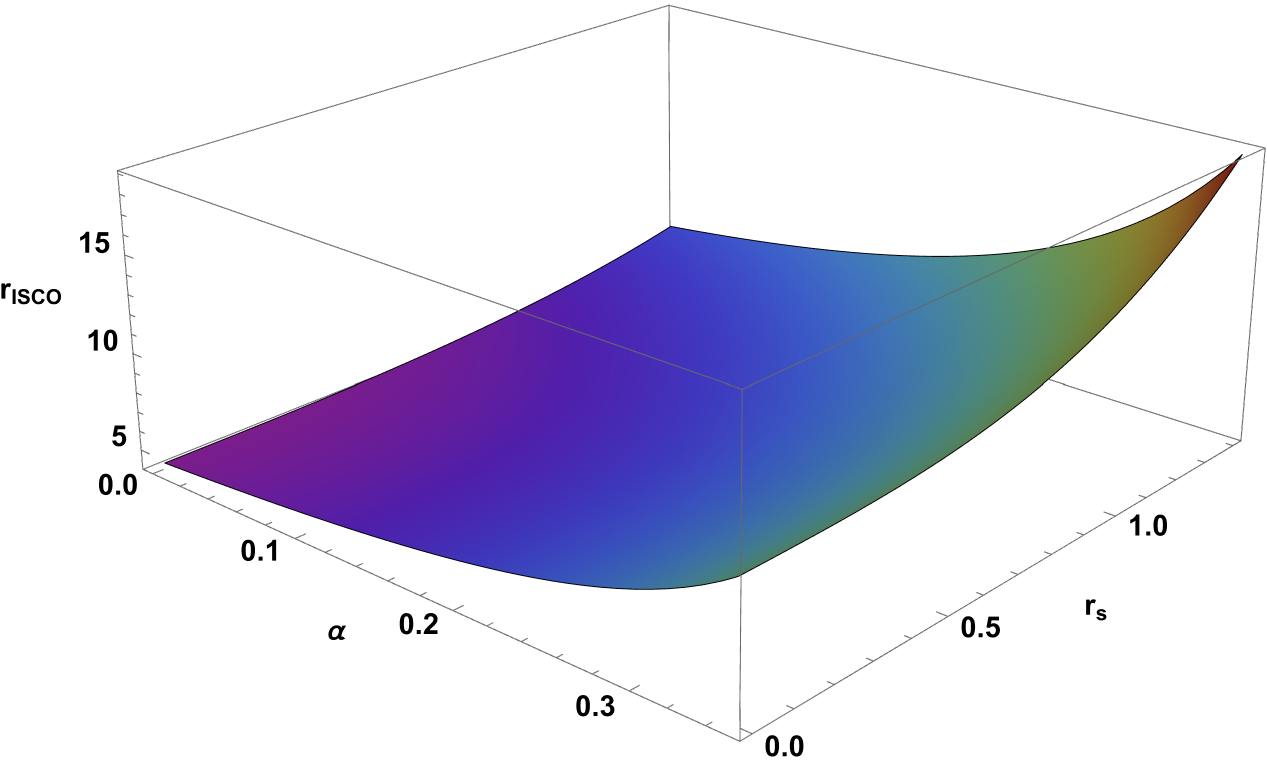}\\
    (i) $b=0.5$\\
    \includegraphics[width=0.85\linewidth]{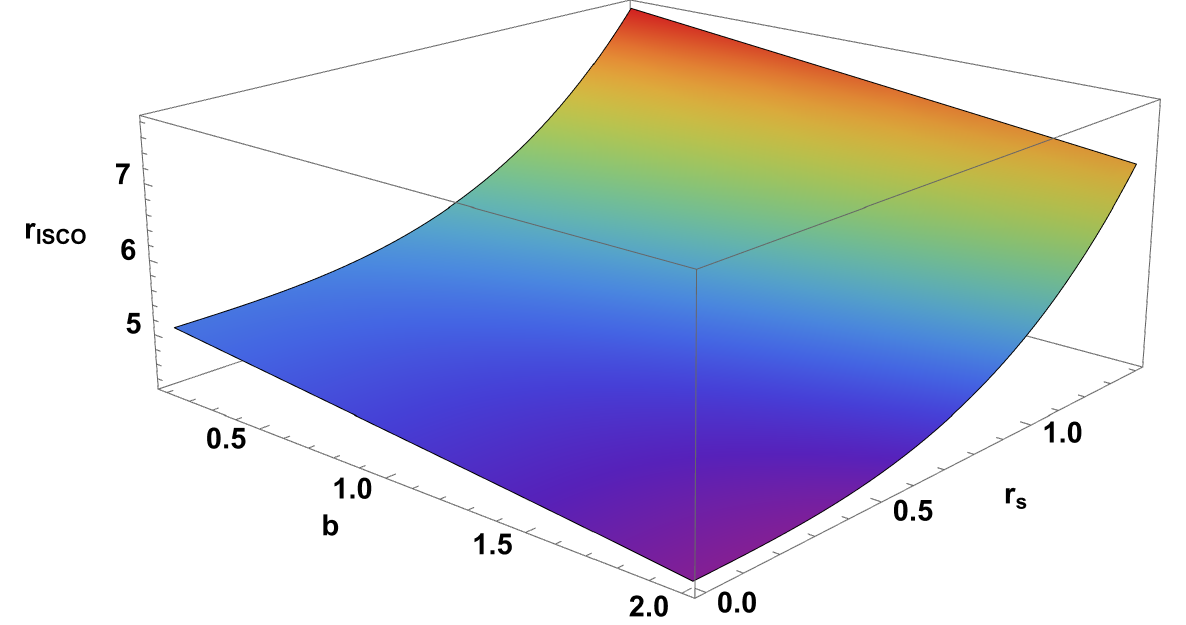}\\
    (ii) $\alpha=0.1$\\
    \caption{\footnotesize 3D plot of the ISCO radius as a function of $(\alpha,r_s)$ and $(b,r_s)$. Here $M=1,\,\rho_s=0.02,\ell_p=10$.}
    \label{fig:ISCO-radius}
\end{figure}

Simplification of the relation (\ref{dd6}) and (\ref{dd7}) using Eq. (\ref{dd5}) results
\begin{widetext}
\begin{align}
\mathcal{L}_\text{specific}&=r\,\sqrt{\frac{\frac{M}{r} + \frac{4\pi \rho_s r_s^3}{r + r_s} 
- \frac{|\alpha| b^2}{r^2}\, {}_2F_1\left(-\frac{1}{2}, -\frac{1}{4}, \frac{3}{4}, -\frac{r^4}{b^4} \right) 
- \frac{|\alpha| r^2}{3b^2}\, {}_2F_1\left( \frac{1}{2}, \frac{3}{4}, \frac{7}{4}, -\frac{r^4}{b^4} \right) 
+ \frac{r^2}{\ell_p^2}}{1-\frac{3M}{r}-8\pi\,\rho_s\,r_s^2\,\mbox{ln} {\left(1+\frac{r_s}{r}\right)}-\frac{4\pi \rho_s r_s^3}{r + r_s}+\frac{2\lvert \alpha\rvert\, b^2}{r^2}\,{}_2F_1\left(-\frac{1}{2},-\frac{1}{4},\frac{3}{4},-\frac{r^4}{b^4}\right)+\frac{|\alpha| r^2}{3b^2} \cdot {}_2F_1\left( \frac{1}{2}, \frac{3}{4}, \frac{7}{4}, -\frac{r^4}{b^4} \right)}},\label{dd8}\\
\mathcal{E}_\text{specific}&=\pm\,\frac{\left(1-\frac{2M}{r}-8\pi\,\rho_s\,r_s^2\,\mbox{ln} {\left(1+\frac{r_s}{r}\right)}+\frac{\lvert \alpha\rvert\, b^2}{r^2}\,{}_2F_1\left(-\frac{1}{2},-\frac{1}{4},\frac{3}{4},-\frac{r^4}{b^4}\right)+\frac{r^2}{\ell^2_p}\right)}{\sqrt{1-\frac{3M}{r}-8\pi\,\rho_s\,r_s^2\,\mbox{ln} {\left(1+\frac{r_s}{r}\right)}-\frac{4\pi \rho_s r_s^3}{r + r_s}+\frac{2\lvert \alpha\rvert\, b^2}{r^2}\,{}_2F_1\left(-\frac{1}{2},-\frac{1}{4},\frac{3}{4},-\frac{r^4}{b^4}\right)+\frac{|\alpha| r^2}{3b^2} \cdot {}_2F_1\left( \frac{1}{2}, \frac{3}{4}, \frac{7}{4}, -\frac{r^4}{b^4} \right)}}.\label{dd9}
\end{align}
\end{widetext}
Here $\mathcal{L}_\text{specific}$ and $\mathcal{E}_\text{specific}$, respectively, represent the specific angular momentum and specific energy of test particles orbiting around the selected BH. 

One can see that BH mass $M$, the curvature radius $\ell_p$, the NCS parameters $(|\alpha|,b)$, and the DMH profile characterized by $(r_s,\rho_s)$, modify these physical quantities associated with time-like particles moving in circular orbits around the BH.

In Figure~\ref{fig:momentum}, we present plots showing the behavior of the specific angular momentum of test particles as a function of the circular orbit radius $r = r_c$, by varying NCS parameters \(\alpha\), \(b\), and the halo radius \(r_s\), while keeping all other parameters fixed. Panels (i) and (ii) illustrate that the specific angular momentum increases with increasing \(\alpha\) and \(r_s\), respectively. In contrast, panel (iii) shows that the angular momentum decreases as the parameter \(b\) increases.

Similarly, in Figure~\ref{fig:energy}, we plot the specific energy of test particles as a function of \(r = r_c\) under variations of the same parameters: \(\alpha\), \(b\), and \(r_s\), with all other parameters held fixed. The qualitative behavior of the specific energy mirrors that of the particular angular momentum, specifically, it increases with \(\alpha\) and \(r_s\), and decreases with \(b\).

\begin{table*}[tbhp]
\centering
\begin{tabular}{|c||c|c|c|c|c|c|c|}
\hline
$\alpha \backslash r_s$ & 0.2 & 0.4 & 0.6 & 0.8 & 1.0 & 1.2 & 1.4 \\
\hline\hline
0.05 & 4.58392 & 4.63625 & 4.77264 & 5.03017 & 5.44766 & 6.06950 & 6.94770 \\
0.10 & 4.94477 & 5.00387 & 5.15855 & 5.45172 & 5.92859 & 6.64069 & 7.64783 \\
0.15 & 5.41530 & 5.48337 & 5.66225 & 6.00260 & 6.55792 & 7.38887 & 8.56482 \\
0.20 & 6.05467 & 6.13501 & 6.34705 & 6.75198 & 7.41442 & 8.40685 & 9.81075 \\
0.25 & 6.97052 & 7.06841 & 7.32788 & 7.82507 & 8.64003 & 9.86129 & 11.5863 \\
0.30 & 8.37846 & 8.50299 & 8.83450 & 9.47165 & 10.5174 & 12.0834 & 14.2906 \\
\hline
\end{tabular}
\caption{\footnotesize Numerical values of ISCO radius $ r_\text{ISCO} $ for varying values of $ \alpha $ and $ r_s $. Here $M=1,\,\rho_s=0.02,\,b=0.5,\,\ell_p=10$.}
\label{tab:3}
\hfill\\
\centering
\begin{tabular}{|c||c|c|c|c|c|c|c|}
\hline
$b \backslash r_s$ & 0.2 & 0.4 & 0.6 & 0.8 & 1.0 & 1.2 & 1.4 \\
\hline\hline
0.2 & 5.0597 & 5.1189 & 5.27394 & 5.56796 & 6.04622 & 6.76019 & 7.76939 \\
0.4 & 4.98311 & 5.04224 & 5.19704 & 5.49049 & 5.96782 & 6.68054 & 7.68837 \\
0.6 & 4.90638 & 4.96546 & 5.12002 & 5.41292 & 5.88932 & 6.60081 & 7.60726 \\
0.8 & 4.82938 & 4.88841 & 5.04275 & 5.33513 & 5.81064 & 6.52092 & 7.52602 \\
1.0 & 4.75184 & 4.81085 & 4.96502 & 5.25695 & 5.73165 & 6.44079 & 7.44461 \\
1.2 & 4.67342 & 4.73244 & 4.88653 & 5.17813 & 5.65215 & 6.36030 & 7.36293 \\
\hline
\end{tabular}
\caption{\footnotesize Numerical values of ISCO radius $ r_\text{ISCO} $ for various values of $b$ and $r_s$.  Here $M=1,\,\rho_s=0.02,\,\alpha=0.1,\,\ell_p=10$.}
\label{tab:4}
\end{table*}

The next important feature of massive test particles traversing around the BH is the innermost stable circular orbits. The ISCO corresponds to the smallest radius at which a test particle can stably orbit a BH. Inside the ISCO, circular orbits become unstable, and particles either plunge into the BH or move outward. The ISCO thus marks the transition between stable circular motion and dynamical instability.

For a circular orbit at radius $r_{c}$, the following conditions must hold:
\begin{itemize}
\item Existence of circular orbit: 
\[\frac{dU_{\text{eff}}}{dr}\Big|_{r=r_{c}} = 0.\]
  
\item Stability of circular orbit:
\[
\frac{d^{2}U_{\text{eff}}}{dr^{2}}\Big|_{r=r_{c}} > 0.\]
  
\item ISCO condition: The ISCO corresponds to the marginally stable orbit, where stability is lost, i.e.,
\[\frac{d^{2}U_{\text{eff}}}{dr^{2}}\Big|_{r=r_{\text{ISCO}}} = 0.\]
\end{itemize}

Substituting the effective potential given in Eq. (\ref{dd1}) into the ISCO condition results in the following polynomial equation:
\begin{align}
3\,f(r)\,f'(r)+r\,f''(r)\,f(r)-2\,r\,(f'(r))^2=0.\label{dd10}
\end{align}

\begin{figure}[ht!]
    \centering
    \includegraphics[width=0.75\linewidth]{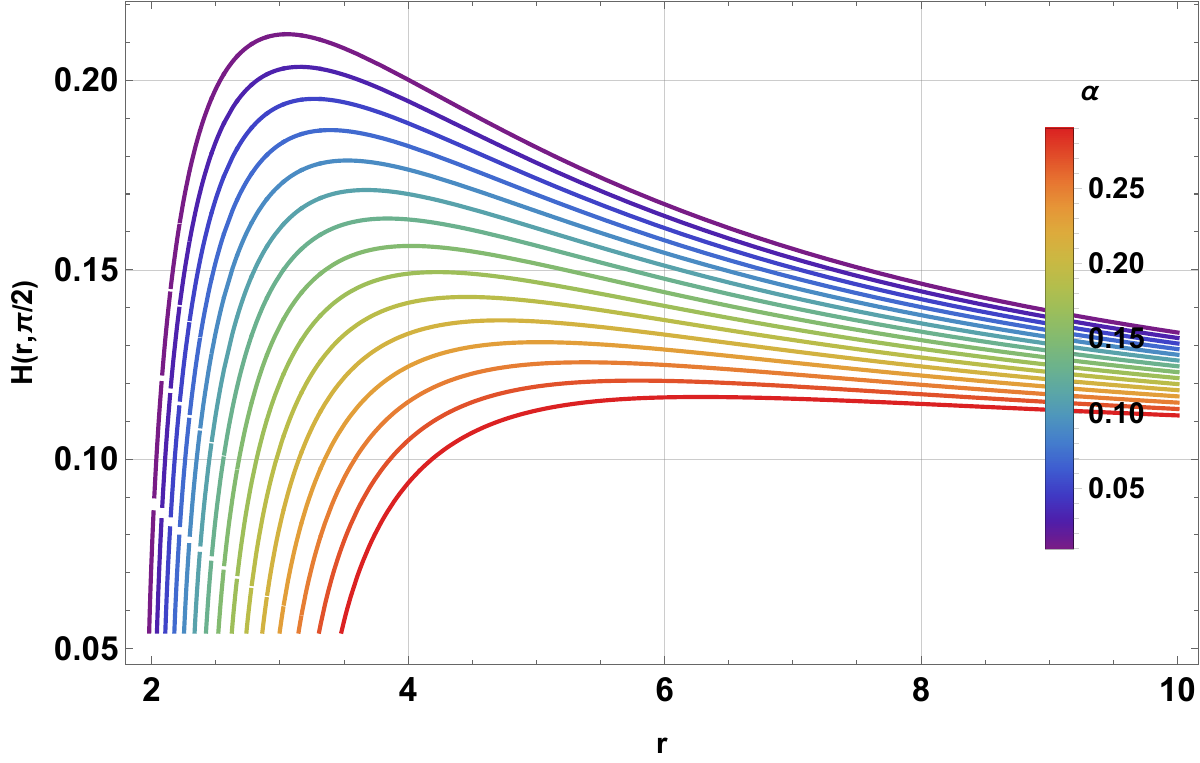}\\
    (i) $b=0.5,\,r_s=0.2$\\
    \includegraphics[width=0.75\linewidth]{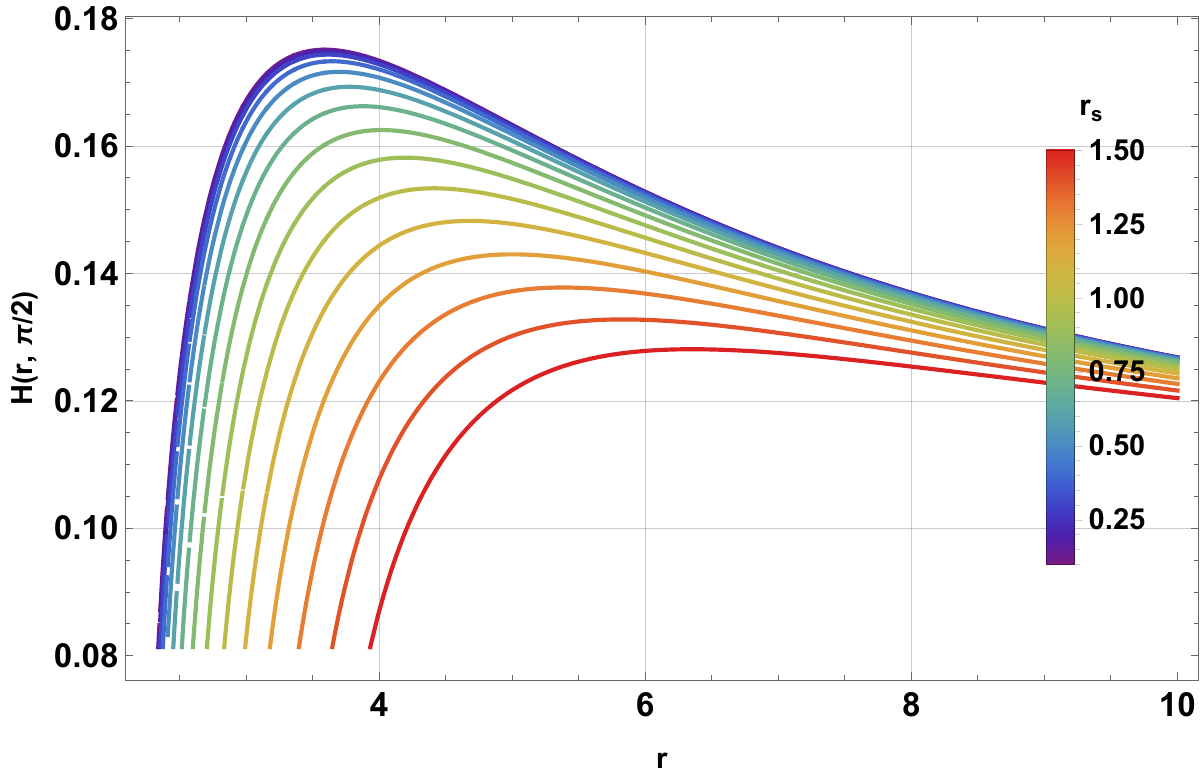}\\
    (ii) $\alpha=0.1,\,b=0.5$\\
    \includegraphics[width=0.75\linewidth]{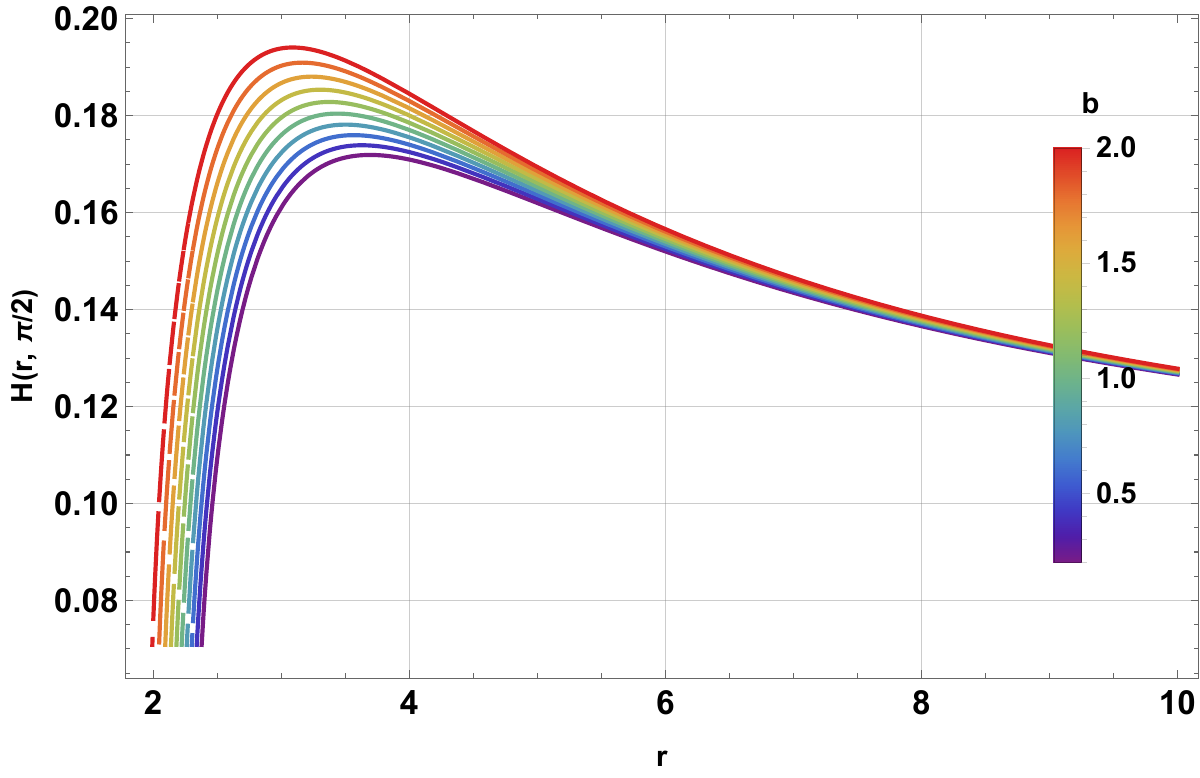}\\
    (iii) $\alpha=0.1,\,r_s=0.2$\\    
    \caption{\footnotesize Behavior of the potential function $H(r, \theta)$ for different values of $\alpha$, $b$ and $r_s$. Here $M=1,\,\rho_s=0.02,\,\ell_p=10,\,\theta=\pi/2$.}
    \label{fig:potential-function}
\end{figure}

Substituting the metric function $f(r)$ into Eq. (\ref{dd10}), one will arrive at a polynomial equation in $r$ whose exact solution is a bit of a challenge. However, the ISCO radius $r=r_\text{ISCO}$ using this polynomial equation can be determined numerically by assigning suitable values to the parameters involved in the equation. 

In Tables \ref{tab:3}-\ref{tab:4}, we present numerical values of ISCO radius by varying NCS parameters $(|\alpha|, b)$ and the DMH radius $r_s$, keeping other fixed values.

In Figure~\ref{fig:ISCO-radius}, we present a 3D plot of the ISCO radius \( r_\text{ISCO} \) as a function of the parameter pairs \( (\alpha, r_s) \) and \( (b, r_s) \).

\subsection{Topological Features of Photon Rings}

Topologically, photon rings are unstable closed null orbits, and this instability is universal: any small perturbation pushes photons either toward the horizon or out to infinity. This unstable character underlies the formation of the BH shadow, since the critical impact parameter associated with photon rings defines the shadow boundary \cite{perlick}. Moreover, recent topological analyses have shown that the number and stability type of photon rings are constrained by index theorems, ensuring that BHs generally possess at least one unstable photon ring \cite{liu2024}. These topological properties connect fundamental aspects of spacetime geometry with observational signatures such as shadows and lensing patterns.

To study the topological property of the light rings, one can introduce a potential function as \cite{BB2,BB3}
\begin{widetext}
 \begin{equation}
H(r,\theta)=\sqrt{-\frac{g_{tt}}{g_{\theta\theta}}}=\frac{\sqrt{1-\frac{2M}{r}-8\pi\,\rho_s\,r_s^2\,\mbox{ln} {\left(1+\frac{r_s}{r}\right)}+\frac{\lvert \alpha\rvert\, b^2}{r^2}\,{}_2F_1\left(-\frac{1}{2},-\frac{1}{4},\frac{3}{4},-\frac{r^4}{b^4}\right)+\frac{r^2}{\ell^2_p}}}{r\,\sin \theta},\label{cc1}
\end{equation}   
\end{widetext}
where the function $H(r, \theta)$ is regular for $r >r_h$, the horizon radius. One can show that the photon sphere radius can occur by the condition $\partial_r H(r, \theta)=0$.

From the above expression (\ref{cc1}), we observe that geometric and physical parameters, such as the BH mass $M$, the string cloud parameters $(\alpha, b)$, the curvature radius $\ell_p$, and the DM profile characterized by $(r_s,\rho_s)$, modify this potential function.

In Figure~\ref{fig:potential-function}, we illustrate the behavior of the potential function $H(r, \theta)$ by varying the parameter \(\alpha, r_s\) and \( b\). In panels (i) and (ii), we observed that as \(\alpha\) and \(r_s\) increase, the potential function reduces its value for a particular angular coordinate $\theta=\pi/2$. In contrast, panel (iii) shows an increase in this potential function with increasing values of parameter $b$.

The key vector field ${\bf v}=(v_r\,,\,v_{\theta})$ using the definition in \cite{AA2,AA3} is given as
\begin{widetext}
\begin{align}
v_r&=-\frac{\left[1-\frac{3M}{r}-8\pi\,\rho_s\,r_s^2\,\mbox{ln} {\left(1+\frac{r_s}{r}\right)}-\frac{4\pi \rho_s r_s^3}{r + r_s}+\frac{2\lvert \alpha\rvert\, b^2}{r^2}\,{}_2F_1\left(-\frac{1}{2},-\frac{1}{4},\frac{3}{4},-\frac{r^4}{b^4}\right)+\frac{|\alpha| r^2}{3b^2} \cdot {}_2F_1\left( \frac{1}{2}, \frac{3}{4}, \frac{7}{4}, -\frac{r^4}{b^4} \right)\right]}{r^2\,\sin\theta},\label{cc2}\\
v_{\theta}&=-\frac{\sqrt{1-\frac{2M}{r}-\rho_s\,r_s^2\,\mbox{ln} {\left(1+\frac{r_s}{r}\right)}+\frac{\lvert \alpha\rvert\, b^2}{r^2}\,{}_2F_1\left(-\frac{1}{2},-\frac{1}{4},\frac{3}{4},-\frac{r^4}{b^4}\right)+\frac{r^2}{\ell^2_p}}}{r^2}\,\frac{\cot \theta}{\sin \theta}.\label{cc3} 
\end{align}  
\end{widetext}
Consequently, the normalized field components read
\begin{align}
n_r=\frac{v_r}{\sqrt{v^2_r+v^2_{\theta}}},\nonumber\\
n_{\theta}=\frac{v_{\theta}}{\sqrt{v^2_r+v^2_{\theta}}}.\label{cc4}   
\end{align}
At $(r,\theta)=(r_{\rm ph},\pi/2)$ one recovers the zero of the unit field ${\bf n}$ as expected.

\begin{figure*}[ht!]
    \centering
    \includegraphics[width=0.31\linewidth]{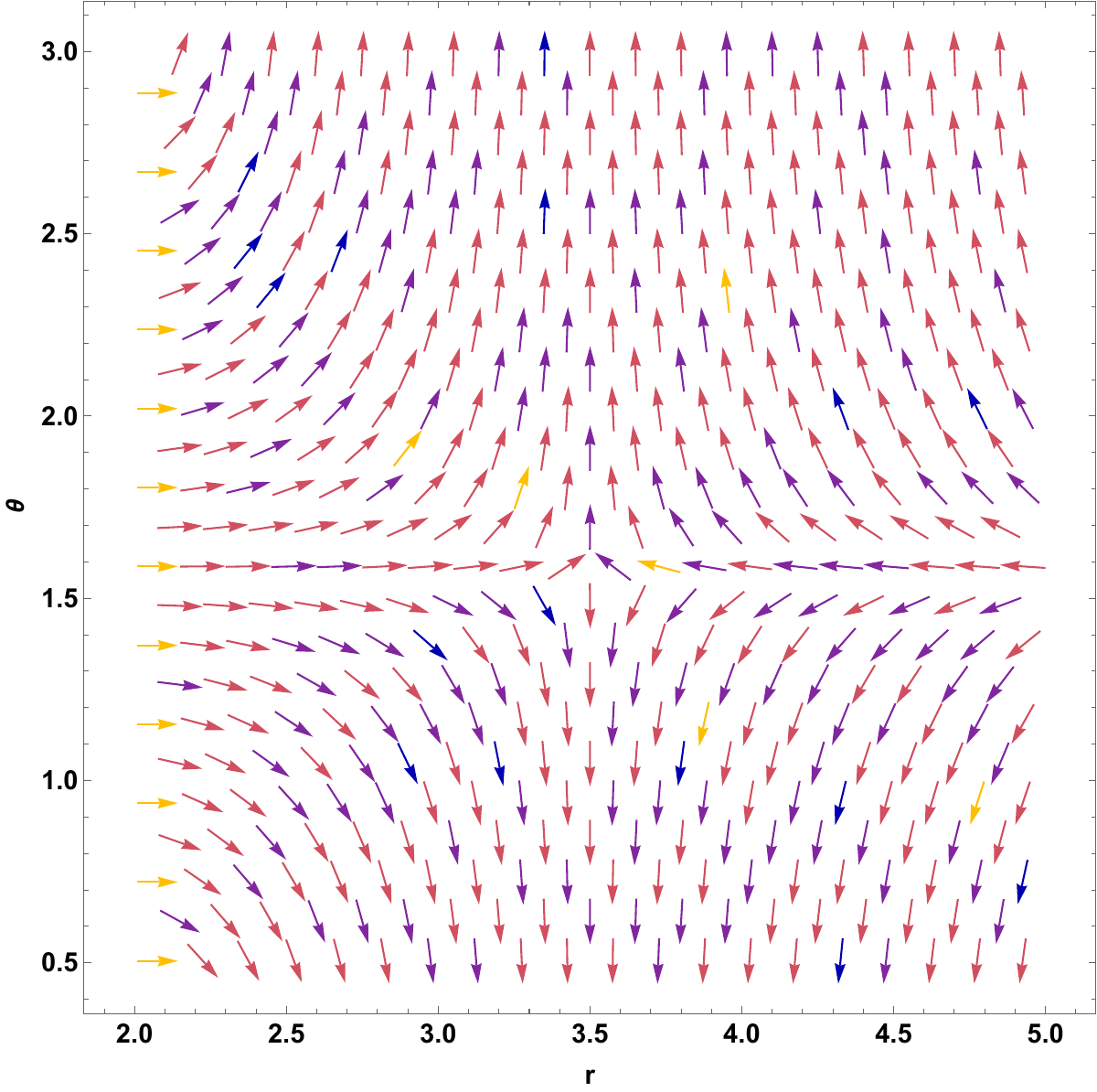}\quad
    \includegraphics[width=0.31\linewidth]{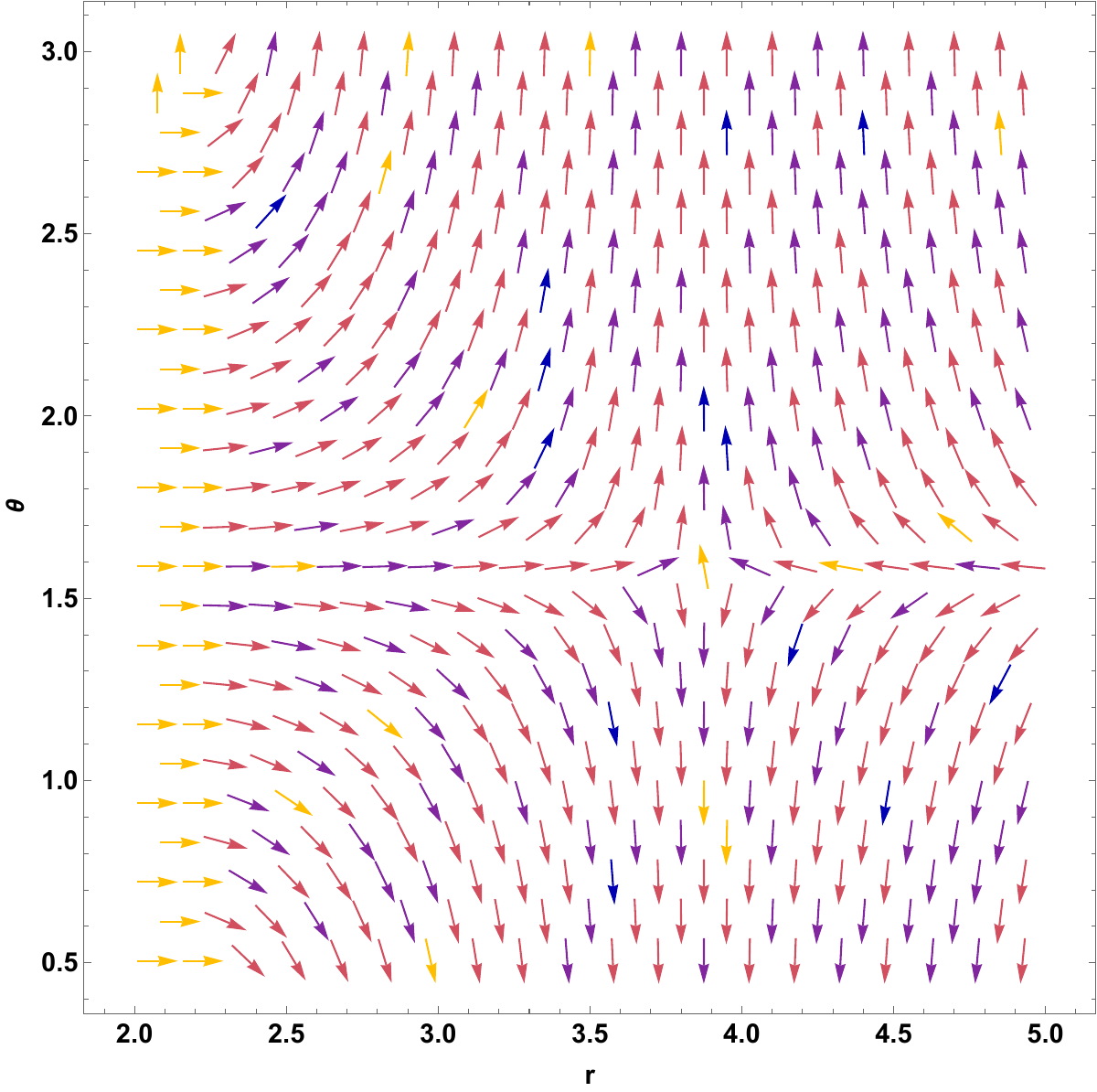}\quad
    \includegraphics[width=0.31\linewidth]{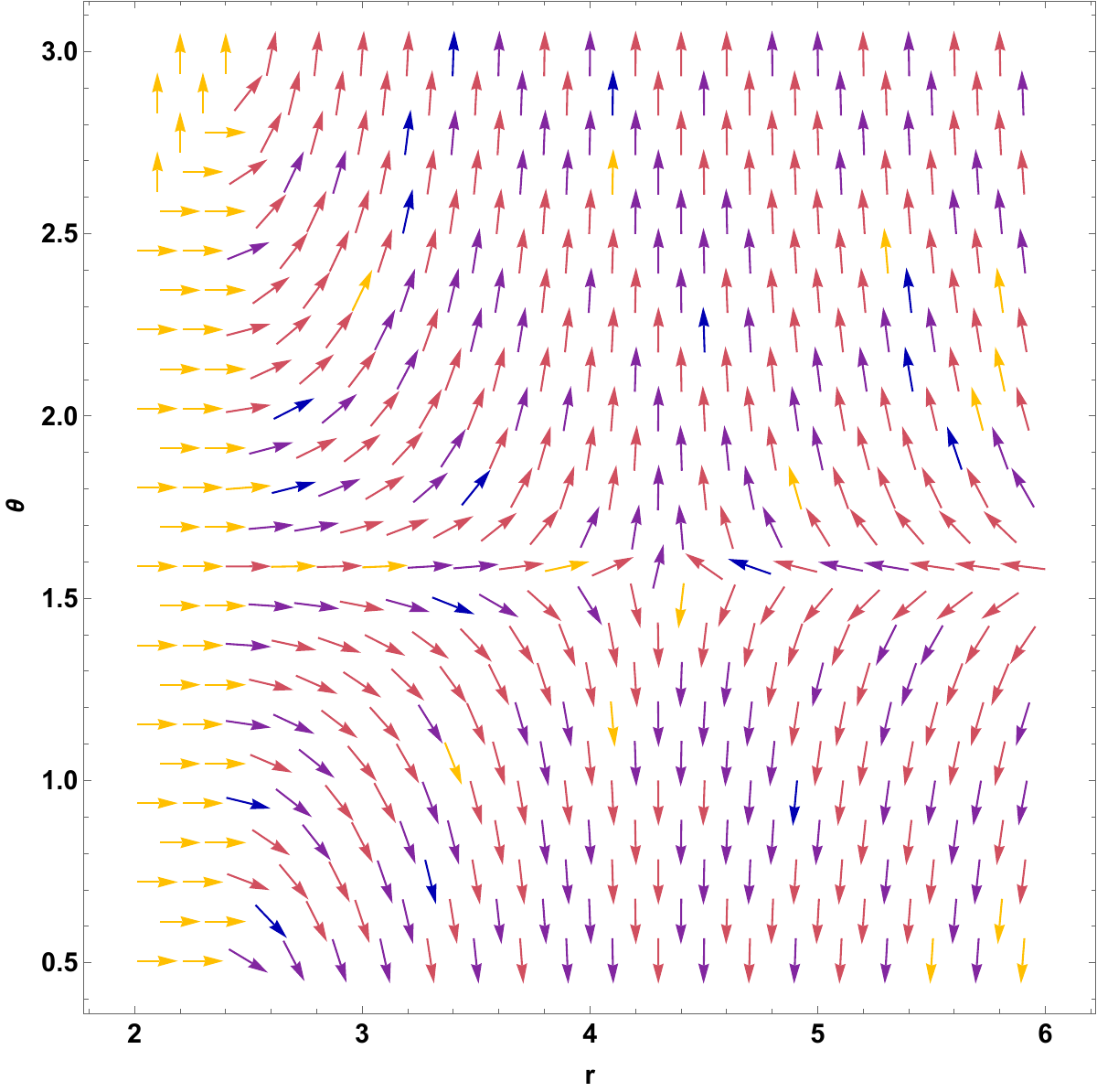}\\
    (i) $\alpha=0.05$ \hspace{4cm} (ii) $\alpha=0.10$ \hspace{4cm} (iii) $\alpha=0.15$
    \caption{\footnotesize The arrows represent the unit vector field ${\bf n}$ on a portion of the $r-\theta$ plane for the BH with different $\alpha$. $M=1\,\ell_p=10,\,\rho_s=0.05,\,b=0.5,\,r_s=0.5$}
    \label{fig:unit-vector-1}
\end{figure*}

\begin{figure*}[ht!]
    \centering
    \includegraphics[width=0.31\linewidth]{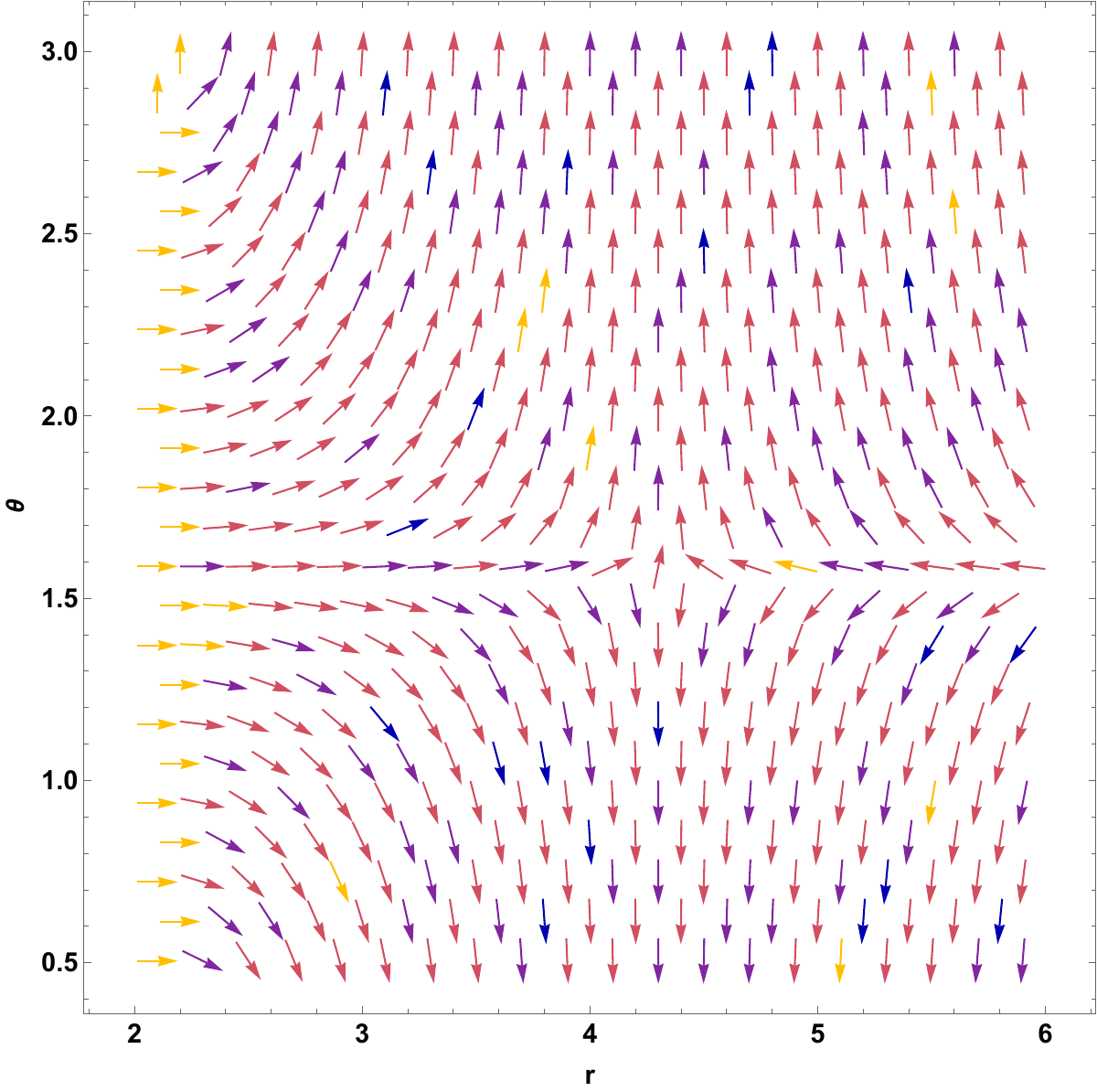}\quad
    \includegraphics[width=0.31\linewidth]{unit-vector-fig-2-b.pdf}\quad
    \includegraphics[width=0.31\linewidth]{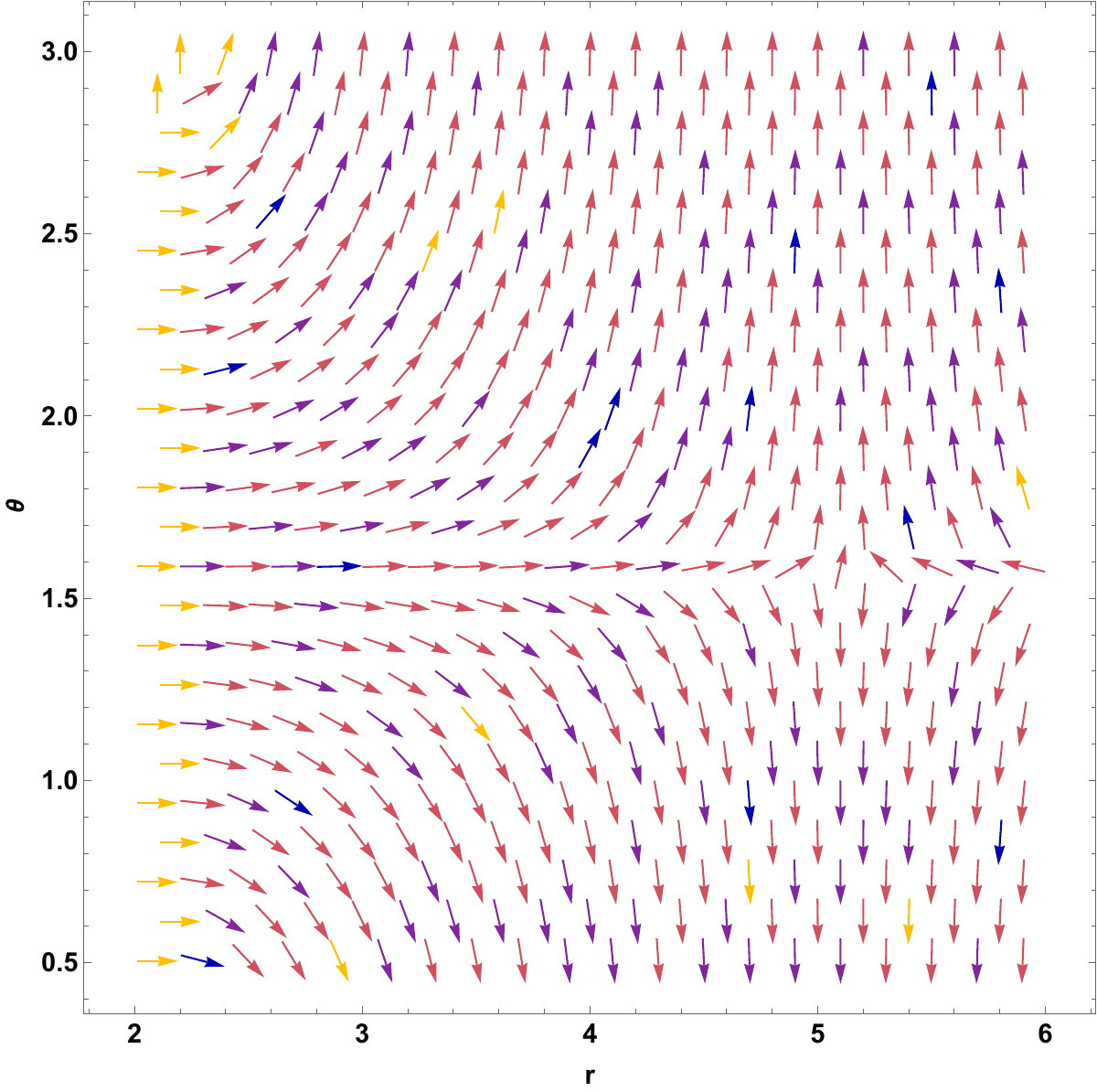}\\
    (i) $r_s=0.3$ \hspace{4cm} (ii) $r_s=0.7$ \hspace{4cm} (iii) $r_s=0.9$
    \caption{\footnotesize The arrows represent the unit vector field ${\bf n}$ on a portion of the $r-\theta$ plane for the BH with different $r_s$. $M=1\,\ell_p=10,\,\rho_s=0.05,\,b=0.5,\,\alpha=0.a$}
    \label{fig:unit-vector-2}
\end{figure*}

\begin{figure*}[ht!]
    \centering
    \includegraphics[width=0.31\linewidth]{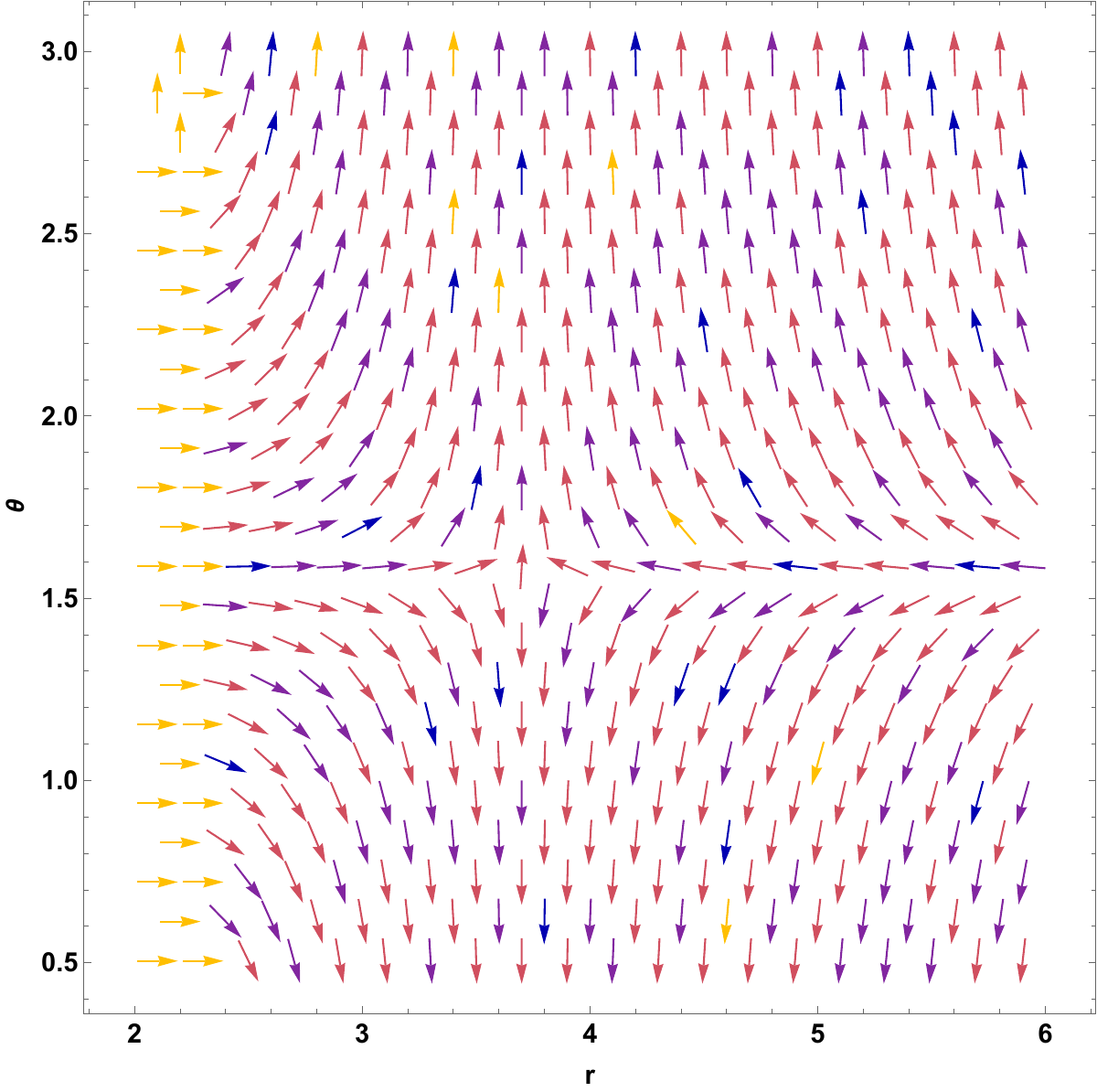}\quad
    \includegraphics[width=0.31\linewidth]{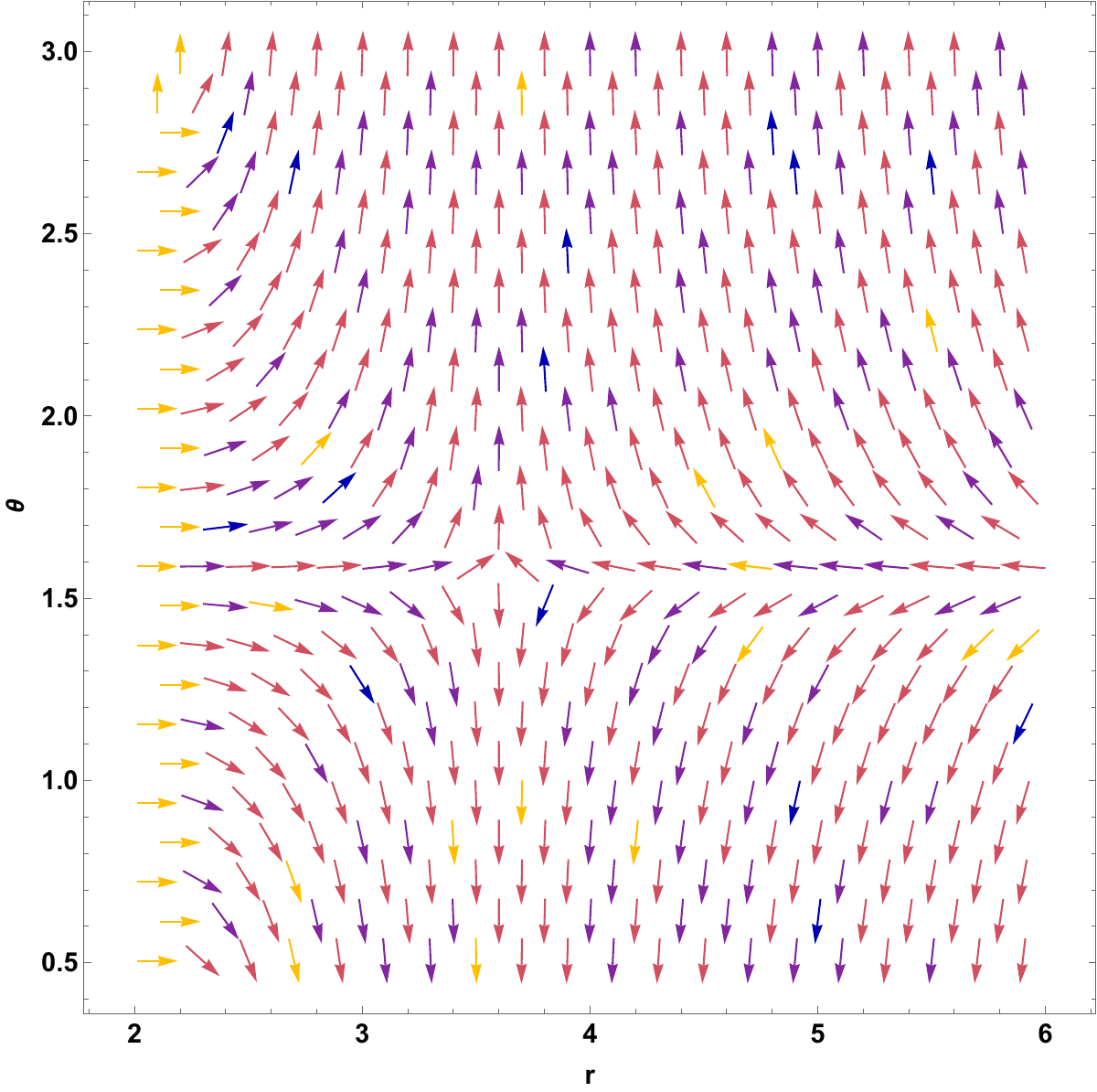}\quad
    \includegraphics[width=0.31\linewidth]{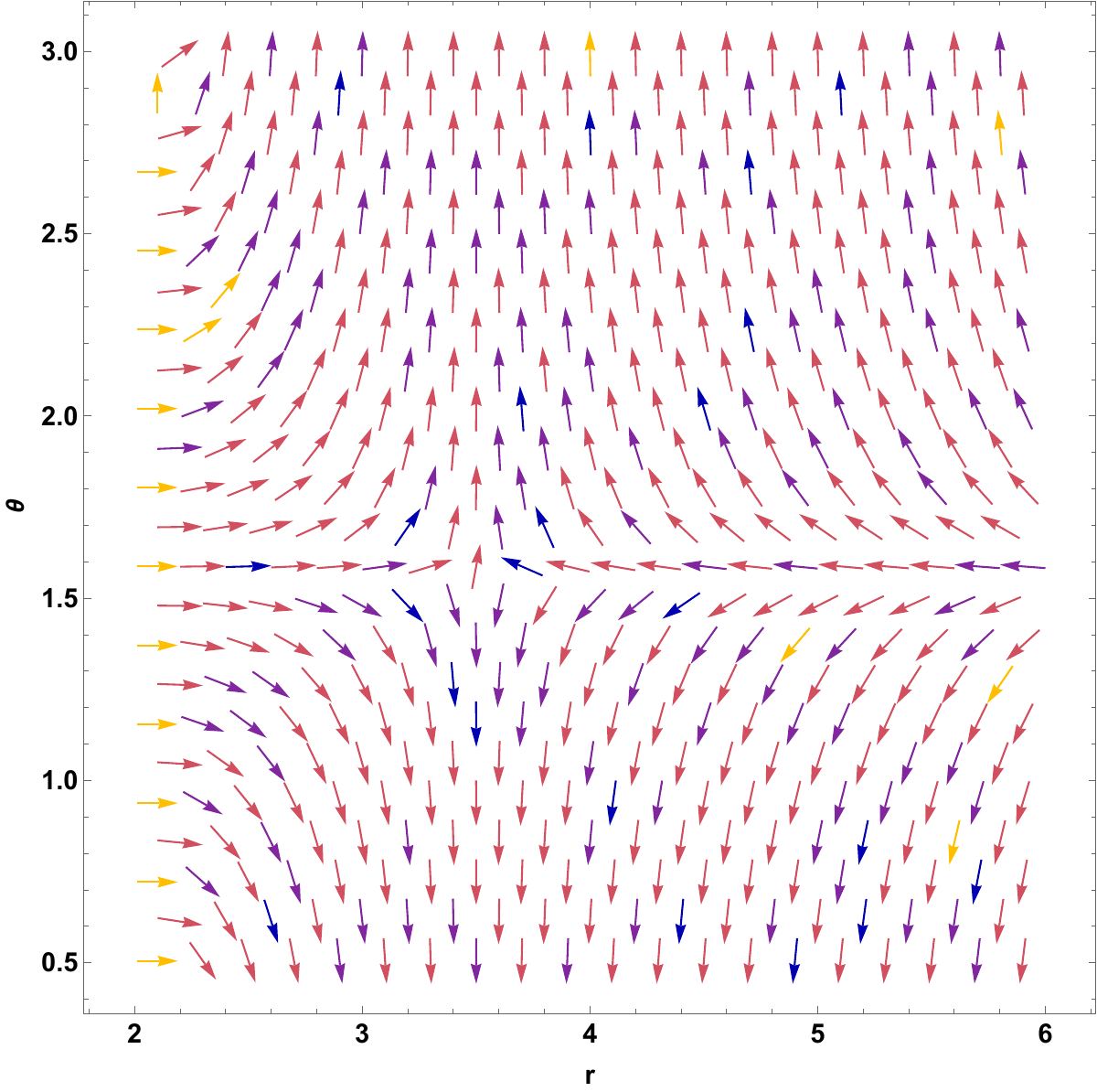}\\
    (i) $b=0.2$ \hspace{4cm} (ii) $b=0.5$  \hspace{4cm} (iii) $b=0.8$
    \caption{\footnotesize The arrows represent the unit vector field ${\bf n}$ on a portion of the $r-\theta$ plane for the BH with different $b$. $M=1\,\ell_p=10,\,\rho_s=0.05,\,\alpha=0.a,\,r_s=0.2$}
    \label{fig:unit-vector-3}
\end{figure*}

In Figure \ref{fig:unit-vector-1} to \ref{fig:unit-vector-3}, we depict the normalized vector field in $r-\theta$ plane for different values of the parameter \(\alpha\), \(r_s\) and \(b\). The arrows represent the normalized vector field for the BH solution. 

\section{Thermodynamics}\label{Sec:IV}

Black-hole thermodynamics intertwines classical geometry, quantum field theory in curved spacetime, and statistical mechanics. The area law and Hawking radiation elevate black holes to genuine thermodynamic systems with entropy and temperature~\cite{Bekenstein1973,Hawking1975}. In asymptotically AdS backgrounds, a well-behaved canonical ensemble emerges and accommodates the Hawking-Page transition between thermal AdS and large AdS black holes~\cite{HawkingPage1983}. In the modern ``black-hole chemistry'' framework, the cosmological constant acts as a pressure and the ADM mass becomes enthalpy, yielding consistent first-law and Smarr relations and enabling Van der Waals-like criticality~\cite{KastorRayTraschen2009,Dolan2011,CveticEtAl2011,KubiznakMann2012,KubiznakMannTeo2017}. Here we adapt this thermodynamic program to the full metric function \eqref{function-1}, in which the cloud of strings (NCS) contributes a scale-dependent hypergeometric piece governed by the parameter $b$. This new structure deforms temperature extrema, heat-capacity divergences, and global phases, while the dark-matter (DM) halo retains its logarithmic imprint from the previous model.

For compactness, we define
\begin{align}
\mathcal{H}(r)&\equiv{}_2F_1\!\left(-\frac12,-\frac14;\frac34;-\frac{r^4}{b^4}\right),\\
\mathcal{G}(r)&\equiv{}_2F_1\!\left(\frac12,\frac34;\frac74;-\frac{r^4}{b^4}\right),  
\end{align}
and $z(r)\equiv-\,r^4/b^4$. Throughout, we work in geometrized units $8\pi G=c=1$.

\subsection{Horizon data and primary thermodynamic quantities}

We begin by extracting the basic thermodynamic variables at the event horizon. The largest real root of $f(r)$ defines the horizon radius $r_h$. Imposing $f(r_h)=0$ solves the mass parameter $M$ (interpreted as enthalpy in the extended framework) in terms of $r_h$.

\paragraph{Mass, temperature, entropy, and volume.}
With the full lapse of Eq.~\eqref{function-1}, the horizon condition gives
\begin{equation}
M(r_h)=\frac{r_h}{2}\!\left[
1-8\pi\rho_s r_s^2\ln\!\Big(1+\frac{r_s}{r_h}\Big)
+\frac{|\alpha|\,b^2}{r_h^2}\,\mathcal{H}(r_h)
+\frac{r_h^2}{\ell_p^2}
\right].
\label{eq:Mass_full}
\end{equation}

\begin{figure*}[tbhp]
\centering
\includegraphics[width=0.40\linewidth]{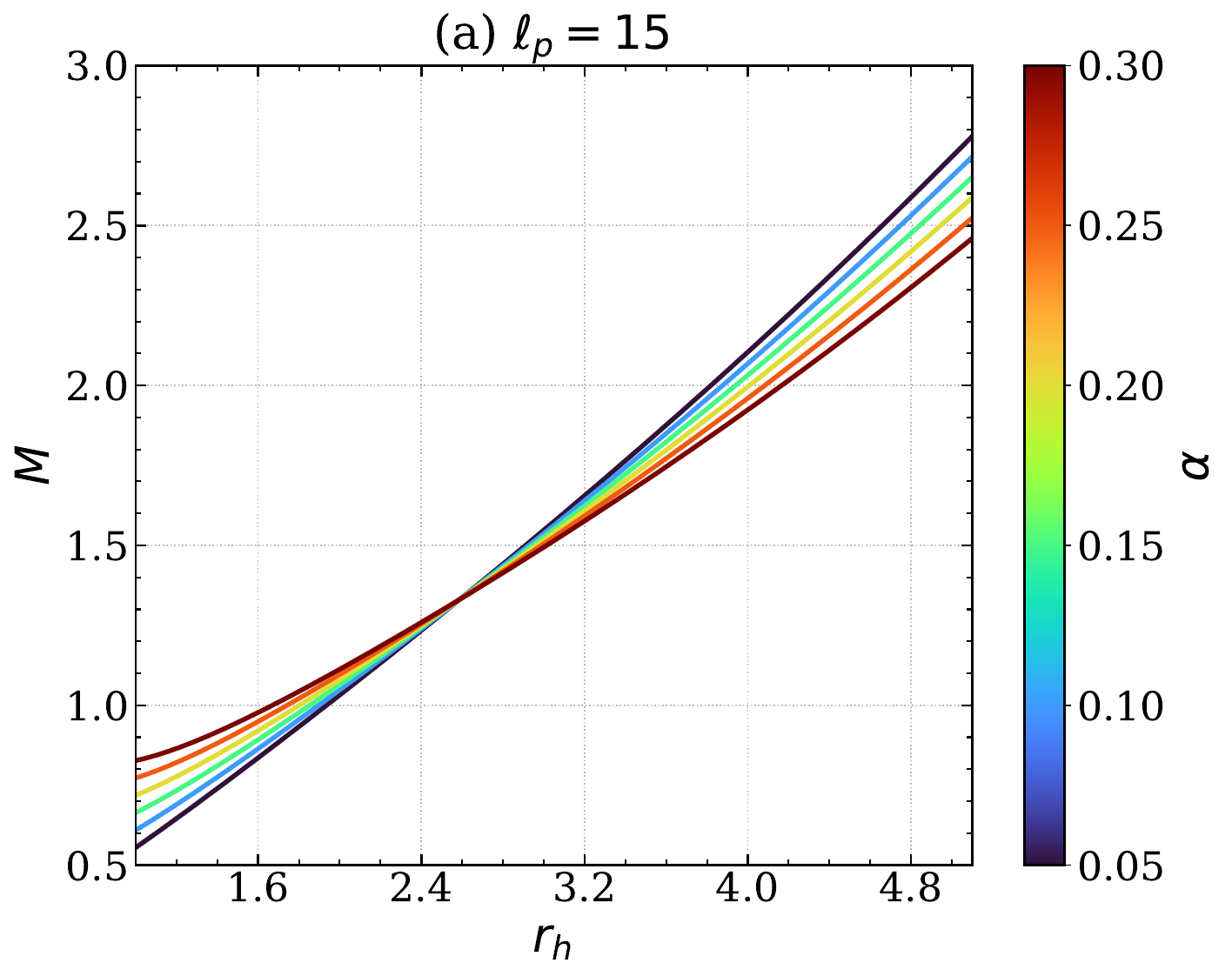}\qquad
\includegraphics[width=0.40\linewidth]{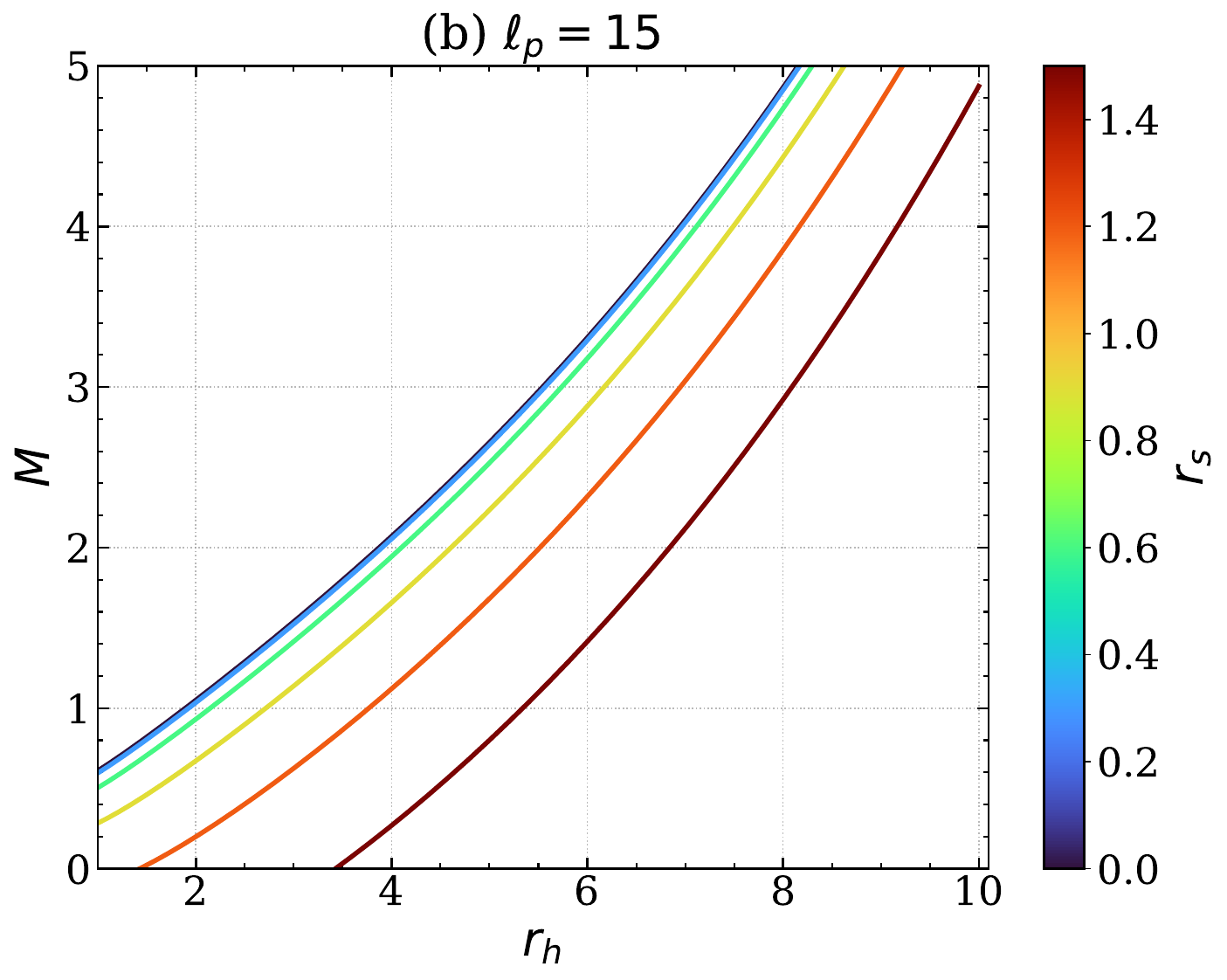}
\includegraphics[width=0.40\linewidth]{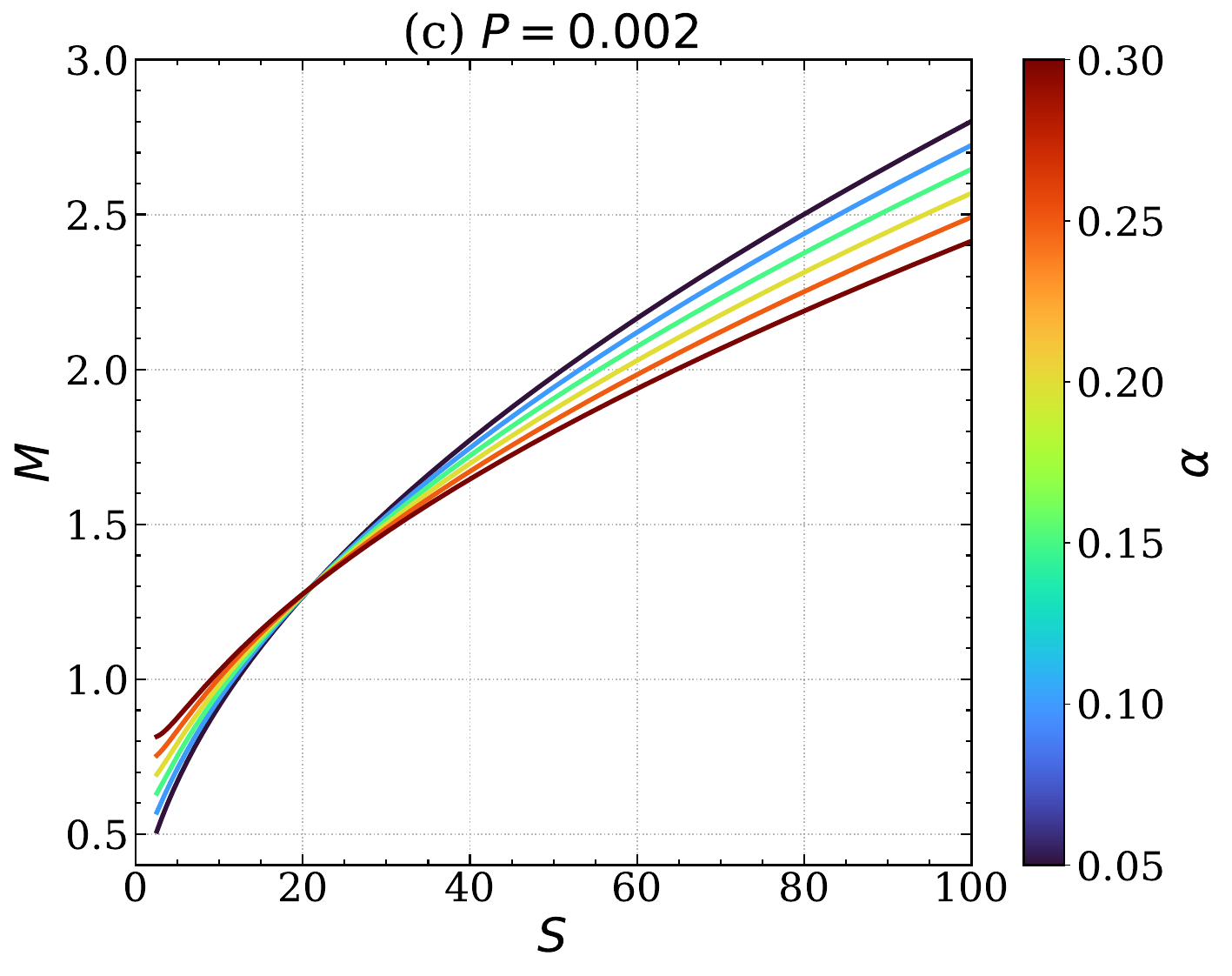}\qquad
\includegraphics[width=0.40\linewidth]{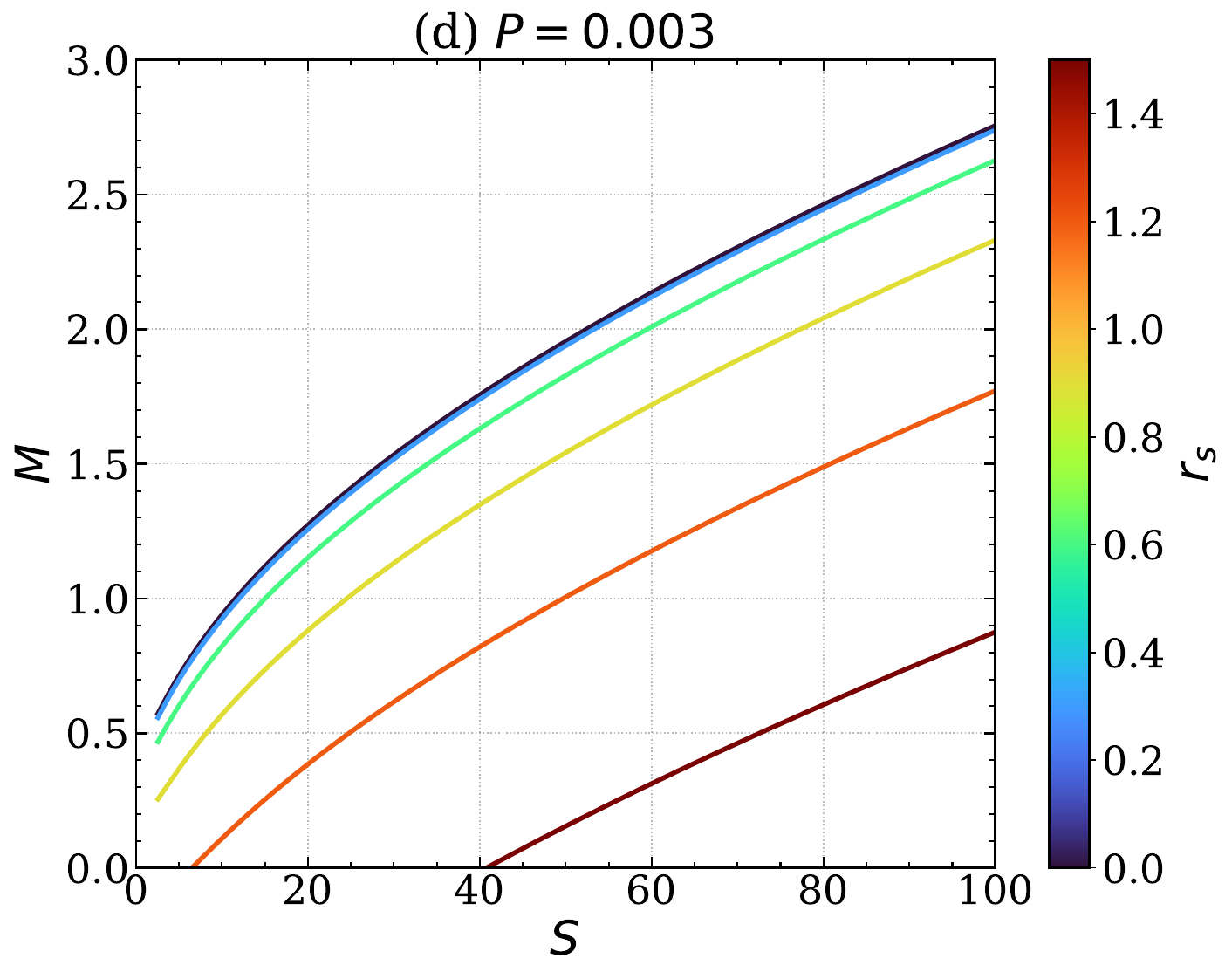}
\caption{\footnotesize
Black-hole enthalpy $M$ for a Schwarzschild-AdS black hole surrounded by a Dehnen-type dark matter halo and a cloud of strings (full model with hypergeometric NCS term; $ 8\pi = 1$). Top row: $M$ vs.\ horizon radius $r_h$; bottom row: $M$ vs.\ entropy $S=\pi r_h^2$. We fix the NCS scale to $b=1.5$.
(a) $\ell_p=15$, $r_s=0.2$, $\rho_s=0.02$, $\alpha\in\{0.05,0.10,0.15,0.20,0.25,0.30\}$.
(b) $\ell_p=15$, $\alpha=0.1$, $\rho_s=0.05$, $r_s\in\{0.0,0.3,0.6,0.9,1.2,1.5\}$.
(c) $P=0.002$ (so $\ell_p=\sqrt{3/P}$), $r_s=0.2$, $\rho_s=0.02$, $\alpha$ as in (a).
(d) $P=0.003$ (so $\ell_p=\sqrt{3/P}$), $\alpha=0.1$, $\rho_s=0.05$, $r_s\in\{0.0,0.3,0.6,0.9,1.2,1.5\}$.
In (a,c) the string-cloud parameter $\alpha$ is scanned at fixed $(r_s,\rho_s)$; in (b)-(d), the halo scale $r_s$ is scanned at fixed $(\alpha,\rho_s)$. The AdS scale is fixed by $\ell_p$ on the top row and by $P=3/\ell_p^2$ on the bottom row.
}
\label{fig:mass-fourpanels}
\end{figure*}

Figure~\ref{fig:mass-fourpanels} shows how the enthalpy responds to the NCS and halo sectors in the full model. 
For fixed $(r_s,\rho_s)$ [panels (a) and (c)], increasing $\alpha$ raises $M$ at small and intermediate radii-due to the positive NCS contribution $\propto |\alpha|\,b^2 r_h^{-2}\,{}_2F_1(-\tfrac12,-\tfrac14;\tfrac34;-r_h^4/b^4)$-while the AdS term dominates at large $r_h$, driving the common cubic growth in $r_h$ and the asymptotic scaling $M\sim P\,S^{3/2}$ when plotted against $S$. 
At fixed $(\alpha,\rho_s)$ [panels (b) and (d)], enlarging the halo scale $r_s$ lowers $M$ in the small-$r_h$ regime through the logarithmic halo piece, with differences gradually suppressed as the AdS contribution $\propto r_h^3/\ell_p^2$ takes over.
Overall, the trends across the four panels are consistent with the temperature and specific-heat systematics: the NCS stiffens the small-$r_h$ sector, the halo depresses it, and AdS controls the large-$r_h$ rise; plotting at fixed pressure makes the enthalpic character explicit in the $S^{3/2}$ tail. 

The Hawking temperature follows from $\kappa=\tfrac12 f'(r_h)$ as $T=f'(r_h)/(4\pi)$. Using
\[
\frac{d}{dr}\!\left[\frac{b^2}{r^2}\,\mathcal{H}(r)\right]
=-\,\frac{2b^2}{r^3}\,\mathcal{H}(r)\;-\;\frac{2r}{3b^2}\,\mathcal{G}(r),
\]
and eliminating $M$ via \eqref{eq:Mass_full}, one finds the compact, $M$-free form
\begin{align}
T(r_h)
&=\frac{1}{4\pi r_h}\Bigg[
1-8\pi\rho_s r_s^2\ln\!\Big(1+\frac{r_s}{r_h}\Big)
+\frac{8\pi\rho_s r_s^3}{r_h+r_s}\notag\\
&-\;|\alpha|\!\left(\frac{b^2}{r_h^2}\,\mathcal{H}(r_h)
+\frac{2r_h^2}{3b^2}\,\mathcal{G}(r_h)\right)
+\frac{3r_h^2}{\ell_p^2}
\Bigg].
\label{eq:T_full}
\end{align}
The Bekenstein-Hawking entropy and the (geometric) thermodynamic volume are~\cite{Bekenstein1973}
\begin{equation}
S=\pi r_h^2,\qquad V=\frac{4\pi}{3}r_h^3.
\label{eq:S_V}
\end{equation}
Compared to the previous model (where the NCS entered as a constant angular deficit $-\alpha$~\cite{Letelier1979}), the only structural novelty is the $r$-dependent NCS sector controlled by $b$, which also contributes through derivatives to $T$.

\subsection{Extended thermodynamics and first law}

Promoting the cosmological constant $\Lambda=-3/\ell_p^2$ to a thermodynamic pressure $P\equiv 3/\ell_p^2$ turns $M$ into enthalpy and augments the first law with a $V\,dP$ term~\cite{KastorRayTraschen2009,Dolan2011,CveticEtAl2011}. In our case, the NCS parameters $(\alpha, b)$ and the DM halo parameters $(\rho_s, r_s)$ naturally act as additional extensive variables with associated work terms.

\paragraph{First law and conjugate potentials.}
Writing $r_h=\sqrt{S/\pi}$, the enthalpy reads
\begin{align}
M(S,P,\alpha,b,\rho_s,r_s)
&=\frac{1}{2}\sqrt{\frac{S}{\pi}}\Bigg[
1-8\pi\rho_s r_s^2\ln\!\Big(1+r_s\sqrt{\frac{\pi}{S}}\Big)\notag\\
&
+\frac{|\alpha|\,b^2\pi}{S}\,
\mathcal{H}\!\left(\sqrt{\frac{S}{\pi}}\right)
+\frac{P}{3}\,S\Bigg],
\label{eq:M_of_S_full}
\end{align}
with $\mathcal{H}\!\left(\sqrt{S/\pi}\right)={}_2F_1\!\big(-\tfrac12,-\tfrac14;\tfrac34;\,-S^2/(\pi^2 b^4)\big)$.
Treating $(\alpha,b,\rho_s,r_s)$ as thermodynamic variables, the extended first law is
\begin{equation}
dM=T\,dS+V\,dP+\Theta_{\alpha}\,d\alpha+\Theta_{b}\,db+\Theta_{\rho}\,d\rho_s+\Theta_{r_s}\,dr_s,
\end{equation}
with conjugates (holding $S,P$ fixed)
\begin{align}
\Theta_{\alpha}&=\left(\frac{\partial M}{\partial \alpha}\right)_{S,P,b,\rho_s,r_s}
=\frac{b^2}{2r_h}\,\mathcal{H}(r_h),
\label{eq:theta_alpha}\\[2pt]
\Theta_{b}&=\left(\frac{\partial M}{\partial b}\right)_{S,P,\alpha,\rho_s,r_s}
=|\alpha|\left[\frac{b}{r_h}\,\mathcal{H}(r_h)
+\frac{r_h^3}{3b^3}\,\mathcal{G}(r_h)\right],
\label{eq:theta_b}\\[2pt]
\Theta_{\rho}&=\left(\frac{\partial M}{\partial \rho_s}\right)_{S,P,\alpha,b,r_s}
=-\,4\pi r_h r_s^2\ln\!\Big(1+\frac{r_s}{r_h}\Big),\label{eq:theta_rho}
\end{align}
\begin{align}
\Theta_{r_s}=\left(\frac{\partial M}{\partial r_s}\right)_{S,P,\alpha,b,\rho_s}
&=-\,4\pi r_h \rho_s\!\notag\\& \times \left[2r_s\ln\!\Big(1+\frac{r_s}{r_h}\Big)
+\frac{r_s^2}{r_h+r_s}\right].
\label{eq:theta_rs}
\end{align}
By construction $T=(\partial M/\partial S)_{P,\dots}$ reproduces \eqref{eq:T_full} and $V=(\partial M/\partial P)_{S,\dots}=4\pi r_h^3/3$. The NCS/DM sectors thus enter as natural work-like deformations of the standard first law.
\begin{figure*}[tbhp]
\centering
{\includegraphics[width=0.40\linewidth]{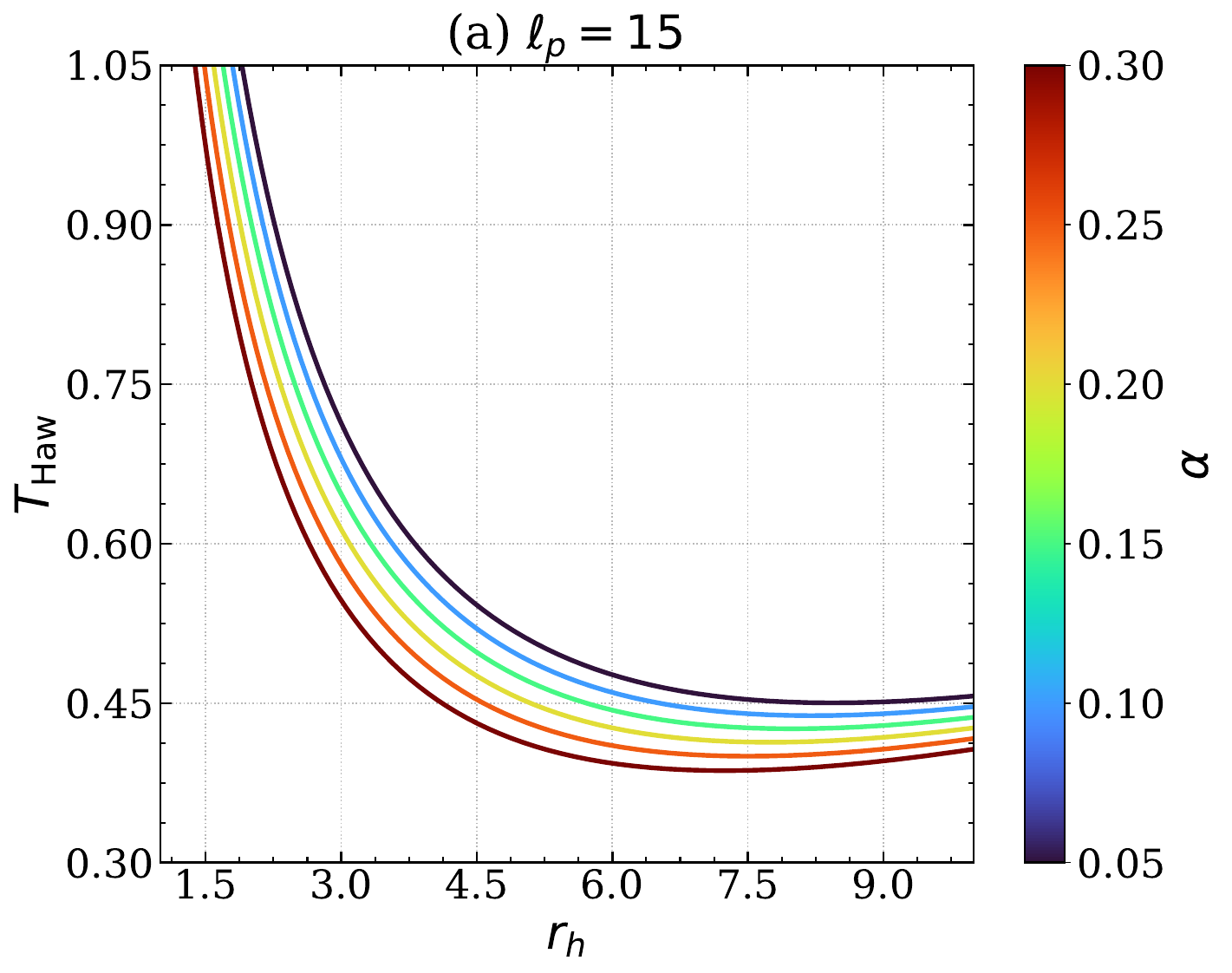}}\qquad
{\includegraphics[width=0.40\linewidth]{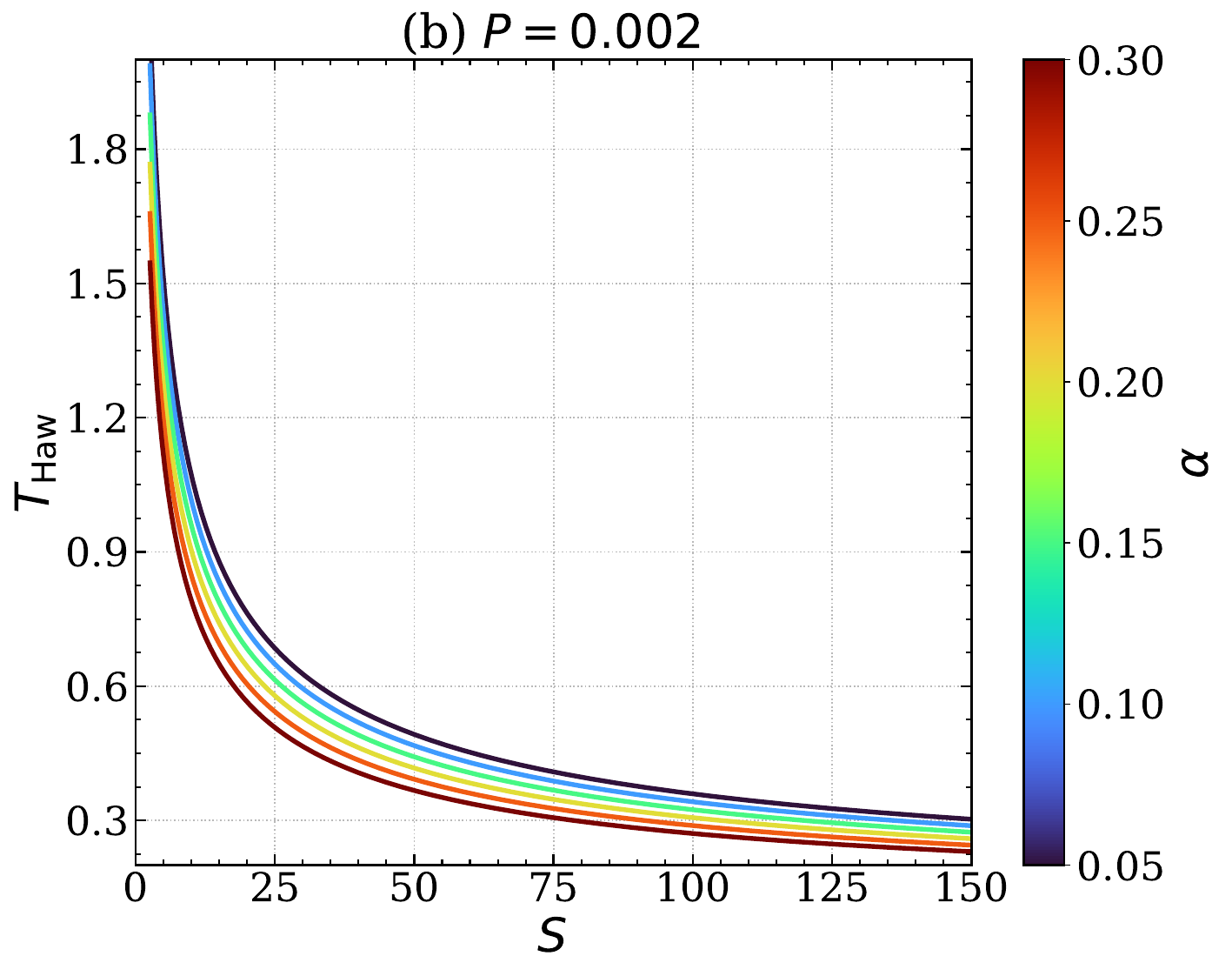}}\\[4pt]
{\includegraphics[width=0.40\linewidth]{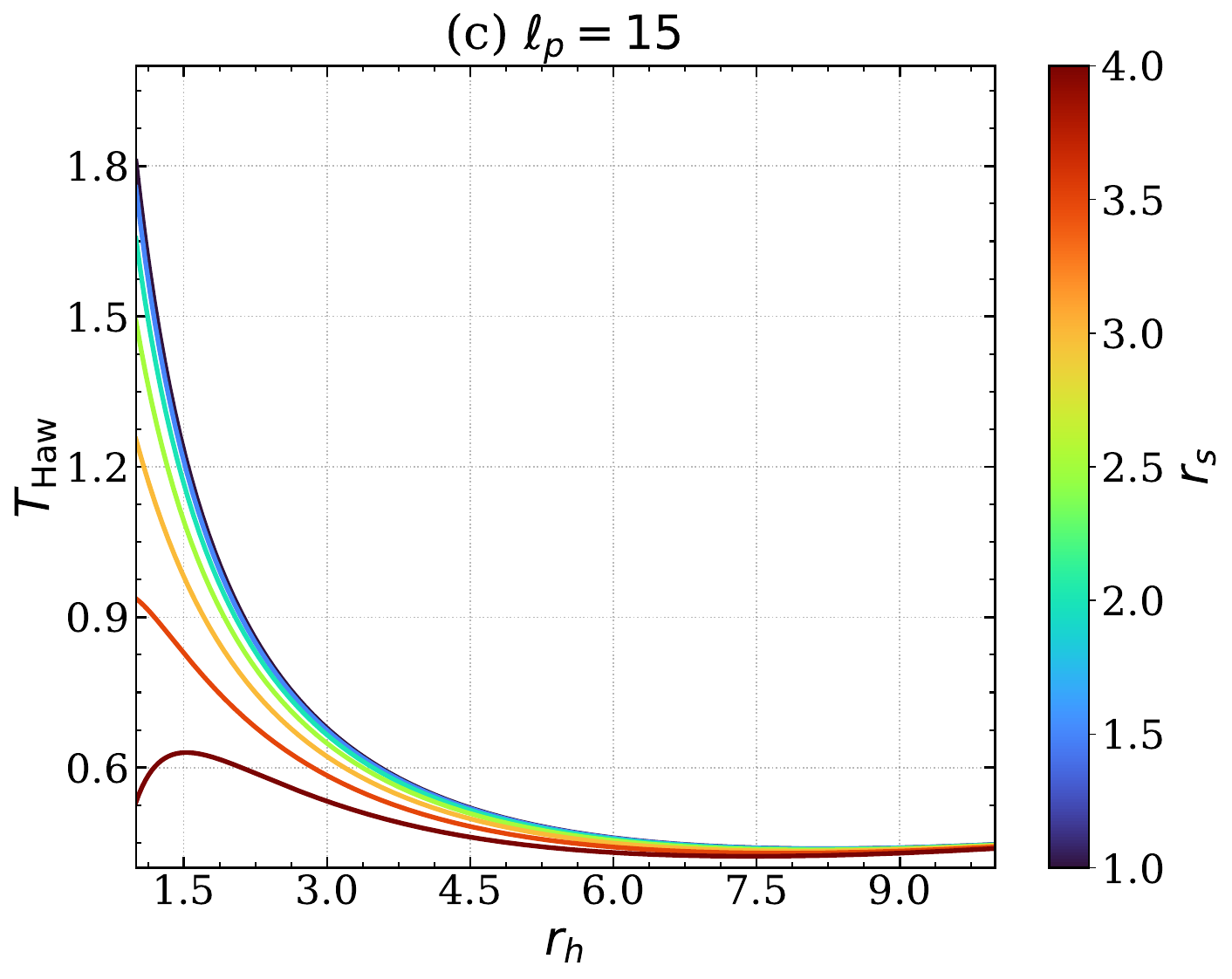}}\qquad
{\includegraphics[width=0.40\linewidth]{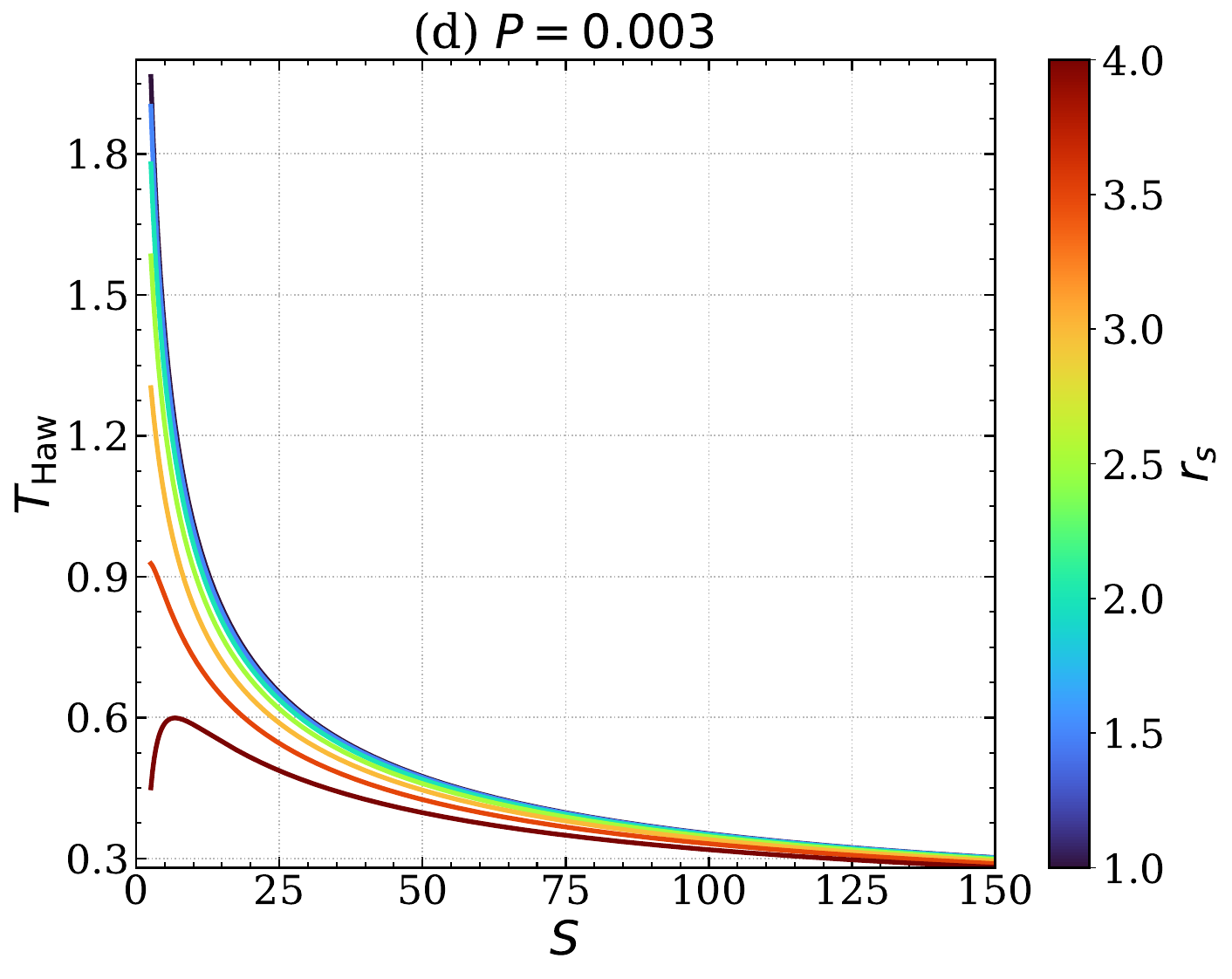}}
\caption{\label{fig:THaw_panels} \footnotesize
Hawking temperature $T_{\rm Haw}$ for a Schwarzschild-AdS black hole with a cloud of strings and a Dehnen-type dark-matter halo, in units with $8\pi=1$ and entropy $S=\pi r_h^2$.
(a) $T_{\rm Haw}$ vs.\ $r_h$ for varying $\alpha\in\{0.05,0.10,0.15,0.20,0.25,0.30\}$ with $r_s=0.2$, $\rho_s=0.02$, and $\ell_p=15$.
(b) $T_{\rm Haw}$ vs.\ $S$ for the same $\alpha$ set with $r_s=0.2$, $\rho_s=0.02$, and fixed pressure $P=0.002$ (so $\ell_p=\sqrt{3/P}$).
(c) $T_{\rm Haw}$ vs.\ $r_h$ for $r_s\in\{1.0,1.5,2.0,2.5,3.0,3.5,4.0\}$ at $\alpha=0.1$, $\rho_s=0.05$, and $\ell_p=15$.
(d) $T_{\rm Haw}$ vs.\ $S$ for the same $r_s$ set at $\alpha=0.1$, $\rho_s=0.05$, and $P=0.003$ (so $\ell_p=\sqrt{3/P}$).
In all panels, we display only the physical branch with $T_{\rm Haw}\ge 0$: each curve starts at a common lower cutoff $r_h\ge r_0$, where $r_0$ is the largest root of $T_{\rm Haw}(r_h)=0$ across the corresponding set of parameters. The colorbar encodes the varied parameter.}
\end{figure*}

\subsection{Hawking temperature and competing scales}

In our units ($8\pi=1$) the Hawking temperature is
\begin{align}
T_{\rm Haw}(r_h)&=\frac{1}{4\pi r_h}\notag\\ &\times\left[
1-\alpha-\rho_s r_s^2\ln\!\Bigl(1+\frac{r_s}{r_h}\Bigr)
+\frac{\rho_s r_s^3}{r_h+r_s}
+\frac{3 r_h^2}{\ell_p^2}
\right].
\label{eq:THaw}
\end{align}
The string cloud and halo terms depress the temperature, whereas the AdS term enforces a linear rise at large $r_h$ ($T_{\rm Haw}\sim (3/4\pi\ell_p^2)\,r_h$). The zero of the bracket defines an extremal radius $r_0$ with $T_{\rm Haw}(r_0)=0$; below $r_0$ the branch is non-thermal, so Figs.~\ref{fig:THaw_panels}(a)-(d) only display $T_{\rm Haw}\ge0$. In Fig.~\ref{fig:THaw_panels}(a) (fixed $\ell_p=15$) and Fig.~\ref{fig:THaw_panels}(b) (fixed $P=0.002$, hence $\ell_p=\sqrt{3/P}$), increasing $\alpha$ uniformly lowers $T_{\rm Haw}$ across the plotted range. The curves can be monotonic or develop a shallow minimum where $dT_{\rm Haw}/dr_h=0$; this turning point signals the locus where the specific heat $C_P$ diverges. In panel (b), plotted against $S=\pi r_h^2$, the AdS contribution makes the large-$S$ tail grow almost linearly, while the matter sector reshapes the small-$S$ region. Figures~\ref{fig:THaw_panels}(c) ($\ell_p=15$) and \ref{fig:THaw_panels}(d) ($P=0.003$) isolate the halo-scale effect. Larger $r_s$ enhances the logarithmic term near the horizon, depressing the small-$r_h$ branch, delaying the thermal onset to $r_h\gtrsim r_0$, and making a turning point more likely. At fixed pressure (panel d), the large-$S$ growth is again governed by AdS, whereas changes in $r_s$ mainly affect the low-$S$ portion. The ``change of direction’’ seen in several curves, either a U-shaped profile or a near-extremal onset at the left edge, is expected from the competition between (i) the string-cloud deficit ($\propto\alpha$), (ii) the halo attraction ($\rho_s,r_s$), and (iii) the AdS heating. Extrema of $T_{\rm Haw}$ track spinodal lines ($dT_{\rm Haw}/dr_h=0$) where $C_P$ diverges, anticipating the stability swap between small and large black holes and the swallow-tail structure of the Gibbs free energy discussed later.

\begin{figure*}[tbph]
  \centering
  \begin{minipage}[c]{0.25\linewidth}
    \includegraphics[width=\linewidth]{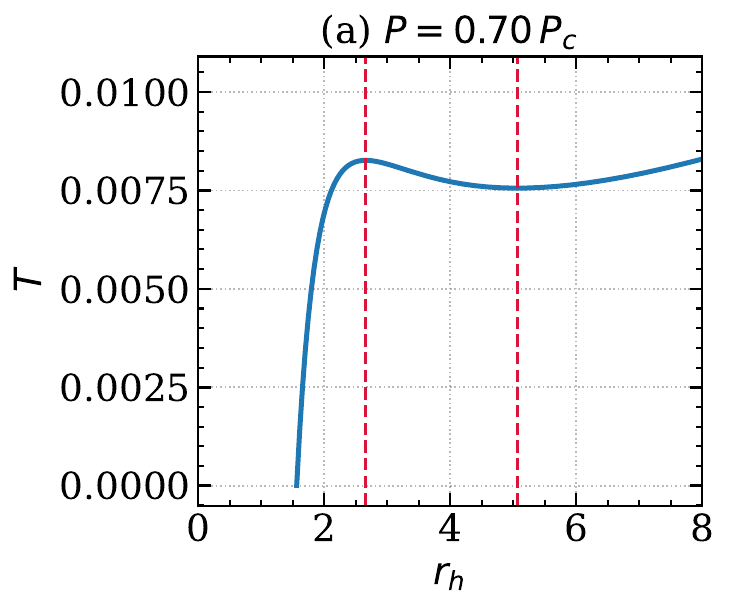}
  \end{minipage}\qquad
  \begin{minipage}[c]{0.25\linewidth}
    \includegraphics[width=\linewidth]{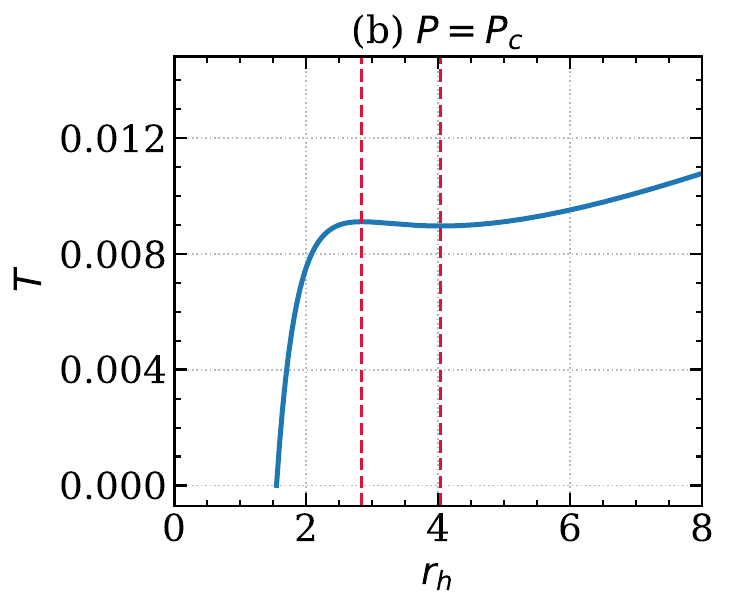}
  \end{minipage}\qquad
  \begin{minipage}[c]{0.25\linewidth}
    \includegraphics[width=\linewidth]{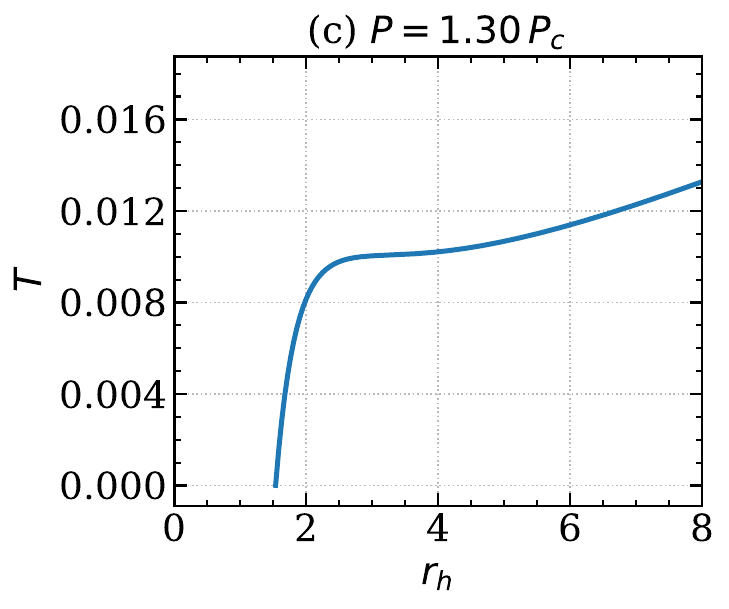}
  \end{minipage}
  \caption{\footnotesize
  Horizon temperature as a function of the horizon radius, $T(r_h)$, at three fixed pressures of the full model (hypergeometric NCS + Dehnen halo + AdS). 
  Critical point for this parameter set: $r_h^\ast=4.0384$, $T_c=0.00896526$, $P_c=0.01304413$. 
  Panels: (a) $P=0.70\,P_c$ (subcritical), (b) $P=P_c$ (critical), (c) $P=1.30\,P_c$ (supercritical). 
  Vertical dashed red lines indicate the spinodal loci where $dT/dr_h=0$ (divergences of the constant–pressure heat capacity $C_P$); for $P<P_c$ (a) two such lines delimit the multivalued region in which a given $T$ corresponds to small/unstable/large black–hole branches, at $P=P_c$ (b) they merge at an inflection point, and for $P>P_c$ (c) no turning point remains and the curve is monotonic. 
  The horizontal axis is restricted to $r_h\le 8$ in all panels.}
  \label{fig:T_vs_rh_threeP}
\end{figure*}

Figure~\ref{fig:T_vs_rh_threeP} makes explicit the relation between horizon kinematics and canonical phases. In the subcritical case [panel (a)], $T(r_h)$ develops two turning points with $dT/dr_h=0$, so that a given temperature intersects the curve three times (small/intermediate/large black holes). The constant–pressure heat capacity satisfies $C_P \propto (dT/dr_h)^{-1}$ along a fixed–$P$ trajectory; hence the outer branches are locally stable ($C_P>0$) while the middle branch is locally unstable ($C_P<0$). At the critical pressure [panel (b)], the two turning points merge into an inflection, $dT/dr_h=0=d^2T/dr_h^2$, where $C_P$ diverges and the small/large branches become indistinguishable. Above criticality [panel (c)], $T(r_h)$ is monotonic and only one stable branch survives. These spinodal loci coincide with the joints where line styles change in the corresponding $G(T)$ plots and underpin the emergence/vanishing of the swallow–tail near $P_c$.

\subsection{Equation of state and criticality}

To analyze phase structure, it is convenient to isolate the non-AdS contribution entering the temperature and rewrite the equation of state $P=P(T,r_h)$.

\paragraph{Isotherms and inflection-point conditions.}
Define
\begin{align}
\mathcal{A}(r)&\equiv
1-8\pi\rho_s r_s^2\ln\!\Big(1+\frac{r_s}{r}\Big)
+\frac{8\pi\rho_s r_s^3}{r+r_s}
\notag\\&-\;|\alpha|\!\left(\frac{b^2}{r^2}\,\mathcal{H}(r)
+\frac{2r^2}{3b^2}\,\mathcal{G}(r)\right),
\label{eq:A_def}
\end{align}
so that \eqref{eq:T_full} becomes $T(r)=\big[\mathcal{A}(r)+P r^2\big]/(4\pi r)$. Hence
\begin{equation}
P(T,r)=\frac{4\pi r\,T-\mathcal{A}(r)}{r^2}.
\label{eq:EoS_full}
\end{equation}
Critical points (if present) satisfy $\left(\partial P/\partial r\right)_T=\left(\partial^2 P/\partial r^2\right)_T=0$, with $\mathcal{A}'(r)$ given by
\begin{align}
\mathcal{A}'(r)
&= \frac{8\pi\rho_s r_s^3}{r(r+r_s)}-\frac{8\pi\rho_s r_s^3}{(r+r_s)^2}
+|\alpha|\Bigg[
\frac{2b^2}{r^3}\,\mathcal{H}(r)
-\frac{2r}{3b^2}\,\mathcal{G}(r)\notag\\
&+\frac{4}{7}\frac{r^5}{b^6}\,{}_2F_1\!\left(\frac{3}{2},\frac{7}{4};\frac{11}{4};-\frac{r^4}{b^4}\right)
\Bigg].
\label{eq:Aprime}
\end{align}
As in charged AdS black holes~\cite{KubiznakMann2012}, first-order small/large-black-hole transitions, if they occur, are governed by an isotherm with an inflection point; the NCS and halo deformations shift both the location and, potentially, the very existence of such criticality.

\begin{figure}[tbhp]
  \centering
  \includegraphics[scale=0.32]{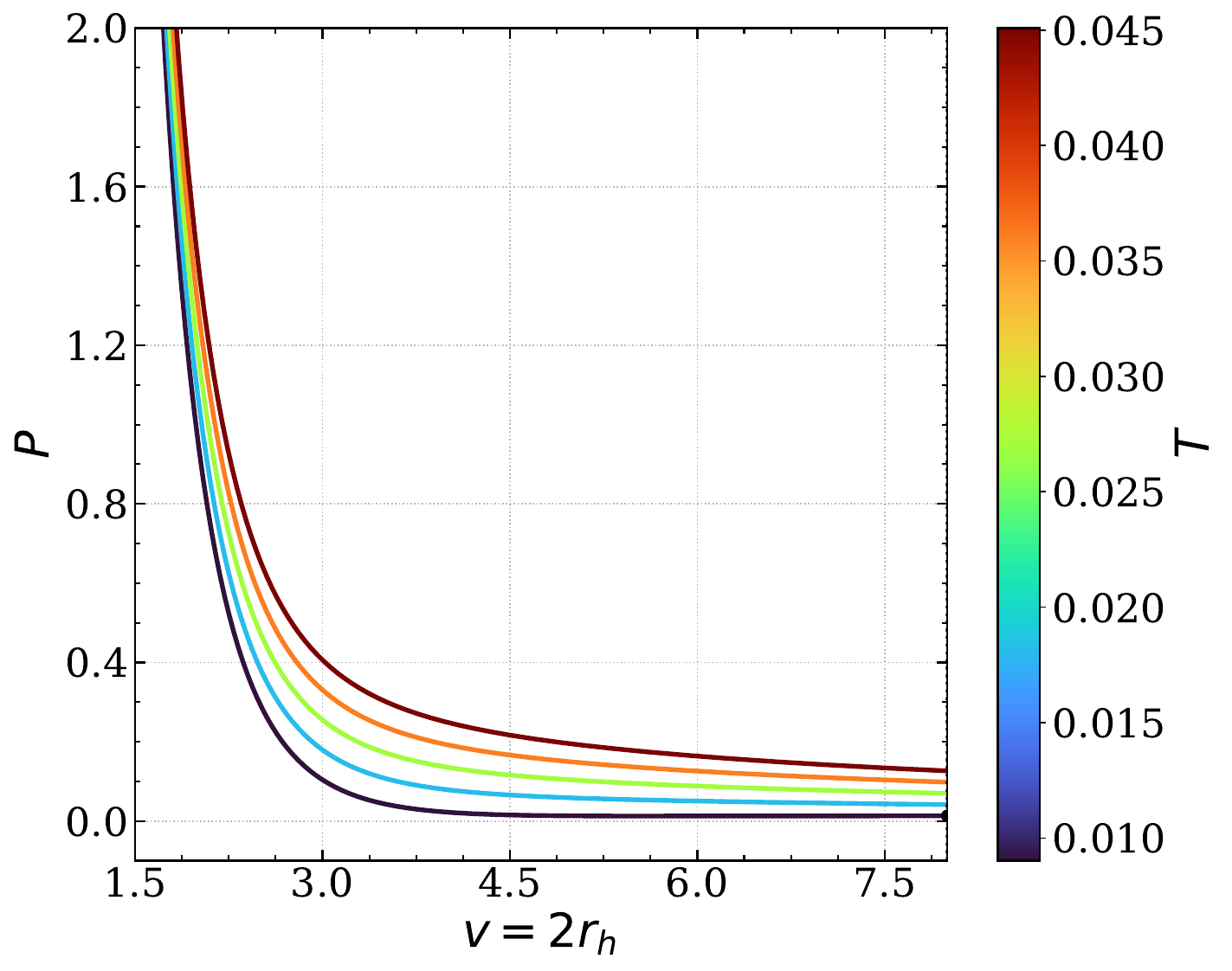}
  \caption{\label{fig:EoS-isotherms} \footnotesize
  Isotherms of the equation of state $P(v)$ for a Schwarzschild-AdS black hole surrounded by a Dehnen-type dark-matter halo and a hypergeometric cloud of strings, in units with $8\pi=1$. The specific volume is $v=2r_h$. Model parameters are fixed to $\alpha=0.75$, $r_s=0.5$, $\rho_s=0.016$, and string-cloud scale $b=1.50$. Colored curves correspond to temperatures taken \emph{relative to the critical temperature} $T_c$ obtained from the inflection-point conditions $\partial_r P=0=\partial_r^2 P$, specifically $T=\{1,2,3,4,5\}\,T_c$. The vertical dotted line marks the critical volume $v^{\ast}=2r_h^{\ast}$ and the black dot indicates $(v^{\ast},P_c)$. Axes ranges emphasize the near-critical region; the colorbar encodes $T$.}
\end{figure}

Within the full model, the isotherms $P(v)$ rise monotonically with $v$ and approach the large-$v$ tail $P\simeq T/v$, as expected from the leading ideal-gas-like contribution of the EoS (Figure \ref{fig:EoS-isotherms}). Because the temperatures shown satisfy $T\ge T_c$, no Van der Waals-type oscillations appear; for $T<T_c$, one would instead see a spinodal segment and a Maxwell construction. The critical point $(v^{\ast},P_c)$, defined by $\partial_r P=0=\partial_r^2 P$ at fixed $T$, is highlighted. Relative to the simplified (deficit-angle) model, the hypergeometric NCS terms together with the halo logarithm make the effective bracket $\mathcal{A}(r)$ less negative in the plotted range, so that $P=(4\pi r T-\mathcal{A})/r^2$ stays positive for all displayed $v$; this explains why the old plot showed predominantly negative $P$ while the present one is fully in $P>0$. Increasing $T$ shifts the isotherms upward and slightly flattens them at large $v$, while the small-$v$ behavior and the location of $(v^{\ast},P_c)$ are controlled by $(r_s,\rho_s)$ and $(\alpha,b)$ through $\mathcal{A}(r)$.

\begin{figure}[tbph]
  \centering
  \includegraphics[width=0.9\linewidth]{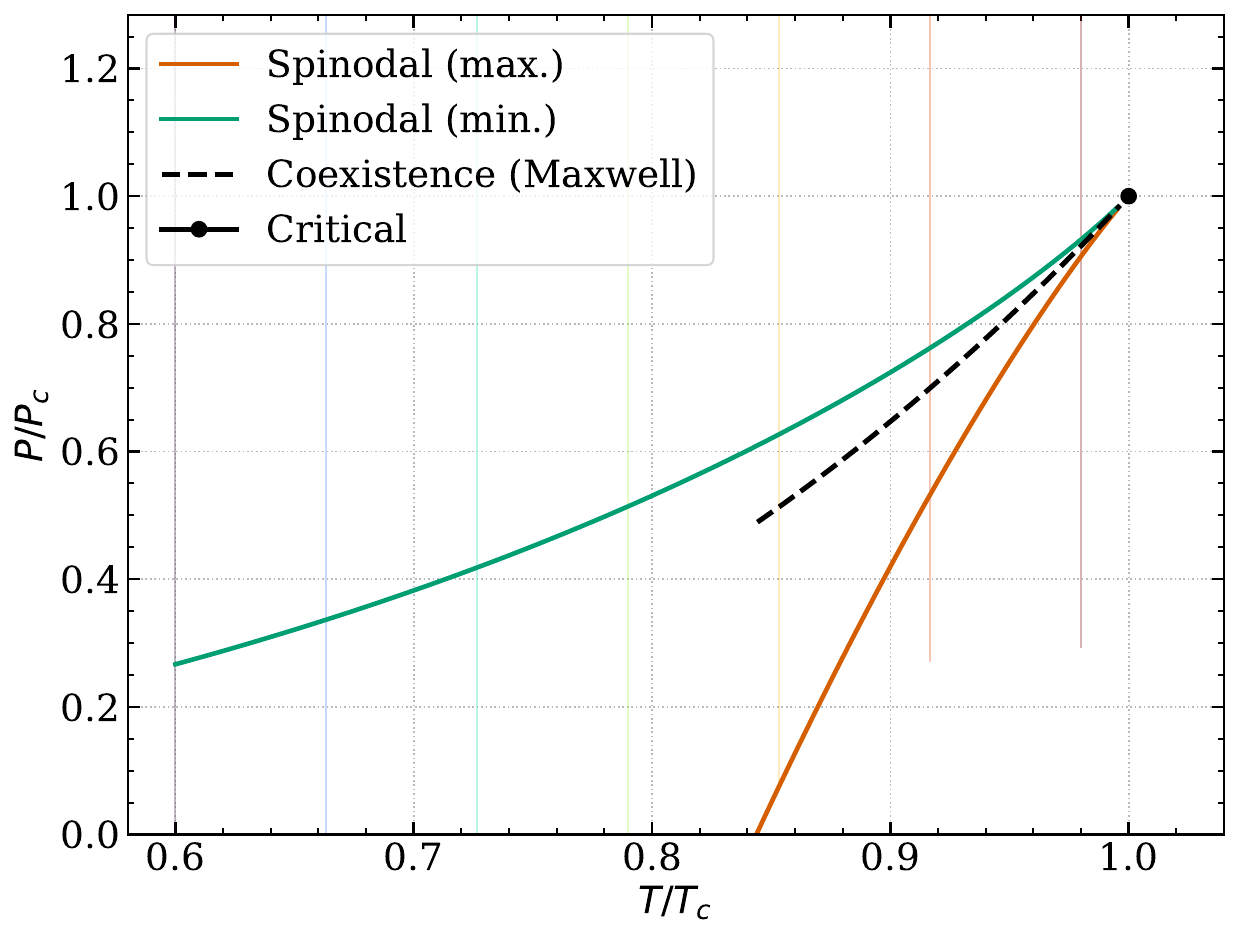}
  \caption{\footnotesize
  Reduced van der Waals $P$–$T$ phase diagram (axes in units of $P_c$ and $T_c$), used here as an orienting map for the canonical black–hole ensemble. The dashed black curve is the liquid–gas coexistence line obtained from the Maxwell equal–area construction (equal Gibbs free energy of the two phases); the solid colored curves bound the spinodal envelope defined by $(\partial P/\partial v)_T\!=\!0$, where mechanical stability is lost. Response functions such as $\kappa_T$ and $C_P$ diverge. The black dot marks the mean–field critical endpoint at $(T/T_c,P/P_c)=(1,1)$ with exponents $\beta\!=\!1/2$, $\gamma\!=\!1$, and $\delta\!=\!3$. Along the coexistence line, the Clapeyron slope $dP/dT\!=\!L/(T\,\Delta v)$ encodes the latent heat $L$ and the volume jump $\Delta v$, which both vanish continuously at criticality.}
  \label{fig:PT_vdw_reduced}
\end{figure}

Pedagogically, Fig.~\ref{fig:PT_vdw_reduced} packages the three geometric ingredients we will reuse for AdS black holes: (i) the Maxwell coexistence locus (black dashed), which in extended thermodynamics tracks the small/large–black–hole first–order transition where $G(T, P)$ has equal branches; (ii) the spinodal envelope (solid) where $(\partial P/\partial v)_T\!=\!0$ and $C_P\!\to\!\infty$, mapping to the $dT/dr_h\!=\!0$ loci delimiting local (in)stability of the black–hole heat capacity; and (iii) the critical point, where the swallow–tail of $G(T)$ shrinks and the order parameter (specific–volume jump, hence horizon–radius jump) vanishes with mean–field scaling. Below $T_c$, the multivalued region between spinodals corresponds to metastable superheated/supercooled phases; above $T_c$, a single supercritical branch persists and sharp features are replaced by broadened response–function peaks (Widom–like crossover), a pattern mirrored by the black–hole ensemble. 

To extract the coexistence line we use Maxwell’s construction on each subcritical isotherm: one determines volumes $v_{\ell}<v_{g}$ such that $P(v_{\ell}, T)=P(v_{g}, T)\equiv P_{\text{coex}}(T)$ and the net area between the isotherm and the plateau $P_{\text{coex}}$ vanishes,
\[
\int_{v_{\ell}}^{v_{g}}\!\big[\,P(v,T)-P_{\text{coex}}(T)\,\big]\,dv=0,
\]
a condition equivalent to equality of Gibbs potentials $\mu_{\ell}(T,P)=\mu_{g}(T,P)$ for the two phases. Along this curve, the Clapeyron relation
\[
\frac{dP}{dT}=\frac{\Delta s}{\Delta v}=\frac{L}{T\,\Delta v}
\]
links the slope $dP/dT$ to the entropy jump $\Delta s=s_{g}-s_{\ell}$ and the specific-volume jump $\Delta v=v_{g}-v_{\ell}$ (with $L=T\,\Delta s$ the latent heat). In the AdS black-hole analogue within extended thermodynamics, we map $v\!\to\!V$ (thermodynamic volume) and $s\!\to\!S$ (Bekenstein–Hawking entropy). Thus,
\[
\left.\frac{dP}{dT}\right|_{\text{BH}}=\frac{\Delta S}{\Delta V},
\]
so the \emph{sign} of the coexistence-line slope directly encodes the relative signs of the jumps $(\Delta S,\Delta V)$ between the small- and large–black-hole branches. In standard cases, $\Delta S>0$ and $\Delta V>0$ (the “large” black hole is more entropic and more voluminous), implying $dP/dT>0$; a negative slope, when it occurs in exotic models, would indicate opposite signs for these jumps.

\subsection{Heat capacity and local stability}

Local thermodynamic stability in the canonical ensemble at fixed pressure is encoded in the sign and divergences of the heat capacity $C_P$.

\paragraph{Spinodal curve and stability windows.}
Using $S=\pi r^2$ and $T(r)=\big[\mathcal{A}(r)+P r^2\big]/(4\pi r)$, we obtain
\[
\left(\frac{\partial T}{\partial r}\right)_P=\frac{1}{4\pi r^2}\left[-\mathcal{A}(r)+r\,\mathcal{A}'(r)-P r^2\right],
\]
and thus
\begin{equation}
C_P
=T\left(\frac{\partial S}{\partial T}\right)_P
=-\frac{\,2\pi r_h^2\,\left(\mathcal{A}(r_h)+P r_h^2 \right)}{\mathcal{A}(r_h)+P r_h^2 - r_h\big[\mathcal{A}'(r_h)+2P r_h\big]}.
\label{eq:Cp_full}
\end{equation}
Poles of $C_P$ mark the spinodal curve separating locally stable ($C_P>0$) and unstable ($C_P<0$) branches. The added $b$-controlled NCS structure and the halo logarithm shift these poles relative to the simpler Schwarzschild-AdS pattern.

\begin{figure*}[tbhp]
\centering
\includegraphics[width=0.40\linewidth]{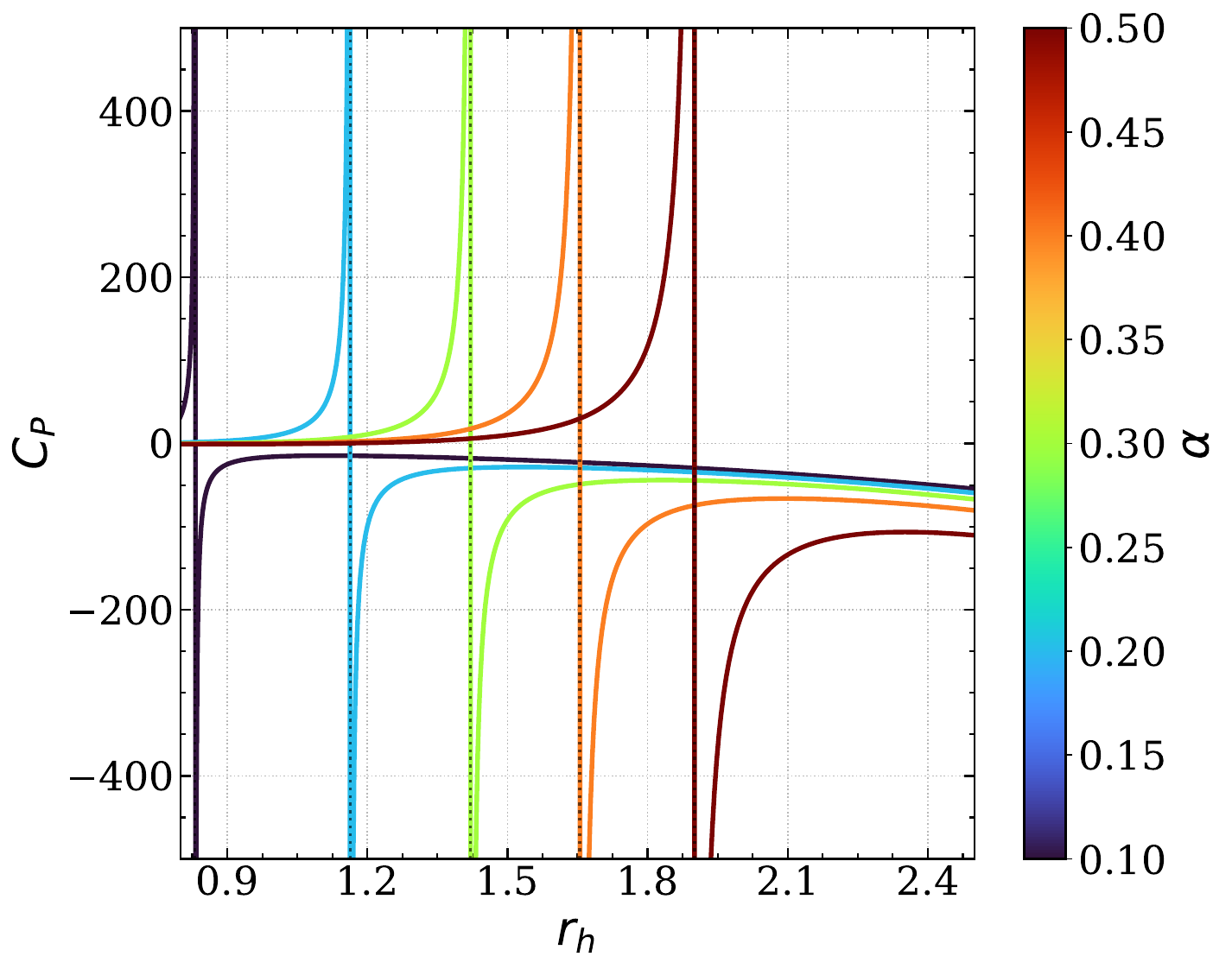}\qquad
\includegraphics[width=0.40\linewidth]{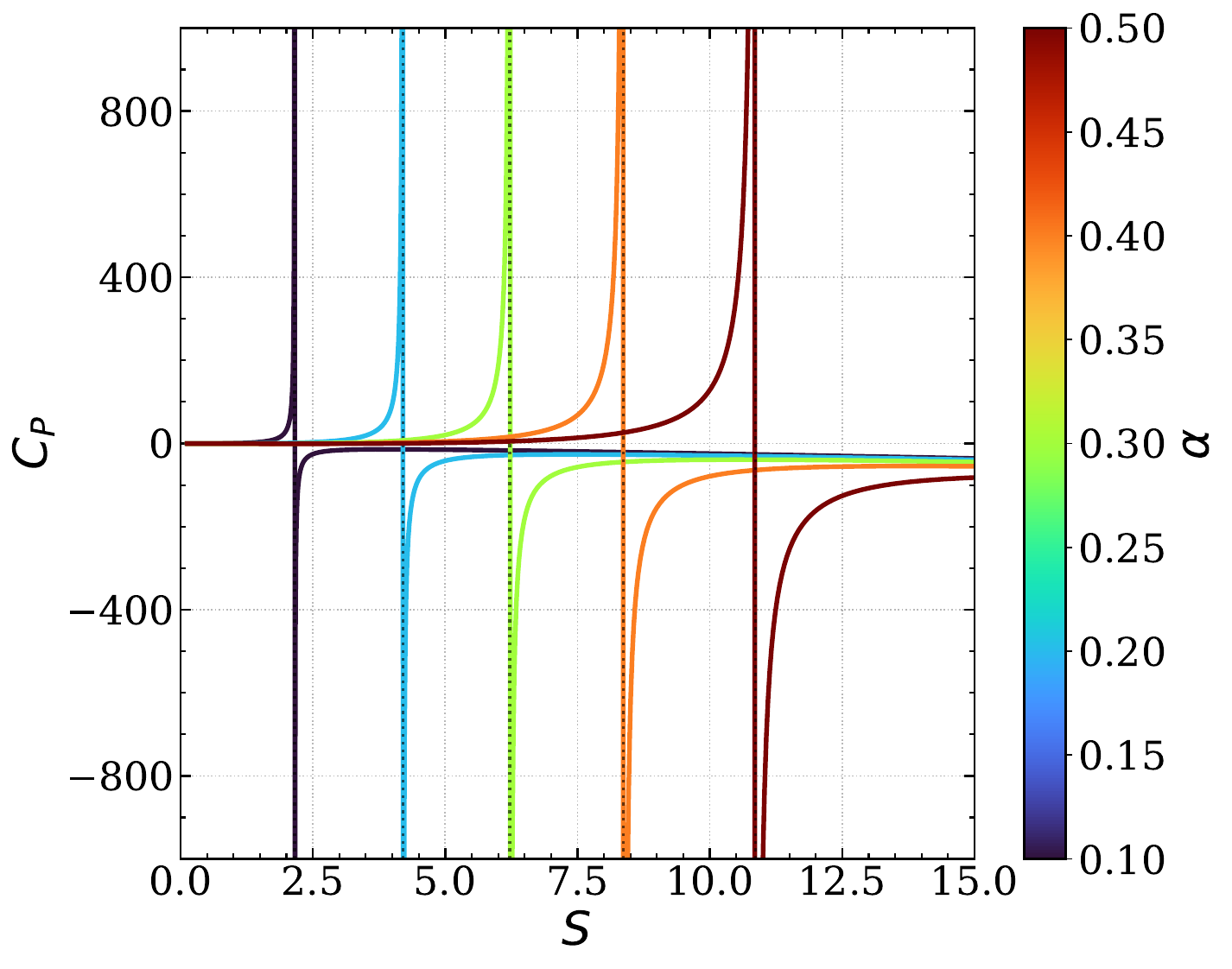}
\caption{\footnotesize
Specific heat at constant pressure $C_P$ for a Schwarzschild-AdS black hole surrounded by a Dehnen-type dark-matter halo and a hypergeometric cloud of strings, in units with $8\pi=1$ and entropy $S=\pi r_h^2$. 
(a) $C_P$ as a function of the horizon radius $r_h$ at fixed AdS length $\ell_p=12$, with halo parameters $r_s=0.5$ and $\rho_s=0.015$ and NCS scale $b=1.50$, varying the string-cloud parameter $\alpha\in\{0.10,0.20,0.30,0.40,0.50\}$. 
(b) $C_P$ as a function of the entropy $S$ at fixed pressure $P=0.012$ (so that $\ell_p=\sqrt{3/P}\simeq 15.81$), again with $r_s=0.5$, $\rho_s=0.015$, and $b=1.50$, scanning the same set of $\alpha$ values. 
Vertical dotted lines mark the divergences of $C_P$, i.e.\ the roots of $dT/dr_h=0$, which separate locally stable ($C_P>0$) and unstable ($C_P<0$) branches.}
\label{fig:cp-alpha}
\end{figure*}

Figures \ref{fig:cp-alpha}(a)-(b) show the constant-pressure heat capacity $C_P$ as functions of $r_h$ and $S=\pi r_h^2$, respectively, for the full model including the Dehnen halo and the hypergeometric NCS sector. The halo parameters are fixed at $(r_s,\rho_s)=(0.5,0.015)$ and the NCS scale at $b=1.50$; the scanned curves correspond to $\alpha\in[0.10,0.50]$. The dotted vertical lines indicate the loci where $dT/dr_h=0$, at which $C_P$ diverges and the local thermodynamic branch changes stability. As $\alpha$ increases, the NCS contribution effectively lowers the near-horizon temperature and shifts the divergence positions along the horizontal axis, reshaping the windows with $C_P>0$ (locally stable) and $C_P<0$ (locally unstable). At large $r_h$ (or large $S$), the AdS term dominates and $C_P$ approaches a positive, slowly varying regime, while the matter-sector deformations (halo and NCS) control the structure and location of the near-critical region. The qualitative pattern is consistent with the standard Van der Waals-like picture: a small-black-hole branch with negative $C_P$, a critical region where $C_P$ blows up, and a large-black-hole branch with positive $C_P$ for sufficiently large $r_h$ (or $S$).

\subsection{Gibbs free energy and Hawking-Page transition}

Global thermodynamic preference at fixed $P$ is captured by the Gibbs free energy $G=M-TS$, whose sign change identifies the Hawking-Page (HP) line~\cite{HawkingPage1983}.

\paragraph{Swallow-tails and HP shift.}
Combining \eqref{eq:Mass_full} and \eqref{eq:T_full} yields
\begin{align}
&G(r_h;P,\alpha,b,\rho_s,r_s)
=\frac{r_h}{4}\Bigg[
\,1-8\pi\rho_s r_s^2\ln\!\Big(1+\frac{r_s}{r_h}\Big)
-\frac{8\pi\rho_s r_s^3}{r_h+r_s}
\notag\\
&\quad
+\,|\alpha|\!\left(\frac{3b^2}{r_h^2}\,\mathcal{H}(r_h)
+\frac{2r_h^2}{3b^2}\,\mathcal{G}(r_h)\right)
-\frac{P}{3}\,r_h^2
\Bigg].
\label{eq:G_full}
\end{align}
For $\alpha=\rho_s=0$, one recovers $G=\tfrac{r_h}{4}(1-P r_h^2/3)$ and the HP point at $r_h=\ell_p$~\cite{HawkingPage1983}. The NCS/DM contributions lower $G$ by a finite amount at fixed $r_h$, shifting the HP temperature and displacing any swallow-tail structure in the $(T, P)$ plane, in line with the black-hole chemistry picture~\cite{KubiznakMann2012}.

\begin{figure*}[tbhp]
  \centering
  \includegraphics[width=0.8\linewidth]{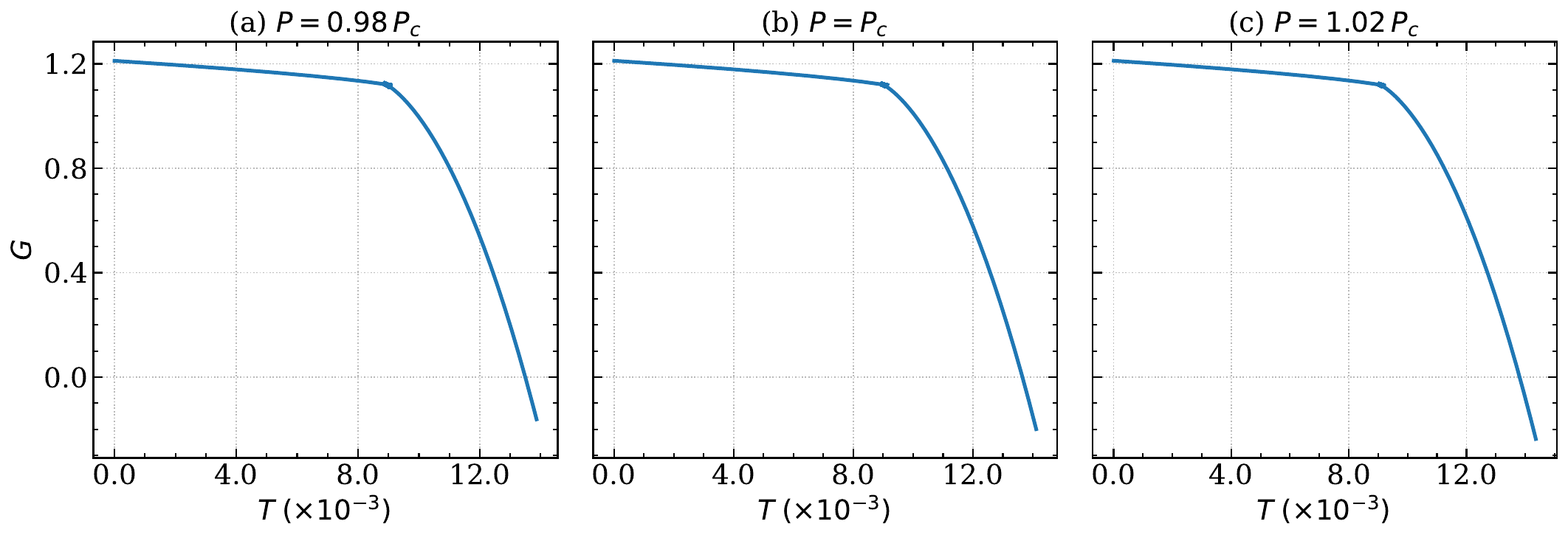}
  \caption{\footnotesize
  Gibbs free energy $G(T)$ at fixed pressures in the full model (hypergeometric NCS + Dehnen halo + AdS).
  We sample three values around criticality: $P=0.98\, P_c$, $P=P_c$, and $P=1.02\, P_c$.
  Continuous (dashed) segments denote locally stable (unstable) branches determined by the sign of $C_P$.
  Turning points in $T(r_h)$—where $dT/dr_h=0$ and $C_P$ diverges—appear as the joints between line styles.}
  \label{fig:G_vs_T_threeP}
\end{figure*}

\paragraph*{Canonical picture near $P_c$.}
Figure~\ref{fig:G_vs_T_threeP} displays the canonical Gibbs potential across three closely spaced pressures around the critical point.
For $P<P_c$ (left panel), $G(T)$ develops the characteristic \emph{swallow–tail}:
Two locally stable branches (small and large black holes, solid lines) coexist over a finite temperature window, separated by an unstable middle branch (dashed).
The first–order transition temperature is read off from the cusp where the two stable branches exchange global optimality. At $P=P_c$ (middle), the swallow–tail shrinks to a \emph{cusp} with continuous first derivative and divergent $C_P$, marking the second–order endpoint.
For $P>P_c$ (right), the multi–valued structure disappears and only a single, everywhere stable branch remains, so no first–order transition is possible.

\begin{figure*}[tbph]
  \centering
  \includegraphics[width=0.40\linewidth]{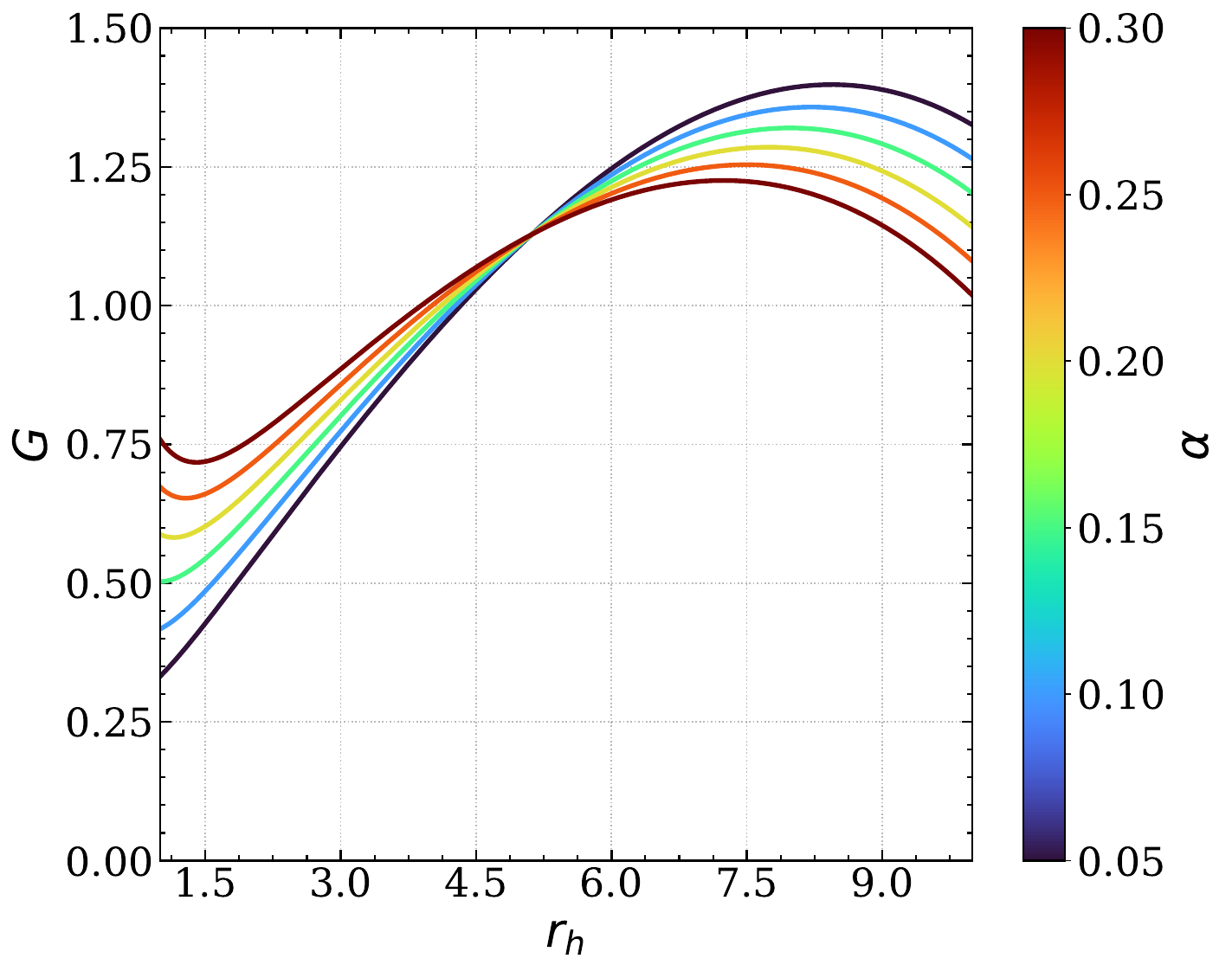}\qquad
  \includegraphics[width=0.40\linewidth]{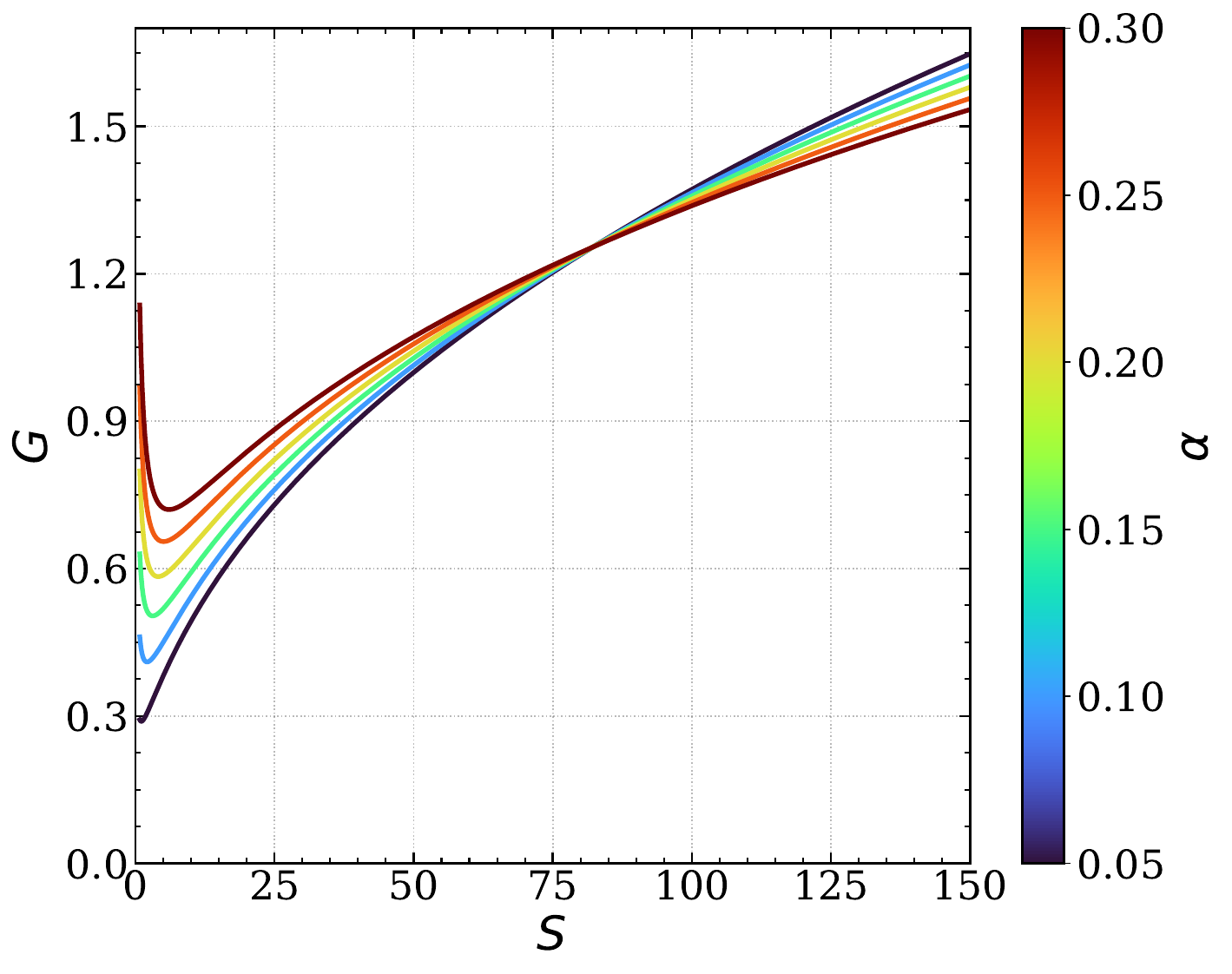}\\[4pt]
  \includegraphics[width=0.40\linewidth]{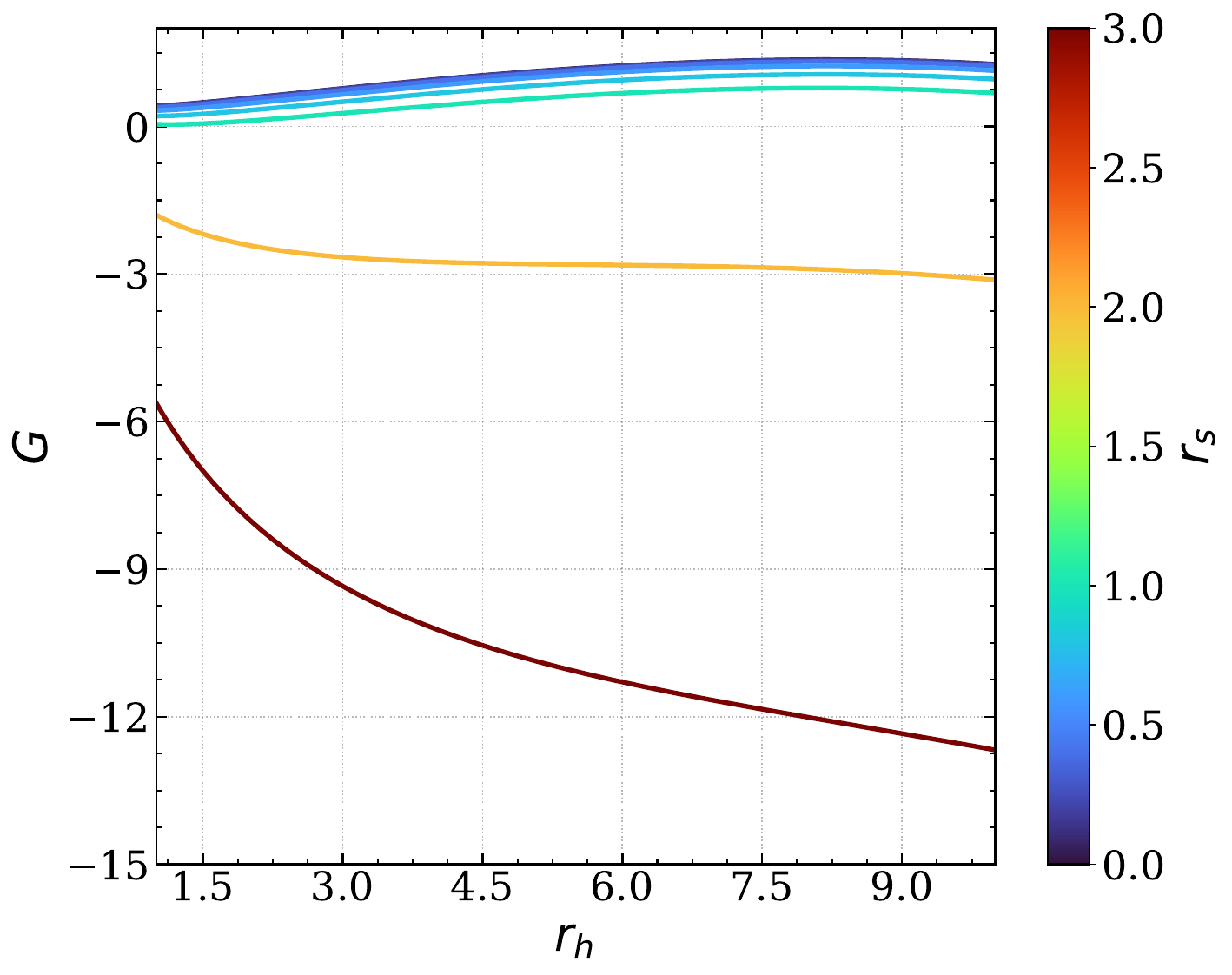}\qquad
  \includegraphics[width=0.40\linewidth]{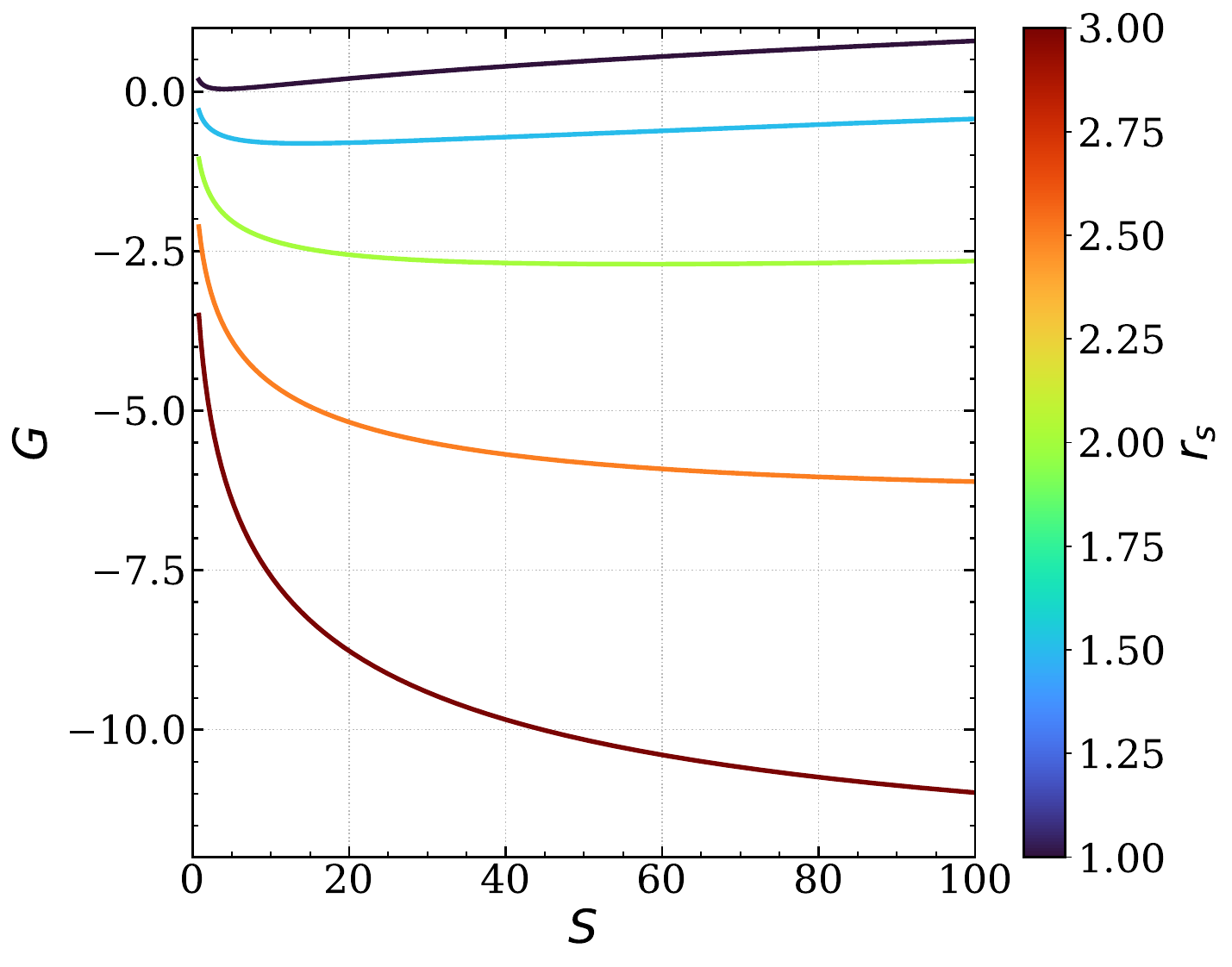}
  \caption{\label{fig:Gibbs-fourpanels}\footnotesize
  Gibbs free energy $G=M-TS$ for the Schwarzschild-AdS black hole with a Dehnen-type dark-matter halo and a cloud of strings in the \emph{complete} model (hypergeometric NCS), using units $8\pi=1$ and $S=\pi r_h^2$. We fix the NCS scale to $b=1.50$.
  (a) $G$ vs.\ $r_h$ at fixed $\ell_p=15$ with $r_s=0.2$ and $\rho_s=0.02$, scanning $\alpha\in\{0.05,0.10,0.15,0.20,0.25,0.30\}$.
  (b) $G$ vs.\ $S$ at fixed pressure $P=0.002$ (so $\ell_p=\sqrt{3/P}$), again with $r_s=0.2$, $\rho_s=0.02$, and the same $\alpha$ set.
  (c) $G$ vs.\ $r_h$ at fixed $\ell_p=15$ with $\alpha=0.1$ and $\rho_s=0.05$, scanning the halo scale $r_s\in\{0.0,0.2,0.4,0.6,0.8,1.0,2.0,3.0\}$.
  (d) $G$ vs.\ $S$ at fixed pressure $P=0.003$ (so $\ell_p=\sqrt{3/P}$) with $\alpha=0.1$, $\rho_s=0.05$, and $r_s\in\{1.0,1.5,2.0,2.5,3.0\}$.
  In all panels, the colorbar encodes the parameter being scanned. The NCS/DM sectors generically lower $G$ at fixed $r_h$ (or $S$), shifting the Hawking-Page crossing $G=0$ and the overall phase structure relative to the pure Schwarzschild-AdS case.}
\end{figure*}

These three regimes match the spinodal structure extracted from $C_P$ in Eq.~\eqref{eq:Cp_full} and from the turning points of $T(r_h)$:
the junctions between solid and dashed segments in Fig.~\ref{fig:G_vs_T_threeP} occur precisely at $dT/dr_h=0$.
The net effect of the hypergeometric NCS sector (controlled by $b$ and $\alpha$) and of the Dehnen halo $(\rho_s,r_s)$ is to shift the location of the spinodal lines and, consequently, the width and position of the swallow–tail. In particular, larger $\alpha$ or larger $r_s$ tend to lower $G$ at fixed $T$, broadening the coexistence interval and pushing the Hawking–Page crossing to lower temperatures, consistently with Figs.~\ref{fig:THaw_panels}–\ref{fig:Gibbs-fourpanels}.

Figure~\ref{fig:T_vs_rh_threeP} displays the horizon–temperature map $T(r_h)$ at three representative pressures of the full model (hypergeometric NCS + Dehnen halo + AdS). In the subcritical case [panel (a)], $T(r_h)$ develops two turning points where $dT/dr_h=0$, so that a given temperature intersects the curve three times. These intersections correspond to a small black hole branch, an intermediate branch, and a large black hole branch. The sign of the constant–pressure heat capacity,
\[
C_P \;=\; T\Big(\frac{\partial S}{\partial T}\Big)_P
\quad\propto\quad \Big(\tfrac{dT}{dr_h}\Big)^{-1},
\]
changes across the turning points: the outer branches have $C_P>0$ (locally stable) and the middle branch has $C_P<0$ (locally unstable).

At the critical pressure [panel (b)], the two turning points merge into an inflection point, with $dT/dr_h=0=d^2T/dr_h^2$. This is the endpoint of the first–order line: $C_P$ diverges there, and the distinction between small and large black holes vanishes continuously. Above criticality [panel (c)], $T(r_h)$ is monotonic, there is no multivalued region, and only a single locally stable branch remains for all $r_h$. These features provide the kinematic origin of the canonical phase structure seen in $G(T)$ at fixed $P$: the swallow–tail for $P<P_c$ arises precisely because $T(r_h)$ is multivalued, its cusp at $P=P_c$ reflects the inflection, and the smooth single–branch behavior for $P>P_c$ follows from the monotonicity of $T(r_h)$. In all cases, the junctions between stability sectors occur at the spinodal loci where $dT/dr_h=0$, i.e., where $C_P$ blows up.

\paragraph{Gibbs free energy.}

Figures~\ref{fig:Gibbs-fourpanels}(a)-(d) summarize the Gibbs free energy $G=M-TS$ as a function of the horizon radius $r_h$ (top row) and the entropy $S=\pi r_h^2$ (bottom row) for representative scans in the string-cloud parameter $\alpha$ and in the halo scale $r_s$.
At fixed $\ell_p$ and fixed $(r_s,\rho_s)$ (Fig.~\ref{fig:Gibbs-fourpanels}(a)), increasing $\alpha$ lowers the entire $G(r_h)$ curve, with the most significant shifts concentrated at small and intermediate radii where the cloud-of-strings (NCS) contribution competes most effectively with the dark-matter (DM) bracket and with the subleading AdS term; as a result, the zero of $G$ (Hawking-Page point) moves to smaller $r_h$ and the near-horizon slope becomes less steep.  In several $\alpha$-curves, $G(r_h)$ displays a shallow minimum before joining the asymptotic AdS branch, an imprint of the familiar small/large-BH competition that will reappear in the canonical ensemble at fixed $P$.  These tendencies are clearly visible in the color-coded family of curves in panel (a). 

When the same $\alpha$-scan is redone at fixed pressure and plotted as $G(S)$ [Fig.~\ref{fig:Gibbs-fourpanels}(b)], the large-$S$ sector is dominated by the AdS contribution, producing the characteristic linear descent of $G$ with $S$, while NCS/DM deformations reshape the small-$S$ portion: one often finds a local maximum-minimum pair, so that $G(S)$ can cross the horizontal axis more than once as $\alpha$ varies.  This multi-crossing pattern heralds the emergence (or sharpening) of a swallow-tail in the usual $G$-$T$ representation at fixed $P$, i.e., a first-order transition between small and large black holes over a finite temperature interval.  In practice, increasing $\alpha$ (at fixed $r_s,\rho_s$) tends to make the negative-$G$ portion wider and to shift the Hawking-Page temperature downward.

Figures \ref{fig:Gibbs-fourpanels} (c)-(d) isolate the effect of the halo scale $r_s$ at fixed $(\alpha,\rho_s)$.  At fixed $\ell_p$ (Fig.~\ref{fig:Gibbs-fourpanels}(c)), enlarging $r_s$ depresses $G(r_h)$ most noticeably at small and intermediate radii, again through the logarithmic DM term in the bracket.  For sufficiently large $r_s$, the minimum of $G$ deepens and moves to slightly larger $r_h$, and the sign change $G=0$ is reached earlier along the curve.  In the fixed-$P$ plot $G(S)$ (Fig.~\ref{fig:Gibbs-fourpanels}(d)), the same trend produces a broader interval with $G<0$ and sharper extrema at small $S$, while the linear AdS tail at large $S$ remains essentially unchanged.  In specific windows of $(r_s, P)$, we observe \emph{exotic} profiles in which $G(S)$ develops two local minima separated by a barrier; this opens the door to reentrant behavior as a function of temperature (or pressure), with the thermodynamically preferred branch switching small $\to$ large $\to$ small BH (or vice versa) across successive first-order lines before the ultimate Hawking-Page transition.  The visual signatures of this mechanism, an incipient or well-formed swallowtail and multiple $G=0$ crossings in the low-$S$ region, can be tracked directly in the families of curves shown in Figs. \ref{fig:Gibbs-fourpanels}(c)-(d). 

Across all panels, strengthening the NCS sector (larger $\alpha$) or enlarging the halo ($r_s$) typically makes $G$ more negative at fixed $(r_h, S)$, shifts the Hawking-Page point, and amplifies the nonconvex region responsible for swallow-tail structures in the canonical ensemble.  The interplay among the AdS term (fixing the linear large-$S$ fall), the DM bracket (governing the small-to-intermediate scale), and the NCS contribution (acting as an effective angular deficit) naturally explains the appearance of multiple extrema and multi-crossing patterns in $G$, including the reentrant scenarios highlighted above.

\subsection{Limiting regimes and qualitative trends}

It is instructive to outline the regimes of large and small black holes, where the NCS/DM effects simplify analytically and the physics is most transparent.

\paragraph{(i) Schwarzschild-AdS with constant NCS as a limit.}
Taking $b\to 0$ with $|\alpha|=\alpha$, the hypergeometric combination collapses to a constant,
\[
\frac{|\alpha|\,b^2}{r^2}\,\mathcal{H}(r)\;\longrightarrow\;-\alpha,
\]
and all formulas reduce to those of the previous model with a deficit angle~\cite{Letelier1979}.

\paragraph{(ii) Large black holes ($r_h\gg r_s,b$).}
Halo corrections decay as $r_s^4/r_h^2$, while the NCS terms are suppressed by powers of $b/r_h$, so
$
T\simeq \frac{1}{4\pi}\big(\frac{1}{r_h}+\frac{3r_h}{\ell_p^2}\big)
$
up to small renormalizations. The HP point is only mildly shifted.

\paragraph{(iii) Small black holes ($r_h\ll r_s$).}
The enhancement $\ln(1+r_s/r_h)\sim\ln(r_s/r_h)$ depresses $T$, pushes the $C_P$ pole to larger radii, and delays the onset of $G<0$. These qualitative features mirror the deformations seen in charged or otherwise dressed AdS black holes~\cite{KubiznakMann2012,Johnson2014,OkcuAydiner2017,WeiLiuPRL2015}.

\section{Conclusions and Outlook}\label{Sec:V}
We have investigated a Schwarzschild-AdS black hole embedded in a Dehnen-type dark-matter (DM) halo and dressed by a \emph{new cloud of strings} (NCS) that contributes a scale-dependent hypergeometric correction to the lapse. By combining null and timelike geodesics, photon sphere and shadow diagnostics, a topological light-ring analysis, and extended-phase-space thermodynamics, we identified a coherent set of signatures that distinguish this geometry from both the vacuum AdS case and configurations with only a DM halo or a constant string deficit.

Geometrically, the NCS parameters $(|\alpha|,b)$ and the halo pair $(\rho_s,r_s)$ deform the metric in complementary ways: the NCS introduces a nonlocal, radius-dependent softening of the strong-field potential well, while the Dehnen halo imprints a characteristic logarithmic attraction. These deformations directly influence the effective potentials for null and timelike motion, controlling all derived observables. In the null sector, we found that the photon sphere radius $r_{\rm ph}$ and the shadow radius $R_s=b_c$ increase monotonically with $|\alpha|$ and with the halo scale $r_s$ at fixed background (cf.\ Figs.~\ref{fig:criticals} and Tabs.~\ref{tab:1}-\ref{tab:2}). Physically, strengthening the NCS or enlarging the halo displaces the unstable circular null orbit outward and enlarges the critical impact parameter. Near-critical trajectories obtained from the full orbital equation \eqref{bb13} corroborate this interpretation: increasing $|\alpha|$ systematically enhances the total bending (smaller exit angles) and \emph{increases} the number of whirls around $r_{\rm ph}$ (Fig.~\ref{fig:multi_alpha_orbits}). The topological diagnostic built from $H(r,\theta)$ confirms the presence of a single unstable equatorial light ring whose location shifts smoothly with $(|\alpha|,b,r_s)$, with no spurious stable rings in the static, spherically symmetric sector.

Consistent with these trends, Fig.~\ref{fig:deflection-capture} compiles two complementary diagnostics of null scattering in the full background. Panel (a) shows the deflection angle $\chi(b)$ for several string-cloud couplings $\alpha$, restricted to $b>b_c(\alpha)$: as $b\!\downarrow\!b_c$ the deflection grows sharply, and larger $|\alpha|$ shifts $b_c$ upward and yields systematically larger $\chi(b)$ at fixed $b$. Panel (b) recasts this information into the capture cross-section $\sigma_{\rm cap}=\pi b_c^2$, which increases monotonically with $|\alpha|$ and, at fixed $\alpha$, with the halo scale $r_s$. Both behaviors quantify how the hypergeometric NCS sector and the Dehnen halo deepen the effective potential for null geodesics, thereby enlarging the capture basin and reinforcing the strong-lensing regime.

For timelike motion, the same competition reshapes circular-orbit energetics: the specific energy and angular momentum increase with $|\alpha|$ and $r_s$, and decrease with $b$. As a consequence, the ISCO radius $r_{\rm ISCO}$ grows with $|\alpha|$ and $r_s$ (Tab.~\ref{tab:3}) and diminishes as $b$ increases (Tab.~\ref{tab:4}). These shifts are astrophysically relevant: a larger $r_{\rm ISCO}$ lowers accretion efficiency and displaces characteristic variability and QPO frequencies, providing an independent observational channel to constrain the model, complementary to shadow measurements.

In the extended thermodynamic framework, the cosmological constant plays the role of pressure, and the ADM mass becomes enthalpy. The NCS and halo sectors enter the first law with well-defined conjugate potentials, while the Hawking temperature acquires explicit hypergeometric and logarithmic contributions (Eq.~\eqref{eq:T_full}). The interplay between AdS heating at large $r_h$ and the matter sectors at small $r_h$ depresses $T_{\rm Haw}$ in the near-horizon regime, generates spinodal lines where $C_P$ diverges, and yields a familiar small/significant black-hole first-order transition with a Gibbs swallow-tail terminating at a critical point (Figs.~\ref{fig:THaw_panels}–\ref{fig:T_vs_rh_threeP}). Compared with constant-deficit models, the scale $b$ shifts both the location and even the existence window of criticality by modulating the NCS contribution in the strong-field region.

Taken together, these results offer clear observational leverage. The monotonic trends of $r_{\rm ph}$ and $R_s$ with $(|\alpha|,r_s)$, combined with the systematic behavior of $r_{\rm ISCO}$, motivate joint constraints that fuse EHT-class shadow observables (shadow size and photon-ring structure) with spectroscopy and timing (relativistic line profiles and QPOs), and, potentially, ringdown features governed by the modified effective potential. Our tables and parameter maps delineate broad regions of $(|\alpha|,b,r_s)$ that are compatible with current shadow radii; forthcoming multi-wavelength and multi-messenger observations are expected to tighten these bounds substantially.

Looking ahead, it is natural to extend the analysis to the rotating case to incorporate frame dragging and quantify corrections to $r_{\rm ph}$, $R_s$, and $r_{\rm ISCO}$; to compute quasinormal modes and greybody factors on the full background, connecting dynamical and thermodynamic stability; to develop lensing observables beyond the near-critical regime (strong-deflection angles and time delays); and to confront the model with M87* and Sgr~A* in a Bayesian pipeline that exploits $R_s$, photon-ring diameter and thickness, and astrophysical priors. Overall, the NCS\,+\, Dehnen-halo deformation imprints scale-dependent, observationally testable signatures on both dynamics (shadows, light rings, ISCO) and thermodynamics (spinodals, critical point) of AdS black holes, providing concrete handles to probe non-vacuum strong-field gravity with present and forthcoming data.

\section*{Acknowledgments}

F.A. acknowledges the Inter University Centre for Astronomy and Astrophysics (IUCAA), Pune, India, for granting a visiting associateship. E. O. Silva acknowledges the support from grants CNPq/306308/2022-3, FAPEMA/UNIVERSAL-06395/22, FAPEMA/APP-12256/22, and (CAPES) - Brazil (Code 001).

\section*{Data Availability}

No data were generated or created in this article.

\section*{Conflicts of interest statement}

The authors declare no conflicts of interest.

\bibliographystyle{apsrev4-2}
%

\end{document}